
\input harvmac
\input epsf.tex
\overfullrule=0mm
\newcount\figno
\figno=0
\def\fig#1#2#3{
\par\begingroup\parindent=0pt\leftskip=1cm\rightskip=1cm\parindent=0pt
\baselineskip=11pt
\global\advance\figno by 1
\midinsert
\epsfxsize=#3
{\bf Fig. \the\figno:} #1\par
\endinsert\endgroup\par
}
\def\figlabel#1{\xdef#1{\the\figno}}
\def\encadremath#1{\vbox{\hrule\hbox{\vrule\kern8pt\vbox{\kern8pt
\hbox{$\displaystyle #1$}\kern8pt}
\kern8pt\vrule}\hrule}}

%
\def\tvi{\vrule height 12pt depth 6pt width 0pt}
\def\tv{\tvi\vrule}
\def\ptvi{\vrule height 8pt depth 6pt width 0pt}
\def\ptv{\ptvi\vrule}

\def\frac#1#2{\scriptstyle{#1 \over #2}}

%
%

\def\CP{{\cal P}}

\def\({ \left( }
\def\){ \right) }
%


\def\IR{\relax{\rm I\kern-.18em R}}
\font\cmss=cmss10 \font\cmsss=cmss10 at 7pt
\def\IZ{\relax\ifmmode\mathchoice
{\hbox{\cmss Z\kern-.4em Z}}{\hbox{\cmss Z\kern-.4em Z}}
{\lower.9pt\hbox{\cmsss Z\kern-.4em Z}}
{\lower1.2pt\hbox{\cmsss Z\kern-.4em Z}}\else{\cmss Z\kern-.4em Z}\fi}
\def\inbar{\,\vrule height1.5ex width.4pt depth0pt}
\def\IB{\relax{\rm I\kern-.18em B}}
\def\IC{\relax\hbox{$\inbar\kern-.3em{\rm C}$}}
\def\ID{\relax{\rm I\kern-.18em D}}
\def\IE{\relax{\rm I\kern-.18em E}}
\def\IF{\relax{\rm I\kern-.18em F}}
\def\IG{\relax\hbox{$\inbar\kern-.3em{\rm G}$}}
\def\IH{\relax{\rm I\kern-.18em H}}
\def\II{\relax{\rm I\kern-.18em I}}
\def\IK{\relax{\rm I\kern-.18em K}}
\def\IL{\relax{\rm I\kern-.18em L}}
\def\IM{\relax{\rm I\kern-.18em M}}
\def\IN{\relax{\rm I\kern-.18em N}}
\def\IO{\relax\hbox{$\inbar\kern-.3em{\rm O}$}}
\def\IP{\relax{\rm I\kern-.18em P}}
\def\IQ{\relax\hbox{$\inbar\kern-.3em{\rm Q}$}}
\def\IGa{\relax\hbox{${\rm I}\kern-.18em\Gamma$}}
\def\IPi{\relax\hbox{${\rm I}\kern-.18em\Pi$}}
\def\ITh{\relax\hbox{$\inbar\kern-.3em\Theta$}}
\def\IOm{\relax\hbox{$\inbar\kern-3.00pt\Omega$}}



\def\bv{\bigg\vert}

\def\Ga{\alpha}
\def\Ge{\epsilon}

\def\Gs{\sigma}


\def\dim{{\rm dim\,}}

\def\cotan{{\rm cotan \, }}

\def\encadre#1{\vbox{\hrule\hbox{\vrule\kern8pt\vbox{\kern8pt#1\kern8pt}
\kern8pt\vrule}\hrule}}
\def\encadremath#1{\vbox{\hrule\hbox{\vrule\kern8pt\vbox{\kern8pt
\hbox{$\displaystyle #1$}\kern8pt}
\kern8pt\vrule}\hrule}}

\Title{SPhT/93-125}
{{\vbox {
\bigskip
\centerline{Laughlin's wave functions, Coulomb gases and}
\centerline{expansions of the discriminant.} }}}
\bigskip
\centerline{P. Di Francesco,}
\medskip
\centerline{M. Gaudin,}
\medskip
\centerline{C. Itzykson,}
\medskip
\centerline{and}
\medskip
\centerline{F. Lesage,}

\bigskip

\centerline{ \it Service de Physique Th\'eorique de Saclay
\footnote*{Laboratoire de la Direction des Sciences
de la Mati\`ere du Commissariat \`a l'Energie Atomique.},}
\centerline{ \it F-91191 Gif sur Yvette Cedex, France}

\vskip .5in

In the context of the fractional quantum Hall effect,
we investigate Laughlin's celebrated ansatz for the ground
state wave function at fractional filling of the lowest Landau level.
Interpreting its normalization in terms of a one component plasma,
we find the effect
of an additional quadrupolar field on the free
energy, and derive estimates for the thermodynamically equivalent
spherical plasma. In a second part, we present various methods for
expanding the wave function in terms of Slater determinants,
and obtain sum rules for the coefficients.
We also address the apparently simpler question of counting the number
of such Slater states
using the theory of integral polytopes.

\noindent
\Date{06/93}


\lref\klein{F. Klein, Lectures on the Icosahedron, Dover (N.Y.), 1956.}

\lref\dunne{G.V. Dunne, UCONN-Preprint 1993}

\lref\qhe{R.E. Prange, S.M. Girvin, "The quantum Hall effect",
Springer, New-York, 1987.}
\lref\multi{
J.M. Caillol, D. Levesque, J.J. Weis, J.P. Hansen, J.Stat. Phys. {\bf
28} (1982) 325.

S.W. de Leeuw, J.W. Perram, Physica {\bf A113} (1982) 546.

P. Choquard, J. Cl\'erouin, Phys. Rev. Lett. {\bf 50} (1983) 2086.

A. Alastuey,  B. Jancovici, J. Physique {\bf 42} (1981) 1.

B. Jancovici, Phys. Rev. Lett. {\bf 46} (1981) 386.

A. Alastuey, Annales de Physique {\bf 11} (1986) 653.}

\lref\EVA{M.A. Evarafov, {\it Asymptotic estimates and entire functions},
Gordon and Breach eds., New York (1961) pp. 9-16 .}

\lref\quatre{I.G. Macdonald, "Symmetric functions and Hall polynomials",
Clarendon Press, Oxford (1979)-pp25 formulae (3.4)-(3.5).}

\lref\cinq{L. Comtet, "Analyse combinatoire", Presses Universitaires
de France, Paris (1970), vol. 1, page 181.}

\lref\refsix{F. J. Dyson, J. Math. Phys. {\bf 3} (1962) 140-156,157-165,
166-175.

J. Gunson, J. Math. Phys. {\bf 3} (1962) 752-753.

K. G. Wilson, J. Math. Phys. {\bf 3} (1962) 1040-1043.
}

\lref\refsept{I.J. Good, J. Math. Phys. {\bf 11} (1970) 1884.}

\lref\caillol{J.M. Caillol,D. Levesque, J.J. Weis, J.P. Hansen
J. Stat. Phys. Vol.28 No2 1982 (325)}

\
\newsec{Introduction}

These notes originate from an  attempt to understand the
normalization -- and other properties -- of the many body fermionic
wave functions suggested by Laughlin as candidates for
the ground states of the fractional quantum Hall effect.
Similar quantities appear in the context of the two--dimensional
classical one--component plasma (sometimes called jellium)
and in statistical problems related to matrices or random
polynomials.

Our work is divided into three parts.  In the first one, after
recalling the simplest cases of quantum wave functions at
odd fractional filling of the lowest Landau level proposed
by Laughlin, we discuss their expansion in terms of Slater
determinants which would allow to obtain- among other things -
their normalization (section 2).  We then record the standard
interpretation of the normalization problem as computing the
partition function of a two dimensional - neutralized - one
component plasma.  In a variant we show that an added quadrupolar
compensating field has no effect on the thermodynamic properties
(at complete filling) except to modify the shape of the
"quantum liquid drop".  It is possible to substitute to the planar
geometry a spherical one with higher symmetry (section 3).
This permits one to obtain bounds for the residual free energy
- in the Coulomb interpretation- using H\"older's and Hadamard's
inequalities.  These bounds can be conjecturally improved by using
the behaviour at zero temperature of some special symmetric
configurations for finitely many particles (up to 42).  We devote
appendix C to an interesting instability of the cuboctahedral
configuration.

Part two is centered around one of the most fundamental
objects of algebra: the discriminant of a polynomial
\eqn\discr{ D= \prod_{i \neq j} (x_i -x_j) \ ,}
where $i$ runs over $N$ distinct values (conveniently chosen as the
integers
$0,1,2,...,N-1$ ) and  $x_i$ are indeterminates which may be thought of
as the roots of the monic polynomial
\eqn\pol{ P(x)= \prod_i (x-x_i)=\sum_{k=0}^N (-1)^k \sigma_k x^{N-k}\ \ ;
\ \ \sigma_0 \equiv 1.}
The discriminant $D$ as well as its powers are symmetric polynomial
functions of the indeterminates $x_i$ and as such admit a unique
polynomial representation in terms of the elementary symmetric
functions $\sigma_k$, a well known fact since the days of Newton.
The subsequent representation for $D$ is unfortunately not very useful
for our purposes. On the other hand the ring of symmetric polynomial
functions with rational coefficients can be viewed as a graded vector
space over $\IQ$. Each homogeneous subspace admits a basis in terms
of Jacobi--Schur functions $ch_Y(x_i)$, indexed by partitions or
Young tableaux. Thus with $s$  a positive integer
\eqn\decschur{(-1)^{sN(N-1)/2} D^s(x_i)=
\sum_{|Y|=sN(N-1)} g_Y^{(s)} \ ch_Y(x_i) \ .}

We were unable to find in the immense literature
on symmetric functions and representations of linear groups a discussion
of such an expansion.
In section 4 we give various expressions for the  coefficients $g_Y$
(eqs. (4.16),(4.20),(4.22) and (4.57)) none of them being very
effective.  We also derive a number of their properties and present
some tabulations in appendix D.
As already admitted, even for $s=1$, we could not find a useful
expression for the general integral coefficients $g_Y^{(1)}$ simple
enough for our purposes.
This is left as a challenge for a clever reader.
However in section 5
we exhibit a remarkable
sum rule
\eqn\verygood{ \sum_Y |g_Y^{(s)}|^2 = {[(2s+1)N]! \over N! ((2s+1)!)^N} }
derived from a formula conjectured by F. Dyson, and proved by Wilson,
Gunson, and Good.

The $g_Y$ coefficients also satisfy an overdetermined linear
system (section 6) which turns out in practice to be the fastest
means of computation for small enough number of particles.  Finally
we consider $q$-specializations in section 7.

We devote the third part to yet another problem, at first sight
elementary, which is to count the number of terms in the
expansion (1.3).  It turns out that this is not trivial at all.
We will elaborate in section 8 and appendix E on our findings
in this direction.

As this work was in progress we received an article by G.W. Dunne
\dunne\ which partly overlaps with ours.

\newsec{From the Laughlin wave function to the classical Coulomb gas}

In the fractional quantum Hall effect one observes plateaux in the
Hall conductance indexed by filling fractions of Landau levels, the
infinitely degenerate non--relativistic energy levels for
a charged
particle interacting with a magnetic field (see for instance
\qhe\ for a review).
Assuming the particles constrained in a transverse plane, and choosing
a suitable scale for the field, an orthogonal complete set of states in
the lowest Landau level is given by
\eqn\compstat{ \psi_l(x) =
{1 \over \sqrt{\pi}} x^l e^{-{x{\bar x} \over 2}}, }
where $x$ is a complex variable and $l=0,1,...$, can be interpreted
as the angular momentum eigenvalue.
The normalization of these wave functions is
\eqn\norm{ \langle \psi_l | \psi_{l'} \rangle =
\int {d^2 x \over \pi} e^{-x{\bar x}} {\bar x}^l x^{l'}=l! \delta_{l,l'}}
For a system of N non--interacting particles with Fermi statistics in
the lowest Landau level we take Slater determinants as an orthogonal
basis of wave functions
\eqn\slat{\psi_{l_0,l_1,...,l_{N-1}}(x_0,...,x_{N-1})={e^{-{1 \over 2}
\sum_{i=0}^{N-1} x_i {\overline{x_i}}} \over
\pi^{N/2}} \left\vert\matrix{x_0^{l_0}&x_0^{l_1}&\cdots&x_0^{l_{N-1}}\cr
x_1^{l_0}&x_1^{l_1}&\cdots&x_1^{l_{N-1}}\cr
.& . & \cdots& . \cr
x_{N-1}^{l_0}&x_{N-1}^{l_1}&\cdots&x_{N-1}^{l_{N-1}}\cr}\right\vert.}
We may assume ordered indices $0 \leq l_0<l_1<...<l_{N-1}$, since
the multiparticle
wave function is antisymmetric in the permutations of the $l$'s.
We have then
\eqn\normn{ \langle \psi_{l_0,...,l_{N-1}}|\psi_{l_0',...,l_{N-1}'}
\rangle = N! \prod_{i=0}^{N-1} l_i! \delta_{l_i,l_i'} }
The state with lowest total angular momentum,
corresponds to the choice $l_i=i$, in which case the Slater determinant
reduces up to a factor to the Vandermonde determinant
\eqn\vand{ \Delta(x_0,...,x_{N-1})=\left\vert\matrix{
1&x_0&\cdots&x_0^{N-1}\cr
1&x_1&\cdots&x_1^{N-1} \cr
.&.&\cdots&. \cr
1&x_{N-1}&\cdots&x_{N-1}^{N-1}\cr}\right\vert=\prod_{i>j}(x_i-x_j)}
We use the notation $D(x_0,...,x_{N-1})$ for the discriminant,
equal up to a sign to the square of the Vandermonde determinant
\eqn\vandisc{ D(x_0,...,x_{N-1})= \prod_{i \neq  j}(x_i-x_j)
=(-1)^{N(N-1) \over 2} \Delta(x_0,...,x_{N-1})^2,}
and identify the normalization of the ``densest state" (filling fraction
$1$) as a partition function through
\eqn\partden{Z_N(1)=\int \prod_{i=0}^{N-1}
({d^2 x_i \over \pi} e^{-x_i {\overline{x_i}}}) \Delta(x){\overline{\Delta(x)}}
=\prod_{j=1}^N j! }

Electrons subject to an intense magnetic field are nevertheless
still interacting via Coulomb forces and also with the
substrate as well as impurities.
To this day it is still a difficult problem to understand the properties
of such a system even at very low temperature.
Nevertheless the remarkable stability of such quantum fluids exhibited in
particular by the existence of the Hall plateaux led Laughlin to assume
simple trial wave functions at fractional fillings. In the simplest
cases of filling fractions of the form $\nu={1 \over 2s+1}$, $s=0,1,...$
(where the odd denominators are required by Fermi statistics), his guess
takes the form
\eqn\guess{ \phi_s(x_0,...,x_{N-1})={1 \over \pi^{N/2}}
\Delta(x_0,...,x_{N-1})^{2s+1} e^{-\sum_{i=0}^{N-1} {x_i {\overline{x_i}} \over
2}},}
such that as $x_i \to x_j$ the wave functions have zeroes of order $2s+1$.
It is then natural to ask the following questions:

\item{(a)}compute the normalization
\eqn\normal{ Z_N(2s+1)=\langle \phi_s|\phi_{s} \rangle=
\int \prod_{i=0}^{N-1}
({d^2 x_i \over \pi} e^{-x_i {\overline{x_i}}}) \left(
\Delta(x){\overline{\Delta(x)}}\right)^{2s+1} ,}
with $Z_N(1)=\langle \phi_1|\phi_1 \rangle$ given above.

\item{(b)} Find the expansion of
Laughlin's wave functions $\phi_s$ on the complete basis of
(free particle) Slater determinants
\eqn\decslat{ \phi_s(x_0,...,x_{N-1})=
\sum_{0 \leq l_0<...<l_{N-1} \leq (2s+1)(N-1)} \ g_{l_0,...,l_{N-1}}^{(s)}
\ \psi_{l_0,...,l_{N-1}}(x_0,...,x_{N-1}),}
or equivalently
\eqn\decslatbis{ \Delta(x_0,...,x_{N-1})^{2s+1}=
\sum_{0 \leq l_0<...<l_{N-1}\leq (2s+1)(N-1)} \ g_{l_0,...,l_{N-1}}^{(s)}
\ \vert x^{l_0}...x^{l_{N-1}} \vert,}
with the short hand notation $\vert x^{l_0}...x^{l_{N-1}} \vert$
for the determinant obtained by replacing $x$ successively by
$x_0,x_{1},...$ in the first, second ... line.

This in turn amounts to
\eqn\amount{\eqalign{
\Delta(x_0,...,x_{N-1})^{2s} &=
(-1)^{{sN(N-1) \over 2}} D(x_0,...,x_{N-1})^s \cr
&=\sum_{0 \leq l_0<...<l_{N-1}\leq (2s+1)(N-1)} \ g_{l_0,...,l_{N-1}}^{(s)}
\ ch_{l_0,...,l_{N-1}}(x_0,...,x_{N-1}) \cr}}
The notation $ch_{...}$ stands for the polynomial characters of the general
linear group $GL_N$, expressed here for a diagonal matrix
$X=diag(x_0,..,x_{N-1})$ (we assume for the time being that none of
the differences among $x$'s vanishes) as
\eqn\caracter{ ch_{l_0,...,l_{N-1}}(x_0,...,x_{N-1})=
{\vert x^{l_0}x^{l_1} ... x^{l_{N-1}} \vert \over
\vert 1 x ...x^{N-1} \vert} .}
{}From L'Hospital's rule this remains meaningful when any subsets
of $x$'s coincide.
We call them general Schur functions -- although they were introduced by
Jacobi -- and shall
express them in terms of the (normalized) traces
\eqn\traces{ \theta_k = {t_k \over k} = {1 \over k} \sum_{i=0}^{N-1}
x_i^k \qquad k>0.}
Parenthetically we will also have use for $t_0=N$. Defining the
sequence
\eqn\sequence{ 0 \leq f_0 \leq f_1 \leq \cdots \leq f_{N-1} \leq 2s(N-1) }
through
\eqn\shifts{ l_i = f_i+ i \qquad i=0,1,...,N-1,}
one can record the increasing sequence $l_i$ in a Young tableau with
$f_0$ boxes in the $N$-th line, $f_1$ boxes in the $N-1$-th line,...,
$f_{N-1}$ boxes in the first line.
We will sometimes switch from the notation $\{ l_i \}$ to
$Y\equiv\{ f_i \}$.
We observe that the expression \caracter\ extends naturally as an
antisymmetric quantity under permutations of  $l_0,..,l_{N-1}$ and
correspondingly for $g_{l_0,...,l_{N-1}}^{(s)}$.

The expansion \decslat\ allows one to perform the integral
in the partition function
\eqn\partfunc{ Z_N(2s+1)=N!
\sum_{0 \leq l_0<...<l_{N-1}\leq (2s+1)(N-1)} \vert
g_{l_0,...,l_{N-1}}^{(s)} \vert^2 \prod_{i=0}^{N-1} l_i! \ .}
This formula is useful only insofar we can obtain some mastery of the
coefficients $g^{(s)}_Y$.

The integral in \normal\ continues to make sense for arbitrary (positive)
exponent $p$ instead of the odd integer $2s+1$, in which case
\eqn\normp{\eqalign{
Z_N(p)&=\int \prod_{i=0}^{N-1}
({d^2 x_i \over \pi} e^{-x_i {\bar {x_i}}}) \left(
\Delta(x){\overline{\Delta(x)}}\right)^{p}\cr
&=\int \prod_{i=0}^{N-1} {d^2 x_i \over \pi}
e^{- \sum_{i=0}^{N-1} x_i {\bar{x_i}}+2p \sum_{0 \leq i<j \leq N-1}
ln|x_i -x_j|} .\cr}}
This suggests two alternative descriptions of our problem.
In the first, consider the set of monic polynomials of degree $N$
(with complex coefficients) identified with $(\IC^N)^{\rm symm}$
and define a probability distribution on this space by considering
the roots as independent Gaussian variables
\eqn\gausvar{ d \mu_N(x_0,..,x_{N-1})= \prod_{i=0}^{N-1}
({d^2 x_i \over \pi} e^{-x_i {\bar {x_i}}}) }
This is indeed symmetric in any permutation of the $x$'s.
The first form of \normp\ at least for $p$ a positive integer
involves the computation of the moments of the distribution of
the module of the discriminant $|D|$, thus for any positive $p$
\eqn\moments{\eqalign{& Z_N(p)=\int_0^{\infty} d \sigma_N(u) \  u^p \cr
&d\sigma_N(u)= \int d\mu_N(x_0,..,x_{N-1}) \delta(|D(x)|-u).}}
For instance, for $p$ integral
\eqn\ajout{\eqalign{Z_2(p)&=p! \ 2^p, \ \ \ d\sigma_2(u)={1\over 2}
du e^{-u/2} \cr
Z_3(p)&=2^{-p}(3p)! \sum_{0\leq a \leq p} 3^{2a}
{{p \choose a}^2\over {3p
\choose 2a}}}}

The second and more useful interpretation of the integral \normp\ is
as a canonical partition function of a one component plasma
(logarithmic repulsive two body potential), a classical two--dimensional
Coulomb fluid, with neutralizing background. In this framework
one can avail oneself of a considerable body of knowledge
in the thermodynamic limit $N \to \infty$ \multi ,
and in particular of some evidence for a first order ``fluid--solid"
transition in the parameter $\beta=2p$ of the order of
\eqn\transit{ \beta_{\rm trans.}= 2 p_{\rm trans.}\simeq 142,}
according to Choquard and Cl\'erouin \multi .
It would be of great interest to be able to understand from first principles
the
mechanism of this transition to a classical Wigner solid.

Following Jancovici and Alastuey, the plasma interpretation justifies
a large $N$ behavior for $ln Z_N(p)$ obtained as follows.
Consider $N$ classical particles of charge $e$ in a disk of radius $R$
interacting via a repulsive two--dimensional Coulomb potential
$$ v_{ij}(x_i-x_j)= {e^2 \over 2 \pi} ln {1 \over |x_i-x_j|},$$
choosing an arbitrary length scale inside the logarithm
(which should anyhow disappear from the final result).
The contribution of the charges to the total energy is
\eqn\charcont{ E_{\rm ch.}= -{e^2 \over 4 \pi} ln |\Delta(x_0,..,x_{N-1})|^2. }
We assume a uniform neutralizing background with density ${N \over \pi R^2}$.
Its self--interaction is
\eqn\selfint{ E_{\rm back.}=\left({N \over \pi R^2}\right)^2
{e^2 \over 2} \int_{|x_1|,|x_2| <R} d^2 x_1 \ d^2 x_2 \ {1 \over 2 \pi}
ln{1 \over |x_1-x_2|} .}
Set
\eqn\useful{ \phi(x)={1 \over 2\pi} \int_{|y|<R} d^2 y \ ln{1 \over |x-y|}=
{R^2-x{\bar x} \over 4} +{R^2 \over 2} ln{1 \over R} \ \ \ |x|\leq R.}
This expression follows from the fact that $-\Delta \phi(x)=1$ for
$|x| \leq R$, while for $|x|=R$, $\phi$ is equal to the
potential created by the total charge $\pi R^2$ located at the center,
i.e. ${\pi R^2 \over 2 \pi}ln{1 \over R}$. After a short
calculation
\eqn\backen{ E_{\rm back.}={N^2e^2 \over 4 \pi R^2} ({1 \over 8}+
{1 \over 2}ln{1 \over R}).}
The last contribution to the energy (ignoring kinetic energy
which factorizes from the partition function) is the attractive
interaction between charges and background
\eqn\attrac{ E_{\rm int.}=-{e^2 N \over \pi R^2} \sum_i \phi(x_i)=
{e^2 N \over 4 \pi R^2} \sum_i x_i {\overline{x_i}}
-{e^2 N^2 \over \pi} ({1 \over 4}+{1 \over 2} ln{1 \over R}).}
Thus the total energy divided by $k_BT$ ($T$ the temperature) is
\eqn\totenerg{\eqalign{ {E \over k_B T}&=
(E_{\rm ch.}+E_{\rm back.}+E_{\rm int.})/k_B T \cr
&={e^2 \over 4 \pi k_B T}({N \over R^2}\sum_i  x_i {\overline{x_i}}
-ln|\Delta(x_0,...,x_{N-1})|^2+ N^2 ln R -{3 \over 4}N^2). \cr}}
We set
\eqn\ajouti{{e^2\over 4 \pi k_B T}={\beta\over 2}=p,\ \
R^2=p N}
in such a way that,
\eqn\ajoutii{{E\over k_B T}
=\sum_i x_i \overline{x_i} -2 p \ ln \vert \Delta (x_0,
..., x_{N-1})\vert + {p N^2\over 2} ln (pN)-{3 p N^2\over 4}.}
The excess free energy $F_{\rm exc.}$ of this system, as compared
to the perfect gas (i.e. deriving only from the kinetic energy),
is given by
\eqn\excessisbad{\eqalign{-{F_{\rm exc.} \over k_B T}&=
ln\int_{|x_i| \leq R} \prod_{i=0}^{N-1} {d^2 x_i \over \pi R^2}
e^{-{E \over k_B T}} \cr
&= -{pN^2 \over 2} ln (pN) +{3 \over 4} p N^2-N ln (pN) +
ln Z_N(p).\cr}}
According to thermodynamics, in the limit of large $N$
(i.e. large $R$) we have extended the integrals to the full plane up to
errors of order at most $\sqrt{N}$ in $F_{\rm exc.}$
(i.e. arising from boundary terms).
Since we expect an extensive
excess free energy, we obtain the estimate
\eqn\extexc{ ln Z_N(p)=
{pN^2 \over 2} ln (pN) -{3 \over 4} p N^2+N ln (pN)+ O(N).}
It is the term of order $N$, namely $-{F_{\rm exc.} \over k_B T}$,
depending on $p$, which is shown to undergo a first order phase transition,
namely the limit
\eqn\phasetrans{ \lim_{N \to \infty}{1 \over N}(ln  Z_N(p)
-{pN^2 \over 2} ln (pN) +{3 \over 4} p N^2-N ln (pN))}
should exist and
have a discontinuous derivative for a value of $p$ close to $71$.
In principle eqn.\extexc\ answers question (a) up to the unknown term of
order $N$. But its status is at best heuristic. It is therefore reassuring
that at least for $p=1$ ($s=0$) where we have an exact result, we can justify
\extexc.
Note that in the thermodynamic reasoning the value of $p$ did not play a
crucial role. We devote appendix A to the asymptotic
evaluation of $ln Z_N(1)$ using Euler--Mac Laurin's formula. It reads
\eqn\finappli{\eqalign{ ln Z_N(1)&= ln \prod_{j=1}^N j! \cr
&={N^2\over 2} ln N-{3\over 4} N^2+N ln N+N({ln 2 \pi \over 2}-1)
+{5\over 12} ln N \cr
&+{1-\gamma+5 ln 2 \pi \over 12}+{\zeta'(2) \over 2 \pi^2}+{1 \over 12 N}
-{1 \over 720 N^2}-{1 \over 360 N^3} +O({1 \over N^4}), \cr}}

We add a few remarks.

\item{(i)} The thermodynamic ansatz suggests that
\eqn\ansth{ ln Z_N(p) -{1 \over p} ln Z_{Np}(1) =O(N),}
while (trivially) H\"older's inequality yields
\eqn\hold{ p \ ln Z_N(1) \leq ln Z_N(p).}

\item{(ii)} One can also obtain a crude upper--bound as follows.
{}From the symmetry of the integrand in $Z_N(p)$ \normp\ we have
\eqn\symuse{ Z_N(p)=N! \int_{|x_0|\geq |x_1|\geq \cdots \geq |x_{N-1}|}
\prod_{i=0}^{N-1} ({d^2 x_i \over \pi} e^{-x_i {\overline{x_i}}})
|\Delta(x_0,...,x_{N-1})|^{2p}.}
In this domain
$$|\Delta(x_0,...,x_{N-1})|^{2p} \leq 2^{pN(N-1)}
\prod_{i=0}^{N-1} |x_i|^{2p(N-i-1)}.$$
The integral over the sector is smaller than the full  integral, hence
\eqn\crudestim{ Z_N(p) \leq N! \ 2^{pN(N-1)} \prod_{k=1}^{N-1} (pk)! \ ,}
where the factor $N! \ 2^{pN(N-1)}$ is probably a gross overestimate.
For further discussion of such bounds see the next section.

\newsec{Coulomb Gases.}

\subsec{Quadrupolar mean field.}

In the interpretation of the norm \normal\ as a
one component plasma canonical partition function,
the gaussian measure in \gausvar\
is due to the  mean potential
created by the neutralizing
background charge.  This potential is not only determined
by the locally constant density, but also by the shape of
the bounding domain which, in the case studied so far, is a disk and
therefore induces a rotational symmetry around the center,
the minimum of the harmonic potential.  Far from this
minimum the global form of the potential has no consequence
on the local correlations at a given density, neither on
the residual thermodynamics.  This suggests to add to the quadratic
$z \bar{z}$ term an harmonic function (equivalently the real part of an
entire function) which does not modify the constant charge density.
To ensure convergence of the partition function we are limited
to the real part of a quadratic polynomial, hence up to a translation,
to a multiple of a quadrupolar mean field
$V=Re z^2$, created by far away charges.

It is rather remarkable that the plasma model
with a quadrupolar component of the mean field can also
be completely solved for $\beta =2 \ (s=0)$.  We give here
a short description of the method and of the corresponding results.
We want to evaluate
the partition function depending on the
intensity of the quadrupolar field that will be called
$th\mu$ ($\mu$ real) so that the total potential
\eqn\qadi{
V(z,\bar{z})=z \bar{z}-{1\over 2} th\mu (z^2+\bar{z}^2)
}
confines the charges. To compute
\eqn\qadii{
Z_N(1;\mu)=\int \prod_j {d^2 z_j\over \pi} e^{-\Sigma z_j \bar{z}_j
+{1\over 2}th \mu(z_j^2+\bar{z}_j^2)} \vert \Delta \vert^2
}
we just need to be able to construct the
polynomials $P_n(z)$ ($z\in \IC$) with real coefficients
verifying the unitary
orthogonal relations
\eqn\qadiii{\eqalign{
&\int \int {d^2 z\over \pi} w(z,\bar{z} ) \ P_n(z) P_m(\bar{z})
=\delta_{nm} \cr
& w(z,\bar{z})=e^{-z \bar{z}+{1\over 2}th\mu (z^2+\bar{z}^2)}\cr
& deg(P_n)=n =(0,1,2,...) \cr
& \lim_{\mu\rightarrow 0} P_n(z)=(n!)^{-1/2} z^n
}}
We start from the identity
\eqn\piden{
\int {d^2 z\over \pi} e^{- Z^\dagger A Z/2} = {2\over \sqrt{
(a+d)^2-4 c \bar{c}}}
}
where $Z,Z^\dagger$ are the column and row vectors
${z \choose \bar{z}}, \ (\bar{z},z)$ and
$A=\pmatrix{a&c\cr \bar{c} & d}$
is a hermitian matrix such that $(a+d)^2\geq 4 c \bar{c}$.  Set
\eqn\pmeasure{
z \bar{z}-{1\over 2} th \mu (z^2 + \bar{z}^2)={1\over 2} Z^\dagger A Z
,}
where
\eqn\pmatr{
A=A^\dagger=\pmatrix{1 & -th\mu \cr -th\mu & 1}
.}
Now introduce the vector $X={x \choose \bar{x}}$ and write the
following identity
\eqn\piiden{\eqalign{
1&={\int {d^2 z\over \pi} \exp -{1\over 2} (Z-A^{-1}X)^\dagger
A (Z-A^{-1} X) \over \int {d^2 z\over \pi} \exp -{1\over 2}
Z^\dagger A Z } \cr
&=< \exp \left( {1\over 2} (X^\dagger Z+Z^\dagger X)-{1\over 2}
X^\dagger A^{-1} X \right) > \cr
& = <\exp (\bar{x}z+x \bar{z}-{1\over 1-(th\mu)^2} (x \bar{x}
+{1\over 2} th\mu (x^2+\bar{x}^2))>
}}
Rescaling $x$ as $x=\sqrt{{1-(th\mu)^2\over th\mu}}
u=\sqrt{{2\over
sh 2\mu}}u$ and
moving the term $\exp (-{x \bar{x} \over 1-(th\mu)^2})
$ in
\piiden\ on the left hand side, we get
\eqn\plside{
\exp {\bar{u} u\over th\mu} = < \exp \left[
\bar{u} z \sqrt{{2\over sh 2\mu}} - {\bar{u}^2\over 2} \right]
\exp \left[ u \bar{z} \sqrt{{2\over sh2 \mu}} - {u^2\over
2} \right] >.}

The r.h.s. of this expression involves generating
functions for monic Hermite polynomials
\eqn\herp{
\exp (x u-{u^2\over 2})=\sum_0^\infty H_n(x) {u^n\over n!}.}
Expanding both sides we have
\eqn\expex{
\sum_{n=0}^{\infty} ({u \bar{u}\over th \mu})^n {1\over n!}
 = \sum_{k,l=0}^\infty {\bar{u}^l u^k \over
l! k!} <H_k (\bar{z} \sqrt{{2\over sh2 \mu}} )
H_l (z \sqrt{{2\over sh2 \mu}}) >}
and since in the average $<...>$ the integral of the denominator
is $ch\mu$, we get
\eqn\exppp{
\int {d^2z \over \pi} e^{-\bar{z} z+{1\over 2} th\mu (z^2+\bar{z}^2)}
H_k(\bar{z} \sqrt{{2\over sh2 \mu}}) H_l (z \sqrt{{
2\over sh2\mu}}) = ch\mu {l!\over (th\mu)^l } \delta_{k,l}
}
The polynomials $P_n$ are therefore given by
\eqn\rghtf{
P_n(z)={(th\mu)^{n/2} H_n({\sqrt{2}z\over \sqrt{sh 2 \mu}}) \over
\sqrt{n!} (ch\mu)^{1/2} }=(ch \mu)^{-(n+1/2)} {z^n\over \sqrt{n!}}+...
}
The first few $P$'s read
\eqn\lisfirst{
\eqalign{
&P_0(z)=(ch\mu)^{-1/2} \cr
&P_1(z)=(ch\mu)^{-3/2} z \cr
&P_2(z)={(ch\mu )^{-5/2}\over \sqrt{2!}} (z^2-sh\mu \ ch\mu) \cr
&P_3(z)={(ch\mu)^{-7/2}\over \sqrt{3!}} (z^3-3 sh\mu \ ch\mu \ z)
}}
Substituting in the Vandermonde determinant $\sqrt{n!} (ch\mu)^{(n+1/2)}
P_n(z)$ for $z^n$ we readily evaluate the partition function as
\eqn\sqi{
{Z_N(1;\mu)\over Z_N(1;0)}=(ch \mu)^{N^2}
}
the superimposed field is contributing
only to the dominant term of the free energy by $N^2 \log ch \mu$,
this term being non extensive.  After subtraction, the residual
energy is therefore independent of $\mu$, that is of the mean field
at large distance as was to be expected.

It is also interesting to compute the local density of the Coulomb
system of N charges in that neutralizing background.  We have
\eqn\sqii{\eqalign{
\rho_N(z,\bar{z})&=w(z,\bar{z}) \sum_{n=0}^{N-1} P_n(z)
P_n(\bar{z}) \cr
&\equiv e^{-z \bar{z} +{1\over 2} th \mu (z^2+\bar{z}^2)}
\sum_{n=0}^{N-1} {(th\mu)^n\over ch\mu \ n!}
\vert H_n({\sqrt{2}z\over \sqrt{sh 2 \mu}})\vert^2 }}
which for $\mu=0$ reduces to
\eqn\sqiii{
\rho_N (z,\bar{z} ) \vert_{\mu=0} = e^{-z \bar{z}} \sum_{n=0}^{N-1}
{(z\bar{z})^n\over n!}
.}
The asymptotic behaviour in $N$ for $\vert z^2 \vert$ of
order $N$ can
easily be found outside a transition region, for which
a finer analysis is required.  Since for fixed $z$
\eqn\sqiv{
\lim_{N\rightarrow \infty} \rho_N (z,\bar{z})\equiv 1 \ \ \forall z,
}
there exists a curve $C_N(\mu)$, being a continuous deformation of the
circle $\vert z \vert^2=N$, such that the asymptotic density
$\rho_N$ is equal to 1 in the interior of the domain bounded by
$C_N$ and zero outside. To evaluate $C_N$ asymptotically, we write that
the variation of density
\eqn\vardens{ \rho_{N+1}(z,{\bar z}) - \rho_N(z,{\bar z})  }
is maximal for $z \in C_N$.

Using the large $N$ asymptotic behaviour of
the Hermite polynomial $H_N$ when ${z \bar{z}\over N}$
is finite
\eqn\haha{
{1\over N} ln{ \vert H_N((e^\phi+e^{-\phi})\sqrt{N}) \vert^2 \over N!}
\simeq Re (2 \phi + e^{-2 \phi}
+1),
}
with the definition of $\phi$
\eqn\hahai{
ch \phi = {z\over \sqrt{2 N sh 2 \mu}} , \ \ \ (Re \phi > 0),
}
the variation of density \vardens\ is written, in the "exponential"
approximation:
\eqn\hahaii{
\exp \left[ -z \bar{z} + {1\over 2} th\mu (z^2+\bar{z}^2)+
N (ln th \mu +1 +Re(2 \phi+e^{-2 \phi})) \right]
}
The maximum over $z$ is given by the vanishing
of the derivatives with respect to $z,{\bar z}$, i.e.
\eqn\hahaiii{-{\bar z} + th \mu z+ N {\partial \phi \over
\partial z} (1-e^{-2\phi})) =0
}
and its complex conjugate.
According to \hahai,
\eqn\hai{
sh \phi {\partial \phi\over \partial z}={1 \over \sqrt{2 N sh 2 \mu}} ,
}
so that \hahaiii\ amounts to
\eqn\haii{
th \mu=\vert e^{-2 \phi} \vert .
}
Putting $\phi=\xi+i \eta$ , the equation of the curve is
\eqn\ecurv{
z=\sqrt{N} (e^\mu \cos \eta + i e^{-\mu} \sin \eta )
}
which is an ellipse with semi-axis $a=\sqrt{N} e^\mu$ ,
$b=\sqrt{N} e^{-\mu}$.  The area of this ellipse is
$\pi N$ independently of $\mu$ the intensity of the quadrupolar
field.  The effect of an added quadrupolar field is therefore to deform
the Fermi liquid disc into an ellipse with the same area.

\subsec{N charges on the sphere.}

The thermodynamic equivalence between the harmonic
plasma and the system of charges on a sphere has been proven
for $\beta=2$ and is presumably valid for all temperatures.
The spherical geometry avoids giving a special status to the
center of the disk thereby increasing the symmetry and allows a
clear definition of the pressure since the potential function
does not depend explicitly on the density.

The Coulomb potential created at point $\hat{\mu} (\theta,\phi )$
by a unit charge located at the north pole ($\theta=0$) of the unit
sphere, is defined, up to a constant, by the Green function
of the spherical Laplacian. The
source term must be supplemented by a neutralizing background which
can be taken to be uniformly distributed (to respect the spherical
symmetry) and the total charge of the source must vanish
on the sphere, a surface without
boundary.  The potential
satisfies therefore the inhomogeneous equation
\eqn\yyi{\nabla_{\hat{u}} V=-2 \pi \delta (\hat{u})+{1\over 2}}
with
\eqn\yyii{\int \delta (\hat{u}) d\sigma =1}
\eqn\yyiii{ d\sigma=d (cos \theta )\wedge d\phi \ , \
\ \ \ \int d\sigma
=4 \pi  }
and the constant ${1\over 2}$ represents the neutralizing
density.  Putting $\eta=cos \theta$, we have
\eqn\yyiv{
{\partial\over \partial \eta} (1-\eta^2) {\partial V\over \partial
\eta}={1\over 2}-2 \delta (1-\eta)}
Up to a constant the solution is,
\eqn\yyv{
V=-{1\over 2}ln {1-\eta \over 2}=-ln \left(
sin{\theta\over 2}\right) =ln ({2\over r})}
where $r$ is the euclidian distance between the charge
and the point on the sphere.  We fix an additive
constant requiring a zero average value of $V(\hat{u})$,
giving the potential
\eqn\yyvi{
V_0=ln ({2\over r})-{1\over 2}=-ln (\sqrt{e} sin {\theta\over 2})
}
which satisfies $\int V_0 d\sigma=0$.  The correspondence
between the spherical system and the original planar one is
obtained by a stereographic projection from the south pole
on the tangent plane to the north pole.  The complex coordinate
in this plane is $2 z$, with
\eqn\yyvii{z=tg {\theta\over 2} e^{i\phi}}
\eqn\yyviii{{d\sigma\over 4 \pi}= {d\bar{z} \wedge dz\over 2 \pi i}
{1\over (1+\vert z\vert^2)^2}={d^2 z\over \pi}{1\over (1+\vert z \vert^2)^2}}
while the angular distance $\gamma_{12}$ between two
points $z_1$ and $z_2$ is given by
\eqn\yyviii{
sin^2{\gamma_{12}\over 2}={\vert z_1-z_2\vert^2\over
(1+\vert z_1\vert ^2) (1+\vert z_2\vert^2)}}
On a sphere with unit area, the partition function
at inverse temperature $\beta=2p$ and with two-body
potential $V(\gamma_{12})$ is defined by the average
\eqn\yyix{\eqalign{
I_N (p)&=\left< \prod_{i<j} sin^{2p} {\gamma_{ij}\over 2}
\right> \cr
<A>&=\int \prod_{i=1}^N {d\sigma_i\over 4 \pi} A}}
For a sphere of area S, the partition function
$I_N$ is to be multiplied by $S^N$.  We can
also write
\eqn\yyx{
I_N(p)=\int \prod_{j=1}^N {d^2 z_j\over \pi}
{1\over (1+z_j\bar{z_j})^{(N-1)p+2}} \prod_{i<j}
\vert z_i-z_j\vert^{2p}.}

With the choice of potential $V_0(\gamma_{12})$ defined
in \yyvi , the canonical partition function is
\eqn\yyxi{
Q_N(p)=e^{{p\over2}N(N-1)} I_N(p).}
We note the exact results
\eqn\yyxii{
I_N(1)={1\over (N!)^{N-1}} \left(
\prod_{p=0}^{N-1} p! \right)^2}
and
\eqn\yyxiii{
I_3(p)={(p!)^3 (3p+1)!\over ((2p+1)!)^3}=\left( {\Gamma (p+1) \over
\Gamma (2 p+2) } \right)^3 \Gamma (3 p+2), \ \ \forall p, \ \ \
Re p>-{2\over 3}.}
The
partition functions on the plane $Z_N(p)$
eqn.\normal, and on the sphere $I_N(p)$
eqn.\yyx, differ only by the weight functions.
They become equivalent in the central region
$\vert z \vert =O({1\over \sqrt{Np}})$.

In appendix A, we find from \yyxi\ and
\yyxii\ the asymptotic free energy per unit charge $F_N(1)$
$(\beta=2)$
\eqn\suplem{
-2 F_N(1)={ln Q_N(1)\over N}={1\over 2} ln 2\pi N-{3\over 2}
+ O({ln N\over N}).}
In contrast with $ln Z_N$ for the neutral plasma,
we note the absence of terms in $N^2 ln N$ and $N^2$
in $ln Q_N$ due
to the choice of potential $V_0$ having a spherical
average zero.
If we want to compare the two systems, planar and
spherical, we have to renormalize the scale of the potential,
which is equivalent to introduce an appropriate continuous
background following the convention of Caillol and al. \multi\
(formula 3.4 in their paper).

The residual free energy is now given by the following
form
\eqn\yyxiv{
\eqalign{
-2 f(1)&=\lim_{N\rightarrow \infty} {1\over N} ln \left(
{Q_N(1)\over \sqrt{N!}}\right) \cr
&=\lim \left( {1\over 2} ln 2\pi N-{3\over 2} \right)
-{1\over 2} (ln N-1) \cr
&= \left( {1\over 2} \log 2\pi -1 \right) = -0.0810614...
}}
which agrees with the residual free energy of the neutral
plasma i.e. the term proportional to $N$ in (2.34).

\vskip 2in

{\noindent  Remark on the canonical pressure}

The following section is devoted to obtaining bounds for $f(p)$
when $p\geq 1$, to confirm the dependence
in N observed for $p=1$.  On the sphere, it is the quantity
\eqn\yyxv{
\left( e^{N(N-1)/2} {1\over \sqrt{N!}}\vert \Delta_N\vert^2 \right)^p,\
\ \vert\Delta_N\vert^2=\prod_{i<j} \sin^2{\gamma_{ij}\over 2}}
that produces the correct extensive property.

Assuming for the time being that for any $p$ the same $ln N$ term
as in formula \suplem, this
gives the free energy per particle for
$N\geq 1$ on a sphere of unit area, $S=1$.
For $S$ arbitrary we
must add $N ln S$ to $ln Q_N$.  Although the energy
of the system of $N$ charges on the sphere is not extensive,
a global pressure denoted by $\tilde{p}$ not to be confused
with the notation $p$, is defined by
$\beta \tilde{p}={\partial (ln Q_N+N ln S)\over \partial S}={N\over S}
=density=\rho$.  The non-trivial term in \suplem\ is not contributing
to the global pressure.  On the other hand if we interpret
the free energy $F_N(p)$ as that of an extensive system
of density $\rho$, eq. \suplem\ gives a supplementary contribution
to the pressure
\eqn\yyxvi{
\rho^2 {\partial\over \partial \rho} \left(
-{1\over 4} log \rho \right) = -{1\over 4} \rho.}
The resulting equation of state $\beta \tilde{p}=\rho (1-{\beta\over4})$
is that of a neutral plasma, a system that probably
cannot remain homogeneous for $\beta \geq 4$.

\subsec{H\"older's and Hadamard's inequalities.}

As we noted in \hold , for the plasma on a plane
H\"older's inequality requires $F_N(\beta)$
to be a non-increasing function of $\beta$, that is the
"potential entropy" $S(\beta )=\beta^2{\partial F\over \partial
\beta}$ is negative.  {}This inequality reads
\eqn\hi{
<a^s>  \ \geq \  (<a>)^s, \ \ \forall s\geq 1}
where
\eqn\hii{
<a>={1\over N} \sum_{j=1}^N a_j, \ \ a_j\geq 0}
is an average value, (possibly a weighted average), of non-negative
quantities.  It follows that $<a^\beta>^{1/\beta}$ is a
non-decreasing function of $\beta$ for $\beta \geq 0$.  We simply
need to apply these relations to $a=e^{-\beta_0 V}$, $s={\beta
\over \beta_0}\geq 1$
\eqn\hiii{
Z_N(\beta )=< e^{-\beta V_N} > \Rightarrow
F_N(\beta)=-{1\over \beta N} ln Z_N (\beta ) \ \ \ \ {\rm
non-increasing}}
{}From this we derive the inequalities
\eqn\hiv{
\eqalign{&F_N(\beta )\leq F_N(2) , \ \ \beta \geq 2 \cr
& F_N(\beta ) \geq F_N(2), \ \ 0\leq \beta \leq 2.}}
Similarly on the sphere, if the residual free energy is defined
for all $p$ by the limit
\eqn\hv{
-2 f(p)=\lim_{N\rightarrow \infty} {1\over N} ln
{Q_N(p)\over (N!)^{p/2}}}
we obtain
\eqn\hvi{\eqalign{
&f(p)\leq f(1) , \ \ p\geq 1\cr
&f(p)\geq f(1) , \ \ 0\leq p \leq 1.}}

Now let us look at Hadamard's inequality to obtain a lower
limit to the free energy.  We start from the identity
\eqn\hvii{
\vert \Delta^2 \vert = \det_{i,j} \left\vert \sum_{n=0}^{N-1} z_j^n
\bar{z}_i^n \right\vert \equiv {\det_{i,j}\left\vert \sum_{n=0}^{N-1}
c_n z_j^n \bar{z_i^n} \right\vert\over
\prod_{n=0}^{N-1} c_n}}
where $c_0, c_1, ... , c_{N-1}$ are $N$ positive indeterminates.
The required inequality reads
\eqn\hviii{
\vert \Delta^2\vert \leq {a_1 \cdots a_{N-1} a_N \over
c_0 c_1 \cdots c_{N-1}} ,}
where the $a_i$'s are the diagonal elements of the
positive matrix, the determinant of which appears in the
numerator of the r.h.s. of \hvii\
\eqn\hix{
a_i=\sum_{n=0}^{N-1} c_n \vert z_i\vert^{2n}, \ \
i=1,2,...,N}
Define the polynomial of degree $N-1$
\eqn\hx{
a(\rho)=\sum_{n=0}^{N-1} c_n \rho^n }
Substituting in the expression \yyx\ for the integral
$I_N(p)$ the inequality \hviii\ and writing $\vert z_j \vert^2
=\rho_j$, we get
\eqn\hxi{
I_N(p)\leq \inf_{[c]} \left[ {\int_0^\infty \prod_{j=1}^N
{a(\rho_j)\over (1+\rho_j)^{p (N-1)}} {d \rho_j\over
(1+\rho_j)^2} \over \prod_{n=0}^{N-1} c_n}\right] }
that is
\eqn\hxii{
I_N(p)\leq \inf_{[c]} {(A_p(c))^N\over \left( \prod_n c_n
\right)^p}}
with the convergent integral $A_p(c)$
given by
\eqn\hxiii{
A_p(c)=\int_0^\infty {a^p (\rho) d\rho\over (1+\rho)^{(N-1) p+2}}}
This integral is a homogeneous polynomial of total degree $p$ in
the $N$ parameters $[c]$. Therefore , for every $p \geq 0$, we have
\eqn\hxiv{
{ln I_N(p)\over N} \leq \inf_{[c]} \left[
ln A_p(c) - {p\over N} \sum_n ln c_n \right] }
The r.h.s. of \hxii\ being homogeneous in $[c]$ of degree
zero, we have to seek the minimum according to the direction
of the vector $c_0,...,c_{N-1}$.  In the first "quadrant"
of $R_N$ $(c_n>0)$, this
continuous positive function tends to infinity on all coordinates
planes.  It therefore has an absolute minimum in an interior
direction, and at that point it verifies
\eqn\hxv{
{1\over A} {\partial A\over \partial c_n} - {p\over N} {1\over c_n}
=0 \ \ ; \ \ n=0,1,...,N-1. }
That is
\eqn\hxvi{
\int_0^\infty {a^{p-1} (\rho) c_n \rho^n d\rho\over
(1+\rho)^{(N-1) p+2} } = {A\over N} \ , \ \ \forall n.}
{}From homogeneity, only
$N-1$ equations are independent
since the sum is given by  Euler's identity.
When $p=1$, the solution is obvious
since \hxvi  gives
\eqn\hxvii{
c_n \int_0^\infty {\rho^n d\rho \over (1+\rho)^{N+1}} = {A\over N}
}
i.e.
\eqn\hxviii{
c_n= A {N-1 \choose n} }
where the constant $A$ is left undetermined.  This yields the polynomial
\eqn\hxix{
a(\rho)\equiv A (1+\rho)^{N-1}}
and we get the following upper bound
\eqn\hxx{\eqalign{
{ln I_N (1)\over N}& \leq ln A - {1\over N} \sum_n ln (A {N-1 \choose n})
\cr &\leq -{1\over N} \sum_{n=0}^{N-1} ln {N-1\choose n}} }
or
\eqn\hxxi{
I_N(1) \leq {\left( \prod_{n=0}^{N-1} n! \right)^2 \over
(N-1!)^N}.}
Let us denote by $B_N(p)$ the Hadamard upper bound (3.49)
(given explicitly by the r.h.s.
of \hxxi\ for $p=1$).  Comparing with the exact result
\yyxii\
\eqn\hxxii{
{B_N(1)\over I_N(1)}= {(N!)^{N-1}\over ((N-1)!)^N}={N^N\over N!}
\simeq {e^N\over \sqrt{2 \pi N}}}
from which we get
\eqn\hxxiii{
{ln B_N\over N}= {ln I_N\over N} +1 + O({ln N\over N})}
This shows that the upper bound of $-2 F_N(1)$ given
by Hadamard's inequality and the exact value are only differing
by 1, meaning from \yyxiv\
\eqn\hxxiv{
f(1)=-{1\over 4} ln 2\pi +{1\over 2} \geq -{1\over 4}
ln 2\pi .}
Let us turn to the general case.
The polynomial $a(\rho)$ given in \hxix\ for $p=1$
is still a solution for arbitrary $p$ up to a multiplicative constant.
More precisely the ansatz
\eqn\hxxv{
c_n=A^{1/p} {N-1\choose n}}
verifies the stationarity equations \hxv . Indeed, after
dividing by $A$,
\eqn\hxxvi{
{N-1 \choose n} \int_0^\infty {(1+\rho)^{(N-1) (p-1)} \rho^n \over
(1+\rho)^{(N-1) p+2} } d\rho = {1\over N}}
Since the dependence on $p$ drops out, this
is an identity in view of our previous discussion.  The only
doubt that remains is whether this stationary point is the absolute minimum.
This is certainly the case in the neighbourhood of $p=1$ by
continuity if we note that the matrix $H_{n,m}=
c_n c_m {\partial\over\partial c_n}{\partial\over \partial c_m}
A_p(c)$ is positive definite and that the function $A_p^N (c)$
is convex in $R_N^+$ hence admits only one minimum on the
plane $\sum_n ln c_n=0$.

The inequality \hxiv\ can now be written
\eqn\hxxvii{\eqalign{
{ln I_N(p)\over N}&\leq ln A -{p\over N} \sum_n
ln (A^{1/p} {N-1\choose n}) \cr & \leq
-{p\over N} \sum_{n=0}^{N-1} ln {N-1 \choose n}}
}
The upper bound for ${ln I_N(p)\over Np}$ is according to
\hxx\ independent of $p$ and the same property holds for
\eqn\hxxviii{
{ln ({Q_N(p)\over (N!)^{p/2}})\over Np}=
{ln I_N(p)\over N p}+{N\over 2}-{1\over 2} ln N+...}
Consequently the residual free energy is bounded by
\eqn\hxxix{
f(1)-{1\over 2} \leq f(p) \leq f(1) , \ \ \ p \geq 1.}
where $f(1)=-{1\over 4} ln 2\pi + {1\over 2}$. Numerically
we get
\eqn\hxxx{
-0.4594692... \leq f(p) \leq 0.0405307}
The lower bound is of no interest for $0<p<1$ since
$f(p) > f(1)$ in the high temperature region (i.e. $p<1$).

\subsec{The zero temperature limit for finite $N$.}

Before going to the thermodynamic limit, i.e. keeping N
finite, the free energy $-{1\over \beta} ln I_N(\beta )$
of the system with $N$ unit charges on the sphere has as a
limiting value, when $\beta \rightarrow \infty$ (or
equivalently $p\rightarrow \infty$), the
absolute minimum of the potential function
\eqn\bi{
{\cal V}_N=-\sum_{i<j} ln (\sin {\gamma_{ij}\over 2})}
or
\eqn\bii{
2 {\cal V}_N=-ln \vert \Delta_N\vert^2 +(N-1) \sum_{j=1}^N ln
(1+z_j \bar{z}_j)}
With the definition \hv\ of the residual free energy, we have
for the finite system
\eqn\biii{
\lim_{p\rightarrow \infty} 2 f(p)={1\over N} \inf (2 {\cal V}_N)
-{N-1\over 2} + {1\over 2N} ln N!.}
This result remains valid in the thermodynamic limit provided
the residual entropy vanishes.  Up to an overall rotation
a stable equilibrium configuration does exist on the
sphere since ${\cal V}_N$ reaches its minimum.  The equilibrium
configurations are solutions
of the $N$ equations ${\partial {\cal V}\over \partial z_j}=0$ (and c.c.)
\eqn\viii{
\sum_{j=1\atop j\neq i}^N {1\over z_i-z_j}-(N-1) {\bar{z}_i \over
1+z_i \bar{z}_i}=0 }
having in the planar limit the simplified form
\eqn\biv{
\sum_{j=1\atop j\neq i}^N {1\over z_i-z_j}-(N-1) \bar{z}_i=0}
{}From \viii\ we get the sum rule
\eqn\bv{
\sum_{j=1}^N {z_j \bar{z}_j\over 1+z_j \bar{z}_j}={N\over 2}}
which reads in polar coordinates
\eqn\bvi{
\sum_j \cos \theta_j=0}
This result is true irrespective of the location of
the pole, therefore we have $\sum_j \hat{u}_j=0$ ($\hat{u}_j$
being the unit vector with polar angle $\theta_j$),
which means that the center of gravity of charge
is the center of the sphere as one would naturally expect.  In
the thermodynamic limit the relation \bv\ allows to show that
the radius of the confinement disk of the plasma is 1 (with
gaussian weight $e^{-N p z \bar{z}}$).

The equations \viii\ are formally invariant under rotation
($z_j\rightarrow \xi_j={\alpha z_j +\gamma\over -\bar{\gamma} z_j
+\bar{\alpha}}, \ \alpha \bar{\alpha}+\gamma \bar{\gamma}=1$).
It is easy to find solutions for
configurations invariant under a finite rotation group.
As an example let us consider the case $N=12$, where 2 configurations
come to mind:

-$12_{(60)}$ : The 12 charges are at the vertices of an icosahedron.
The order of the symmetry group is 60.

-$12_{(24)}$ : The 12 charges are at the vertices of a cuboctahedron
(center of the edges of a cube). The order of the symmetry group
is 24.

{}From symmetry we see that the electric field created by
the 11 charges on the twelfth located at the pole is zero:
$\sum_j {1\over \xi_j}=0$ and the equations \viii\ are all
satisfied.  Turning to the question of stability
one finds that the configuration with highest
symmetry group has lowest energy.  In appendix B we
show that the configuration of the cuboctahedron is unstable.
By a similar method we can show that the icosahedron is stable,
that is the Hessian is semi-positive.  The Hessian cannot be
strictly positive since it admits three zero eigenvalues
corresponding to the three generators of global rotations.
The equation for the vibration modes $\lambda , (\lambda=\omega^2)$
is written as
\eqn\bvi{
\sum_{j\atop j\neq i} {\delta_i-\delta_j\over (z_i-z_j)^2}
+ (N-1) (\bar{\delta}_i(1-\lambda)-\bar{z}_i^2 \delta_i)=0}
for the $2N$ eigenvalues $(\delta_i , \bar{\delta}_i)$.
We exhibit in appendix B one ternary unstable mode of the
cuboctahedron.

Let us briefly present the method of calculation of the
potential function ${\cal V}_N$ for some symmetric polyhedral
configurations belonging to the octahedral group or the icosahedral
group with $N=8,12,14,20,30,42$.  We will also give a
table of residual energies to compare with the previous
bounds.

Given formula \bi\  we only need to know
distances between the vertices of the polyhedron, only
3 distinct ones in the case of the icosahedron. For higher
configurations it is more appropriate to compute systematically
the discriminant $\Delta^2$ (formula \bii\ ) knowing the
invariant polynomials (having roots $z_j$ in the complex
plane corresponding to the vertices on the sphere).  Let us
take for instance the configuration $12_{(24)}$ of
mid-edges of a cube (or the cuboctahedron).  Klein \klein\
gives the following polynomial
\eqn\bvii{\eqalign{
f(z)&=z^{12}+1-33 (z^8+z^4)\cr
&\equiv (z^4+1)(z^4-1)(z^4-a^{-4})}}
with
$$a^2+a^{-2}=6, \ \ a^{-1}-a=2$$
We get $a^2=3-\sqrt{8}$
and $a^{-2}=3+\sqrt{8}$.  We have $\Delta^2=\prod_{j=1}^{12}
f'(z_j)$.  The roots appear in four quartets.  We give the
value $f'$ for a typical root, then the product over the
quartet
$$\eqalign{
f'(e^{i\pi /4})&=4 e^{-i \pi /4} (1+a^4)(1+a^{-4})=4 e^{-i \pi /4} 6^2\cr
\prod_{4terms} f'(e^{i \pi /4})&=(4 \times 6^2)^4 \cr
f'(a)&=4 a^3 (a^4+1)(a^4-a^{-4})=4a^3\times 6 a^2\times 6 \times -2
\sqrt{8} \cr
\prod_{4terms} f'(a)&=(a^5 \ 2^3 \sqrt{8}\times  6^2)^4 \cr
\prod_{4terms} f'(a^{-1})&=(a^{-5} \ 2^3 \sqrt{8}\times 6^2)^4 \cr
\Delta^2&=(4\times 6^2 \times 2^6 \times 2^3 \times 6^4)^4=
(2^{17} \ 3^6)^4 \cr
{1\over 12} ln \vert \Delta \vert^2&=6.125058601 }$$
also
$$\eqalign{
(N-1) \sum_j ln (1+\vert z_j \vert^2)&=11 \times 4 ( ln2+
ln (4+\sqrt{8})(4-\sqrt{8}))\cr &=11 \times 2^4 ln 2
}$$
$$
{1\over 12} (N-1) \sum_j ln (1+\vert z_j \vert^2)=10.16615865...$$
$$
e^{-2 V_{12}}={\vert \Delta \vert^2\over \prod_j (1+\vert z_j
\vert^2)^{11}} = {(2^{17} \ 3^6)^4\over 2^{16 \times 11}}
= (2^{-9} 3^2)^{12}$$
$$
{2 {\cal V}_{12}\over 12}=-2 ln 3+9 ln 2=4.041100047$$

The same method applies in the following cases:

-$12_{(60)}$: Vertices of an icosahedron

\eqn\bviii{f(z)= z(z^{10}+11 z^5-1)\equiv z(z^5-a^5)(z^5+a^{-5})}
with
$$\eqalign{&
a^{-5}-a^5=11 \cr
&a=2 \cos {2 \pi\over 5}=2 \xi-1 \cr
&a^{-1}=2 \cos {\pi \over 5}=2 \xi \cr
&\vert \Delta \vert^2=5^{25}
\cr & 2 {\cal V}_{12}=55 ln 5-25 ln 5=30 ln 5.}$$

-$20_{(60)}$: Vertices of the dodecahedron

\eqn\bix{f(z)=z^{20}+1-228 (z^{15}-z^5)+494 z^{10}\equiv
(z^5+a^5)(z^5-a^{-5})(z^5+b^5)(z^5-b^{-5}).}
with
$$\eqalign{&
a^{-5}-a^5=114+50 \sqrt{5} \cr
& b^{-5}-b^5=114-50 \sqrt{5} \cr
& 0<a<b<1, \cr
& \vert \Delta \vert^2=2^{60} 3^{10} 5^{95} \cr
& 2 {\cal V}_{12}=190 ln (a+a^{-1})(b+b^{-1})-
(60 ln 2+10 ln 3+95 ln 5).}$$

-$30_{(60)}$: Mid-edges of the dodecahedron (or the
icosahedron).

\eqn\bx{
\eqalign{f(z)&=z^{30}+1+522 (z^{25}-z^5)-10005 (z^{20}+z^{10})\cr
&\equiv (z^{10}+1)(z^5-a^5)(z^5+a^{-5})(z^5+b^5)(z^5-b^{-5})}}
with
$$\eqalign{&a^{-5}-a^5=261+5^3 \sqrt{5}\cr
&b^{-5}-b^5=-261+5^3 \sqrt{5} \cr
&\vert \Delta \vert^2=2^{90} 3^{60} 5^{205} \cr
& 2 {\cal V}_{12}=290 ln 2(a+a^{-1})(b+b^{-1})-(90 ln 2
+60 ln 3+205 ln 5)}$$

-$42_{(60)}$: Configuration with the 12 vertices and the 30
mid-edges of the icosahedron.

$$\vert \Delta \vert^2=2^{90} 3^{60} 5^{305}$$
For this mixed configuration the discriminant is obtained by
multiplying $\vert \Delta_{12} \vert^2 \times \vert \Delta_{30}
\vert^2$ by the factor $\prod_{11f} f_{30}(z_i)$, the product
being on
the 11 roots of $f_{12}$ which gives us the extra
factor
$$\eqalign{&
(11^2+4)^5 (261+5^3 \sqrt{5}-11)^{10} (-261+5^3 \sqrt{5}+11)^{10}
= 5^{75}\cr &
2 V_{12}=410 ln 2(a+a^{-1})(b+b^{-1})-(100 ln 5+60 ln 3
+90 ln 2)}$$
The results for the free energy per particle ${{\cal V}\over N}$ and the
residual free energy $2 f(\infty)={2{\cal V}\over N}-{N-1\over 2}
+{1\over 2N} ln N!$ are shown in the following table.
We note that the approximate value $f(\infty )\simeq -0.32$
is in the limits defined in \hxxx\ namely : $-0.46 < -0.32 < 0.04$.
Finally H\"older's inequality restricts the width of the
interval from $0.5$ to
approximately $0.37$.
$$\vbox{\offinterlineskip
\halign{\tv\quad # & \quad\tv \quad
# & \quad \tv \quad  # & \quad \tv \quad  #& \quad \tv \quad  #
& \quad \tv \quad  #
&  \quad \tv #\cr
\noalign{\hrule}
\tvi $N$ & $N^{-1} ln \vert \Delta \vert^2 $& $X$
&$ F=N^{-1} {\cal V}$ &
$Y$& $-f$ &\cr
\tvi $8_{(24)}$ & & & 1.1280580 & 1.4186061 & 0.290548& \cr
\tvi $12_{(24)}$ & 6.1250586 & 10.166158 & 2.0205500 &
2.3335996 & 0.313049 &\cr
\tvi $12_{(60)}$ & 3.3529956 & 7.376590 & 2.0117973 &
2.3335996 & 0.3218023 &\cr
\tvi $20_{(60)}$ & 10.273577 & 18.081646 & 3.904034 &
4.2208047 & 0.3167702 &\cr
\tvi $30_{(60)}$ & 15.274491 & 27.880218 & 6.3028631 &
6.627848 & 0.3249848 &\cr
\tvi $42_{(60)}$ & 14.742346 & 33.202164 & 9.2299092 &
9.548976 & 0.3190676&
\cr
\noalign{\hrule} }} $$
\centerline{\bf Table I: Residual free energy at zero temperature on the}
\centerline{\bf sphere for some symmetrical configurations.}
\centerline{ (
with $X=(1-N^{-1}) \Sigma ln (1+\vert z \vert^2)$ and
$Y={1\over 4} (N-1-N^{-1} ln N!)$)}
\vskip 1cm

As already mentionned, we show in appendix B that for $N=12$
the configuration of the cuboctahedron is unstable.
Its potential energy is larger by an amount
$0.008752...$ than the one of the stable icosahedron configuration.
This energy difference
can be interpreted as an "activation energy"
associated to the cuboctahedral configuration
as compared to the
icosahedron.  It is interesting to note that the associated inverse
temperature $\beta_c=2 p_c=1/0.0087...\simeq 114$ is of
the order of magnitude of the estimated fusion temperature
corresponding to the transition from a quasi-solid
phase to a disordered phase.

\newsec{Expansion in characters}

We return to the expansion
of Laughlin's wave functions in terms of Slater determinants.
Using
the notations of section 2 we would like
to get a better understanding of the coefficients $g_Y^{(s)}$.
For short, when $s=1$ or $p=3$ corresponding to filling
fraction $1 \over 3$ we set $g_Y^{(1)} \equiv g_Y$.

Consider the monic polynomial of degree N, $P(x)$ with
roots $x_i$, $0 \leq i \leq N-1$,
\eqn\sI{P(x)=\prod_{i=0}^{N-1} (x-x_i)=\sum_{k=0}^N (-1)^k
\sigma_k x^{N-k} \ \ \ , \sigma_0 \equiv 1.}
It has a double root if and only if its discriminant vanishes
or equivalently if it has a common root with its derivative
\eqn\sII{P'(x)=\sum_{k=0}^{N-1} (-1)^k \sigma_k (N-k) x^{N-k-1}.}
The discriminant, as a symmetric function of the roots is obtained by
eliminating $x$ between \sI\ and \sII .  A systematic way to
perform this elimination is to think of the powers of $x$ as
independent variables and to consider the polynomials,
\eqn\sIII{\eqalign{ x^m P(x),   \ \ \  0 \leq m \leq N-2, \cr
		    x^n P'(x),  \ \ \  0 \leq n \leq N-1,}}
as $2N-1$ linear forms in the $2N-1$ variables
$x^0,x^1,\cdots,x^{2N-2}$. A root common to $P(x)$ and $P'(x)$
then requires the vanishing of the corresponding determinant
which is therefore proportional to the discriminant, in fact
with our conventions equal up to a sign.  For instance for the
classical cases,
\eqn\sIV{\eqalign{P(x)&= x^2-\sigma_1 x+\sigma_2 \cr
		   D&=\left\vert \matrix{1 & -\sigma_1 & \sigma_2 \cr
			      2 & -\sigma_1 & 0 \cr
			      0 & 2 & -\sigma_1 } \right\vert =
			      4 \sigma_2-\sigma_1^2=
			      -\Delta^2 .}}
\eqn\sV{\eqalign{P(x)&=x^3-\sigma_1 x^2+\sigma_2 x-\sigma_3 \cr
D&=\left\vert \matrix{1 & -\sigma_1 & \sigma_2 & -\sigma_3 & 0 \cr
	   0 & 1 & -\sigma_1 & \sigma_2 & -\sigma_3 \cr
	   3 & -2\sigma_1 & \sigma_2 & 0 & 0 \cr
	   0 & 3 & -2\sigma_1 & \sigma_2 & 0 \cr
	   0 & 0 & 3 & -2\sigma_1 & \sigma_2 }\right\vert \cr &=
           -\sigma_1^2 \sigma_2^2-18 \sigma_1 \sigma_2
          \sigma_3+4 \sigma_1^3 \sigma_3+4 \sigma_2^3
	  +27 \sigma_3^2=-\Delta^2 \ .} }
and in general,
\eqn\sVI{D=\left\vert \matrix{1 & -\sigma_1 & \sigma_2 & . &
. & (-1)^N \sigma_N & 0 & . \cr
0 & 1 & -\sigma_1 & . & . &
(-1)^{N-1} \sigma_{N-1} & (-1)^N \sigma_N & . \cr
\vdots & \vdots & \ddots & . & & \vdots & & . \cr
N & -(N-1) \sigma_1 & (N-2) \sigma_2 & . & .
 & 0 & \cdots & . \cr
0 & N & -(N-1) \sigma_1 & . & \cdots & (-1)^{N-1} \sigma_{N-1}
& \cdots & . \cr
\vdots & \ddots & & & \ddots & \vdots & \cdots & . \cr
0 & \cdots & \cdots & & N & -(N-1) \sigma_1 & \cdots &
(-1)^{N-1} \sigma_{N-1}} \right\vert \ . }

Alternatively knowing that
\eqn\sVII{P'(x)=\sum_{i=0}^{N-1} (x-x_0) \cdots \widehat{(x-x_i)}
\cdots (x-x_{N-1})}
we have,
$P'(x_i)=\prod_{j; j \neq i} (x_i-x_j)$,
and,
\eqn\sVIII{D(x_0, \cdots ,x_{N-1})=\prod_{0 \leq i \leq N-1}
\ \prod_{j; j\neq i} (x_i-x_j)=\prod_{i=0}^{N-1} P'(x_i) \ .}

In particular for $N$-th roots of unity
$x_i=exp({i \sqrt{-1} \ 2 \pi \over N})$
, $0 \leq i \leq N-1$,
\eqn\sIX{\eqalign{P(x)&=x^N-1 \cr
		  D(x_0, \cdots , x_{N-1})&=N^N \prod_{i=0}^{N-1}
		  (x_i)^{N-1}=(-1)^{N-1} N^N}  }
One verifies that this does agree up to a sign with the
above determinant.

{}From equation \vandisc\ we have also
\eqn\sX{\eqalign{&(-1)^{N (N-1) \over 2} D(x_0, \cdots, x_{N-1}) =
\Delta^2(x_0, \cdots , x_{N-1}) \cr &=
\left\vert \matrix{1 & 1 & \cdots & 1 \cr
	 x_0 & x_1 & \cdots & x_{N-1} \cr
	 \vdots & \ddots & \ddots & \vdots \cr
	 x_0^{N-1} & x_1^{N-1} & \cdots & x_{N-1}^{N-1}
	 }\right\vert \times
	 \left\vert \matrix{1 & x_0 & \cdots & x_0^{N-1} \cr
		  1 & x_1 & \cdots & x_1^{N-1} \cr
		  \vdots & \ddots & \ddots & \vdots \cr
		  1 & x_{N-1} & \cdots & x_{N-1}^{N-1}}\right\vert \cr
		  & \ \cr &
	=\left\vert \matrix{t_0 & t_1 & \cdots & t_{N-1} \cr
		  t_1 & t_2 & \cdots & t_N \cr
		  \vdots & \ddots & \ddots & \vdots \cr
		  t_{N-1} & \cdots & \cdots & t_{2N-2}}\right\vert
}}
where,
\eqn\sXI{\eqalign{& t_k=x_0^k+x_1^k+ \cdots +x_{N-1}^k \cr
		  & {P'(x)\over P(x)} = \sum_{k=0, \vert x
		  \vert > \vert x_i \vert}^{\infty}
		   {t_k \over x^{k+1}}}}
In particular $t_0=N$. For a generalization (due to Shiota)
of formula \sX\ for
all $s$  see appendix C.
In the example of $N$-th roots of
unity $t_k=0$ except when $k$ is a multiple of $N$, in which case it
is equal to $N$, the above reduces to
\eqn\sXII{(-1)^{N(N-1) \over 2} D= N^N
\left\vert \matrix{1 & 0 & 0 & \cdots & 0 \cr
	 0 & \cdots & \cdots & 0 & 1 \cr
	 0 & \cdots & \cdots & 1 & 0 \cr
	 \vdots & \ddots & & \ddots & \cdots \cr
	 0 & 1 & 0 & \cdots & \cdots }\right\vert
	 =N^N (-1)^{(N-1) (N-2)\over
	 2}}
in agreement with \sIX .

Let us apply \sX\
to give a concise - but not very effective -
formula  for the coefficients $g_Y$ in the expansion over
Schur functions.  For this purpose we recall Frobenius reciprocity
formula expressing a product of $t$'s as a combination of
Schur functions $ch_Y$ having coefficients $\chi_Y$, the characters
of the symmetric group of order $\vert Y \vert$ pertaining to the
same set of Young tableaux.  Recall that $\chi_Y$ is only a class
function on $S_{\vert Y \vert}$.  We denote the classes $(1)^{\alpha_1}
, (2)^{\alpha_2}, \cdots ,$ i.e. $\alpha_1$ cycles of length 1,
$\alpha_2$ cycles of length 2 , etc... then
\eqn\sXIII{t_1^{\alpha_1}t_2^{\alpha_2} \cdots=
\sum_{Y: \vert Y \vert=\sum k \alpha_k} \chi_Y ((1)^{\alpha_1}
(2)^{\alpha_2} \cdots) \ ch_Y (x_0, \cdots , x_{N-1})}
For each class $(1)^{\alpha_1}(2)^{\alpha_2} \cdots$ we have
a number - in fact a relative integer - that we write
$\chi_Y((1)^{\alpha_1} (2)^{\alpha_2}
\cdots)$.  Define symbols $(1) (2) \cdots$ that can be
multiplied and linearly combined (with coefficients in \IZ\ , \IQ\
,\IR\ or \IC).  Attribute to (k) the degree k and consider any
linear combination of fixed degree $\vert Y \vert$.  Then we
can extend the definition of $\chi_Y$ by linearity
\eqn\sXIV{\chi_Y (\sum_{\alpha_1+2 \alpha_2+\cdots=\vert Y \vert}
a_{\alpha_1,\alpha_2,\cdots} (1)^{\alpha_1} (2)^{\alpha_2} \cdots)
=\sum a_{\alpha_1, \alpha_2,\cdots} \chi_Y ((1)^{\alpha_1}
(2)^{\alpha_2} \cdots )}
If the coefficients are integers so will be the value of $\chi_Y$.
With this notation and for $s=1$, $g^{(1)}_Y=g_Y$, we find
by combining \sX\ and \sXIII ,
\eqn\sXV{\eqalign{&\Delta^2(x_0,\cdots ,x_{N-1})=
\cr &\sum_{\vert Y \vert=N(N-1)}
\chi_Y \left\vert \matrix{N & (1) & (2) & \cdots & (N-1) \cr
		(1) & (2) & (3) & \cdots & (N) \cr
		\vdots & \ddots & \ddots & & \vdots \cr
		(N-1) & \cdots & & \cdots & (2N-2) }\right\vert
		ch_Y(x_0,\cdots , x_{N-1}) ,}}
where the sum on the r.h.s. extends over Young tableaux with $N(N-1)$
boxes, {\it at most N lines} (otherwise $ch_Y(x_0, ... , x_{N-1})$
vanishes) {\it and at most $2(N-1)$ columns}
(since the leading term in $\Delta^3(x_0,\cdots ,x_{N-1})$
is $\Delta(x_0^3,\cdots ,
x_{N-1}^3)$ ).
 We call these Young tableaux
{\it weakly admissible} and generalize this notion for arbitrary $s \geq 1$
to those tableaux of area $s N(N-1)$ contained in a rectangle
of size $2 s (N-1) \times N$.  The reason for the qualifier {\it weakly}
is that some of the tableaux have necessarily vanishing coefficients
as will be shown below.  In any case for (weakly) admissible
tableaux,
\eqn\sXVI{g_Y = \chi_Y \left\vert \matrix{N & (1) & \cdots & (N-1) \cr
  				(1) & (2) & \cdots & (N) \cr
				\vdots & \ddots & & \vdots \cr
				(N-1) & (N) & \cdots & (2N-2)}
				\right\vert \ .}
Even though this formula solves in principle the problem of finding
an explicit expression in the case $s=1$ it is not very manageable.
Indeed the r.h.s. is a linear combination of $N!$ values of
characters of $S_{N(N-1)}$, a formidable task to evaluate.  It
is however possible to dispense altogether with the characters
$\chi_Y$ of the symmetric group as follows.  The characters
$ch_Y$ of the linear group are expressed in terms of the normalized
traces $\theta_k={t_k\over k}$ \traces\ as
\eqn\nouv{\eqalign{ch_{Y(f_0,...,f_N)}& =\left\vert
\matrix{p_{f_{N-1}}(\theta_1)& \cdots & * \cr
	  & \ddots & \cr
	  * & & p_{f_0}(\theta .)}\right\vert  \cr
	  &= \sum_{\Sigma_{k\geq 1} k \alpha_k=\vert Y \vert}
	  \chi_Y ((1)^{\alpha_1} (2)^{\alpha_2} ...)
	  {\theta_1^{\alpha_1}\over \alpha_1 !}
	  {\theta_1^{\alpha_2}\over \alpha_2 !}
	  \cdots }}
Here $p_n(\theta_1)$ stands for the elementary Schur functions,
the characters of the n-th symmetric power given by
\eqn\nouvi{p_n(\theta .)=\sum_{\Sigma_{k \geq 1} k \alpha_k=n}
{\theta_1^{\alpha_1}\over \alpha_1 !}
{\theta_2^{\alpha_2}\over \alpha_2 !}
\cdots }
The $*$ in the determinant indicate that moving a unit step
to the right (left) the index of $p$ increases (decreases) by
unity, and one agrees that $p_0=1$ as well as $p_n=0$ for
n negative.  The second part of \nouv\ (from which \nouvi\ follows)
is obtained by inverting eq. \sXIII\ using the orthogonality
and reality of the $\chi_Y$'s.  For short set $\partial_k
\equiv {\partial\over \partial\theta_k}$ in such a way that
\eqn\nouvii{\partial_k p_n(\theta .)=p_{n-k} (\theta .)}
Using these notations we rewrite \sXVI\ as
\eqn\nouviii{g_Y=\left\vert
\matrix{N & \partial_1 & \cdots & \partial_{N-1} \cr
	\partial_1 & \partial_2 & \cdots & \partial_N \cr
	\vdots & \ddots & \ddots & \vdots \cr
	\partial_{N-1} & \partial_N & \cdots & \partial_{2N-2}}
	\right\vert \times \left\vert
\matrix{p_{f_{N-1}}(\theta .) & & & * \cr
	 & \ddots & & \cr
	 & & \ddots & \cr
	 * & & & p_{f_0}(\theta .)}\right\vert }
This gives a more manageable way of computing these
coefficients.  Notice that here we treat the $\theta$'s
as independent variables and that the amount of derivatives
is such as the right hand side is a pure number (indeed
an integer).

Finally if we observe that
\eqn\nouviv{exp \sum_1^\infty z^k \theta_k \ p_n(\theta .)=
{1\over n!} \left( {\partial\over \partial z} \right)^n exp
\sum_1^\infty z^k \theta_k}
we can even dispense with the $p_n$'s and reach a simple
expression
\eqn\nouvv{g_Y=\left\vert \matrix{
{\partial_{x_{N-1}}^{l_{N-1}-N+1}\over (l_{N-1}-N+1)!}  &
{ \partial_{x_{N-1}}^{l_{N-1}-N+2}\over (l_{N-1}-N+2)!} &
. &
{\partial_{x_{N-1}}^{l_{N-1}}\over l_{N-1}!} \cr
\vdots & \ddots & . & \vdots \cr
\vdots & \ddots & . & \vdots \cr
{\partial_{x_0}^{l_0-N+1}\over (l_0-N+1)!} &
{\partial_{x_0}^{l_0-N+2}\over (l_0-N+2)!} &
. &
{\partial_{x_0}^{l_0}\over l_0!} }
\right\vert \times \left\vert
\matrix{t_0 & t_1 & . & t_{N-1} \cr
	\vdots & \ddots & . & \vdots \cr
	\vdots & \ddots & . & \vdots \cr
	t_{N-1} & \cdots & . & t_{2N-2}}
	\right\vert}
As before $t_0=N$ and on the right hand side
the terms ${1\over k!}\partial_x^k$
where $k<0$ or $k>2N-2$ are set equal to zero.

{}From these expressions we derive a number of properties,
starting from the most obvious ones which follow
immediately from the definition, with $p=2s+1$

\noindent{\bf Property 0}

{\it
\item{(i)} All coefficients $g_Y^{(s)}$ are integers.
\vskip .3cm

\item{(ii)} $g^{(s)}_{l_0,... l_{N-1}}=
g^{(s)}_{p(N-1)-l_{N-1},
p(N-1)-l_{N-2},... p(N-1)-l_0}$
\vskip .3cm

\item{(iii)} $g^{(s),N+1}_{l_0,... l_N=pN}=g^{(s),N}_{l_0,
... l_{N-1}}$  ,
\vskip 0.2cm
$ g^{(s),N+1}_{0, l_0+p,...
l_{N-1}+p}=g^{(s),N}_{l_0,... l_{N-1}}$
\foot{where the
extra index, N or N+1 refers to the number of variables.}.
}
\vskip 1cm

Assume that $s=1$ and suppose that $Y$ is contained in an
$N \times N$ square (Its area being $N(N-1)$ this is indeed
possible). This requires $f_{N-1} \leq N$ or $l_{N-1} \leq
2N-1$.  If this is the case $Y$ and its symmetric
$\tilde{Y}$ with rows
and columns interchanged, are both weakly admissible.
Since $\chi_Y$ and $\chi_{\tilde{Y}}$ differ only by the
sign of the permutation and since a cycle $(k)$ has parity
$(-1)^k$, we have,
\eqn\sXVII{g_{\tilde{Y}}=\chi_Y
\left\vert \matrix{N & (1) & -(2) & \cdots & (-1)^{N-2} (N-1) \cr
	 (1) & -(2) & (3) & \cdots & (-1)^{N-1}(N) \cr
	 \vdots & \ddots & & \vdots & \vdots \cr
	 (-1)^{N-2} (N-1) & \cdots & & \cdots & -(2N-2) }
	 \right\vert \ . }
If we set with A and B integers,
\eqn\sXVIII{g_Y=AN+B}
then,
\eqn\sXIX{(-1)^N g_{\tilde{Y}} = -AN+B}
consequently $g_Y-(-1)^N g_{\tilde{Y}}=2AN$ and,

\noindent{\bf Property 1}

{\it
For $Y$ and $\tilde{Y}$ weakly admissible $g_Y \equiv (-1)^N
g_{\tilde{Y}} \ \ mod \ 2N$.

\vskip .2in
\noindent {\bf Example:}

\def\tvp{\vrule height 3.2pt depth 1pt} 
\def\thp{\vrule height 0.4pt width 0.45em}
\def\cc#1{\hfill#1\hfill}
\setbox2=\vbox{\offinterlineskip
\cleartabs
\+ \thp&\thp&\cr
\+ \tvp\cc{}&\tvp\cc{}&\tvp\cr
\+ \thp&\thp&\cr}
\setbox11=\vbox{\offinterlineskip
\cleartabs
\+ \thp&\cr
\+ \tvp\cc{}&\tvp\cr
\+ \thp&\cr
\+ \tvp\cc{}&\tvp\cr
\+ \thp&\cr}
\eqn\sXX{\Delta^2(x_0,x_1)=ch_{\box2}(x_0,x_1)-3 ch_{\box11} (x_0,x_1)}
}
and $1\equiv -3 \ mod \ 4$. For more convincing cases one may look
at the tables in appendix D.

Now we have a slight refinement of Property 0.  Suppose that $p=2s+1$ is
an odd prime, then from Fermat's theorem $(x_i-x_j)^p\equiv x_i^p
 -x_j^p \ \ mod \ p$.  Thus,
\eqn\sXXI{\Delta^p(x_0, \cdots , x_{N-1})\equiv \Delta (
x_0^p, \cdots , x_{N-1}^p) \ \ mod \ p}
and

\noindent{\bf Property 2}

{\it
Apart from $g^{(s)}_{0,p,2p,\cdots , (N-1)p}=1$, all coefficients
$g^{(s)}_{l_0,\cdots , l_{N-1}}$ vanish mod p if $p=2s+1$ is
prime.
}

That $g^{(s)}_{0,p,\cdots ,(N-1)p}=1$ follows from the definition
\decslat\ there being a unique term $x_0^0 x_1^p \cdots x_{N-1}^{p(N-1)}$
with coefficient 1 in the expansion of $\left[ \prod_{i>j} (x_i-x_j)
\right]^p$.  In particular Property 2 applies for $s=1$ where apart from
the leading one, all coefficients are multiples of 3 (see appendix D).
One may generalize Property 2 when $p$ is composite using that if $p=p_1 p_2$
with $p_1$ prime then $\Delta(x_0,\cdots ,x_{N-1})^p\equiv
\Delta(x_0^{p_1}, \cdots , x_{N-1}^{p_1})^{p_2} \ \ mod \ p_1$.

The
expansion \decslat\ over Slater determinants ranges from the most
{\it extended} one with the sequence of $l$'s being $(0,p,2p,\cdots ,
(N-1)p)$ to the most {\it compact} one leaving a huge hole around
the origin (and {\it physically} very unlikely) with every level
from $s(N-1)$ to $(s+1)(N-1)$ occupied i.e. the sequence
\eqn\XXII{l_0=s(N-1), \ \ l_1=s(N-1)+1, \cdots \ , l_{N-1}=(s+1)(N-1)}
For general $s$ let us describe the admissible sequences of $l$'s
i.e. those for which the corresponding coefficient $g^{(s)}_{l_0,
\cdots , l_{N-1}}$ is not a priori zero.  Subtracting successively
$0,1, \cdots,N-1$ we get the admissible sequence of $f_0 \leq f_1
\leq \cdots \leq f_{N-1}$ of Young tableaux.

\noindent{\bf Property 3}

{\it
For $s \geq 1$ the set of admissible Young tableaux is
of the form:

$f_k=2sk+n_{k+1}-n_k, \ \ 0\leq k \leq N-1$

or equivalently,

$l_k=(2s+1) k +n_{k+1}-n_k$

with non negative integers $n_k$, such that

$n_0=n_N=0$ and $0\leq n_k \leq
{1\over 2}(n_{k+1}+n_{k-1})+s$.
}

The last inequality follows from the requirement that $f_k \geq
f_{k-1}$.  Clearly all admissible Y are weakly admissible since
$f_0\leq \cdots \leq f_{N-1} \leq 2s(N-1)-n_{N-1}$ and the last bound is
smaller or equal to $2s(N-1)$.

To prove this property we consider for odd $p$
\eqn\XXIII{\Delta(x_0, \cdots , x_{N-1})^p=
\prod_{i>j} (x_i-x_j)^p=\prod_{i>j}(1-{x_j\over x_i})^p
\times x_{N-1}^{p(N-1)} x_{N-2}^{p(N-2)}\cdots x_0^0 .}
For $i>j$ write ${x_j\over x_i}={x_{i-1}\over x_{i}}
\times {x_{i-2}\over x_{i-1}}\cdots {x_j\over x_{j+1}}$
and expand the last product in \XXIII .  One obtains a sum
of terms of the form
\eqn\XXIV{\eqalign{({x_{N-2}\over x_{N-1}})^{n_{N-1}} ({x_{N-3}\over
x_{N-2}})^{n_{N-2}} \cdots ({x_0\over x_1})^{n_1} \times
x_{N-1}^{p(N-1)} x_{N-2}^{p(N-2)} \cdots x_0^{0} \cr \equiv
x_{N-1}^{p(N-1)-n_{N-1}}x_{N-2}^{p(N-2)+n_{N-1}-n_{N-2}}
\cdots x_k^{pk+n_{k+1}-n_k}\cdots x_0^{n_1}}}
with integer coefficients and
non-negative powers.
Both sides in \XXIII\ being antisymmetric we will find on the
right hand side for each monomial the antisymmetric sum
of its permutations.  To identify these determinants it
is sufficient to retain those monomials for which the
exponents $pk+n_{k+1}-n_k$ form a strictly increasing
sequence.  This is Property 3.  We denote by $A_N^{(s)}$ the number
of admissible tableaux.

For $2 \leq N \leq 29$ we present in the following table the total
number $A_N^{(1)}\equiv A_N$
of admissible tableaux for $p=3$, $s=1$.
\vfill\eject
$$\vbox{\offinterlineskip
\halign{\ptv\quad # & \quad\ptv \quad
# & \quad \ptv \qquad  # &  \quad \ptv #\cr 
\noalign{\hrule}
\ptvi ${\rm N}$ &$ {\rm Number}\ A_N $& ${\rm Ratio}\
A_N / A_{N-1} $& \cr
\noalign{\hrule}
\ptvi $2$  &$2$ & $2.00000$ &\cr
\ptvi $3$  &$5$ & $2.50000$ &\cr
\ptvi $4$  &$16$ & $3.20000$ &\cr
\ptvi $5$  &$59$ & $3.68750$ &\cr
\ptvi $6$  &$247$ & $4.18644$ &\cr
\ptvi $7$  &$1111$ & $4.49797$ &\cr
\ptvi $8$  &$5302$ & $4.77227$ &\cr
\ptvi $9$  &$26376$ & $4.97472$ &\cr
\ptvi $10$ &$135670$ & $5.1436$ &\cr
\ptvi $11$ &$716542$ & $5.28150$ &\cr
\ptvi $12$ &$3868142$ & $5.39834$ &\cr
\ptvi $13$ &$21265884$ & $5.49769$ &\cr
\ptvi $14$ &$118741369$ & $5.58365$ &\cr
\ptvi $15$ &$671906876$ & $5.65857$ &\cr
\ptvi $16$ &$3846342253$ & $5.72451$ &\cr
\ptvi $17$ &$22243294360$ & $5.78297$ &\cr
\ptvi $18$ &$129793088770$ & $5.83515$ &\cr
\ptvi $19$ &$763444949789$ & $5.88201$ &\cr
\ptvi $20$ &$4522896682789$ & $5.92432$ &\cr
\ptvi $21$ &$26968749517543$ & $5.96271$ &\cr
\ptvi $22$ &$161750625450884$ & $5.99770$ &\cr
\ptvi $23$ &$975311942386969$ & $6.02972$ &\cr
\ptvi $24$ &$5909549998347426$ & $6.05913$ &\cr
\ptvi $25$ &$35966989049703188$ & $6.08624$ &\cr
\ptvi $26$ &$219805620101524672$ & $6.11131$ &\cr
\ptvi $27$ &$1348411202206287872$ & $6.13456$ &\cr
\ptvi $28$ &$8301060310442080256$ & $6.15617$ &\cr
\ptvi $29$ &$51270095697195679744$ & $6.17633$ &\cr
\noalign{\hrule} }} $$
\centerline{\bf Table II: Number $A_N$ of admissible tableaux for s=1.}

\vskip 1cm

The last
column gives the ratio of successive terms.  The above
reasoning does not however insure that this is exactly the total
number of terms in the expansion of $\Delta^2$ in characters
as some coefficients might still vanish.  However experience
up to N=5 (appendix D) seems to indicate that these accidents
do not happen.  It is curious to notice that the condition
$n_k \leq {1\over 2}(n_{k+1} +n_{k-1})+s$ is a discrete generalization
of subharmonic functions.

The number of admissible
tableaux is smaller or equal to the number of weakly admissible
ones
$$\Pi_{N,2s(N-1)} (sN(N-1))$$
where $\Pi_{N,M}(f)=\Pi_{M,N}(f)$ stands for the number of
partitions of the integer $f$ in at most $N$ (non-empty)
parts of size at most equal to $M$. Equivalently this is the
number of tableaux in a $M\times N$ rectangle of {\it area} $f$.
It is given in terms of the generating function,
\eqn\XXVI{\sum_{N,f\geq 0} \Pi_{N,M}(f) \ x^N q^f=
{1 \over (1-x) (1-xq) (1-xq^2)\cdots (1-xq^M)}}
Recall that in terms of an infinite sequence of variables
$\theta_1, \theta_2, \cdots$ the (elementary) Schur polynomials
(in the $\theta$'s) are defined through,
\eqn\XXVII{\exp\sum_1^\infty x^n \theta_n=\sum_0^\infty x^n p_n
(\theta .)}
Since
\eqn\XXVIII{{1\over\prod_{k=0}^M(1-xq^k)}=exp\sum_1^\infty
x^n {1\over n} \ {1-q^{n(M+1)}\over 1-q^n}}
we find,
\eqn\XXVIII{\sum_{f\geq 0} \Pi_{N,M}(f)q^f=p_N(\theta_k=
{1\over k} \ {1-q^{k(M+1)}\over 1-q^k})}
An elementary {\it fermion-boson equivalence} gives an alternative
generating function for the restricted number of partitions, namely
\eqn\XXIX{\sum_{N,f\geq 0}\Pi_{N,M}(f)y^Nq^{f+{N(N-1)\over 2}}=
(1+y) (1+yq) \cdots (1+yq^{N+M-1})}
Hence
\eqn\XXX{\sum \Pi_{N,M}(f)q^{f+{N(N-1)\over 2}}=
p_N(\theta_k={(-1)^k\over k}{1-q^{k(N+M)}\over 1-q^k})}
thus

\noindent{\bf Property 4}

{\it
The number of admissible tableaux is smaller than the
number of weakly admissible ones given by $\Pi_{N,2s(N-1)}
(sN(N-1))$.}

Property 3 suggests to write the expansion over characters using
the operator
\eqn\oper{\Delta (\tau_i)=\prod_{1\leq i\leq j\leq N-1} (1-
\tau_i \tau_{i+1} \cdots \tau_j)}
where the ``box shifting" operator\foot{The elementary action of
$\tau_i$ on a Young tableau is to move a box from line $N-i$ to line
$N-i+1$.  The actions of various $\tau$'s commute with each other.}
$\tau_i$ acts on characters as
\eqn\charbox{\tau_i \ ch_{l_0,...,l_{N-1}} = ch_{l_0,...,l_{i-1}+1,
l_i-1,...,l_{N-1}}.}
The expansion of $\Delta^{2s}$ over the characters is also obtained
by retaining the admissible part of the action of $\Delta^{2s+1}(\tau_i)$
on the leading character $ch_{0,(2s+1),2(2s+1),...,(2s+1)(N-1)}$,
namely that of the monomials $\tau_1^{n_1} ...\tau_{N-1}^{n_{N-1}}$
which produce a strictly increasing sequence of indices
$n_1 < (2s+1)-n_1+n_2<...<(2s+1)^{N-2}-n_{N-2}+n_{N-1}<
(2s+1)^{N-1} -n_{N-1}$, i.e.
\eqn\eplic{\eqalign{
\Delta^{2s+1}(\tau_i) \bigg\vert_{adm} &ch_{0,(2s+1),...,(2s+1)(N-1)}\cr=&
\sum_{n_1,...,n_{N-1}\atop admissible} C_{\{ n_1,...,n_{N-1}\} }^{(s)}
\left[ \tau_1^{n_1} \cdots \tau_{N-1}^{n_{N-1}} ch_{0,(2s+1),...(2s+1)
(N-1)} \right] \cr =&
\sum_Y g_Y^{(s)} ch_Y }}
where the $n$'s in the summation are as in Property 3, and
$C_{\{ n \} }^{(s)}=g_Y^{(s)}$ if $Y$ is obtained from the
tableau $\{ 0,(2s+1),...,(2s+1)(N-1) \} $ by action of a
monomial $\tau_1^{n_1}...\tau_{N-1}^{n_{N-1}}$.

This provides us with another compact formula for the $g_Y^{(s)}$,
by straightforward expansion of $\Delta (\tau_i )^{2s+1}$:
\eqn\encore{
g_Y^{(s)}=C_{\{ n\} }^{(s)}=\sum_{m_{ij}=m_{ji}\geq 0 \atop
1\leq i<j\leq N-1} (-1)^{\Sigma_i n_i+\Sigma_{i<j} m_{ij}}
\prod_{i<j} {2s+1 \choose m_{ij}} \prod_{i=1}^{N-1}
{2s+1\choose n_i-\sum_{k<i<j} m_{kj}-\sum_{l\neq i} m_{il}},
}
where the sum over $m_{ij}=m_{ji}$ is restricted to the values for which
the combinatorial factors make sense (${2s+1\choose n}$,for
$0\leq n \leq 2s+1$) we list here the first two cases

N=2:
\eqn\grii{g_Y^{(s)}=C_n^{(s)}=(-1)^n {2s+1 \choose n} ; \ \ 0\leq n \leq s}

N=3:
\eqn\griii{
g_Y^{(s)}=C_{n_1,n_2}^{(s)}=(-1)^{n_1+n_2} \sum_{m\geq 0}
(-1)^m {2s+1\choose m} {2s+1\choose n_1-m} {2s+1 \choose n_2-m}
}
with respectively $Y=\{ n,2s+1-n\} , \ 0\leq n\leq s$ for N=2 and
$Y=\{ n_1,2s+1-n_1+n_2,4s+2-n_2\} , \ 0\leq n_1,n_2; \ 2n_1-n_2\leq
2s, \ 2n_2-n_1 \leq 2s$.  More coefficients are presented in appendix D.
The expression \encore, although increasingly complicated with $N$,
provides us with a factorization property of the
$g_Y^{(s)}$

{\noindent \bf Property 5}

{\it For $g_Y^{(s)}=C_{\{ n_0,...,n_{N-1}\}}^{(s)}$ as above, we
have the factorization property:
\eqn\facto{C_{\{ n_0,...,n_{j-1},0,n_{j+1},...,n_{N-1}\}}^{(s)}=
C_{\{ n_0,...,n_{j-1}\}}^{(s)} \times C_{\{ n_{j+1},...,n_{N-1}\}}^{(s)}
}
}

This is a straightforward consequence of eqn \encore, as the vanishing
of $n_j$ implies that of all $m_{kl}, \ k\leq j\leq l$, therefore
the sum over $m_{ij}\geq 0$ breaks up into a product of two sums
over $m_{kl}$ with respectively $k<l<j$ and $j<k<l$.

Let us derive for $p=3$ yet an alternative expression for the
coefficients $g_Y$ starting from equation \sVIII\ which can be
rewritten
\eqn\XXXI{(-1)^{N(N-1)\over 2} \Delta^3(x_0, \cdots , x_{N-1})
= \vert P'(x) x^0 , P'(x) x^1 , \cdots , P'(x) x^{N-1} \vert }
Recall that this is understood as a determinant with $x_0$
substituted in the first line, $x_1$ in the second, ...
Introduce an other set of variables $\mu_0,\cdots , \mu_{N-1}$.
One readily sees that
\eqn\XXXII{\eqalign{
(-1)^{N(N-1)\over 2}& \Delta^3(x_0, \cdots , x_{N-1})
= \cr &{\partial^N \over \partial \mu_0 \partial \mu_1 \cdots
\partial \mu_{N-1} } \vert P(\mu ) \mu^0 , \cdots , P(\mu )
\mu^{N-1}\vert \bigg{\vert}_{\mu_k=x_k}}}
since $x_0, \cdots , x_{N-1}$ are the roots of $P$.  Set for
simplicity $\phi_m \equiv (-1)^{N-m} \sigma_{N-m}$, ($
\phi_N=\sigma_0\equiv 1$) in such a way that
\eqn\XXXIII{P(x)=\sum_{0 \leq k \leq N} (-1)^k \sigma_k
x^{N-k} \equiv \sum_{0 \leq m \leq N} \phi_m x^m }
\eqn\XXXIV{\eqalign{&(-1)^{N(N-1)\over 2}\Delta^3(x_0, \cdots
, x_{N-1})  \cr & ={\partial^N\over \partial\mu_0 \cdots
\partial\mu_{N-1} } \sum_{m_0,\cdots , m_{N-1}}
\phi_{m_0} \phi_{m_1-1} \cdots \phi_{m_{N-1}-N+1} \ \
\vert \mu^{m_0} \cdots \mu^{m_N-1} \vert \bigg{\vert}_{\mu_k=x_k}\cr
&={\partial^N\over \partial\mu_0 \cdots \partial\mu_{N-1}}
\sum_{m_0< \cdots < m_{N-1}} \left\vert
\matrix{\phi_{m_0} & \phi_{m_0-1} & . & \phi_{m_0-N+1} \cr
	\phi_{m_1} & \phi_{m_1-1} & . & \phi_{m_1-N+1} \cr
	\vdots     & \ddots       & . & \vdots         \cr
	\phi_{m_{N-1}}& \phi_{m_{N-1}-1} & . &
	\phi_{m_{N-1}-N+1}} \right\vert	\times
	\vert \mu^{m_0} ... \mu^{m_{N-1}} \vert
	\bigg{\vert}_{\mu_k
	=x_k} \cr  & \  \cr
	&=\sum_{m_0 < \cdots < m_{N-1}} m_0... m_{N-1}
	\left\vert
\matrix{\phi_{m_0} & \phi_{m_0-1} & . & \phi_{m_0-N+1} \cr
	\vdots & \ddots & . & \ddots \cr
	\vdots & \ddots & . & \vdots \cr
	\phi_{m_{N-1}} & \phi_{m_{N-1}-1} & . &
	\phi_{m_{N-1}-N+1} }\right\vert
	\times \vert x^{m_0-1} ... x^{m_{N-1}-1}\vert}}
We now change $m_k \rightarrow m_k +1$,  substitute
$\phi_m \rightarrow (-1)^{N-m} \sigma_{N-m}$, and collect
the signs, with the result that
\eqn\XXXV{\eqalign{\Delta^3(x_0,\cdots ,& x_{N-1})=
\sum_{m_0 < \cdots
< m_{N-1}} (-1)^{\Sigma m_k} (m_0+1) \cdots (m_{N-1}+1)\times \cr &
\times \left\vert \matrix{\sigma_{N-1-m_0} & \sigma_{N-m_0}& .
& \sigma_{2(N-1)-m_0} \cr
\vdots & \ddots & . & \vdots \cr
\vdots & \ddots & . & \vdots \cr
\sigma_{N-1-m_{N-1}} & \cdots & . &
\sigma_{2(N-1)-m_{N-1}}}
\right\vert \vert x^{m_0} ... x^{m_{N-1}} \vert }}
Set
\eqn\XXXVI{\eqalign{&m_k=f_k+k \ ; \ \ 0\leq k \leq N-1 ,\cr
&\sum m_k=\sum f_k + {N(N-1)\over 2}, \cr
&\prod_k (m_k+1)=\prod_k (f_k+k+1).
}}
Define the Young tableaux
\eqn\XXXVII{Y(f)=\{ 0\leq f_0 \leq f_1 \leq \cdots \leq f_N
\leq N-1 \} }
and $Y(f'')$ given by the sequence $0\leq f_0'' \leq
... \leq f_{N-1}'' \leq N-1$ with
\eqn\XXXVIII{\eqalign{&f''_{N-1}=N-1-m_0=N-1-f_0 \cr
			&\vdots   \hskip 5cm  \vdots \cr
		      &f''_{N-k}=N-1+k-m_k=N-1-f_k \cr
		      &\vdots \hskip 5cm   \vdots \cr
		      &f''_0=2(N-1)-m_{N-1}=N-1-f_{N-1}.}}
Then
\eqn\XXXIX{{\vert x^{m_0} \cdots x^{m_{N-1}} \vert \over
\Delta(x_0, \cdots , x_{N-1})} = ch_{Y(f)} (x_0, \cdots ,
x_{N-1})}
while according to a standard property of characters \quatre ,
\eqn\XXXX{\eqalign{&
\left\vert
\matrix{\sigma_{N-1-m_0} & \cdots & . & \sigma_{2(N-1)-m_0} \cr
	\vdots &           \ddots & . & \vdots \cr
	\vdots & \ddots & . & \vdots \cr
	\sigma_{N-1-m_{N-1}} & \cdots & . &
	\sigma_{2(N-1)-m_{N-1}}}
\right\vert \equiv \cr & \ \cr & \left\vert
\matrix{\sigma_{f''_{N-1}} & \sigma_{f''_{N-1}+1} & . &
\sigma_{f''_{N-1}+N-1} \cr
\sigma_{f''_{N-2}-1} & \cdots & . & \sigma_{f''_{N-2}+N-2} \cr
\vdots & \ddots & . & \vdots \cr
\sigma_{f''_0-(N-1)}& \cdots & . & \sigma_{f''_0} }
\right\vert=ch_{\widetilde{Y(f'')}} (x_0, ...,x_{N-1})
}
}
where $Y\rightarrow \widetilde{Y}$ corresponds to the
transposition (i.e. the interchange of lines and columns)
of Young tableaux (a notation already used) in which elementary
Schur polynomials ($p_k\equiv$ trace of $k$-symmetric power)
get interchanged with elementary symmetric functions
($\sigma_k\equiv$ trace of $k$-antisymmetric power).  This
is recorded in the following picture where $Y(f)$ and
$Y^c(f')\equiv \widetilde{Y(f'')}$ are two complementary parts
of an $N\times (N-1)$ rectangle.

\fig{Notations used for Young tableaux}{tableau1.eps}{5cm}
\figlabel\tab

This gives the final result
\eqn\final{\eqalign{&\Delta^2(x_0, \cdots , x_{N-1})=
\sum_{Y \in N\times (N-1)} (-1)^{{N(N-1)\over 2}+\sum f_k}
\cr & \times
(f_0+1) \cdots (f_{N-1}+N) \ ch_Y(x_0 , \cdots , x_{N-1})
\ ch_{Y^c} (x_0, \cdots , x_{N-1})}}
where the short hand notation $Y\in N\times (N-1)$ indicates
that the tableau is inscribed in the rectangle $N\times (N-1)$
and if Y is given by the sequence of $0\leq f_1 \leq \cdots
\leq f_{N-1}\leq N-1$,  $Y^c$ corresponds to the
complementary sequence $f_0'=0\leq f_1'\leq \cdots \leq
f_{N-1}' \leq N$.

As a (trivial) example using the notation $<l_{0}, l_{1}
\cdots , l_{N-1}>$ for $ch_{f_{0}=l_{0}, \cdots , f_{N-1}=l_{N-1}-N+1}$
we have,

\def\tvp{\vrule height 3.2pt depth 1pt} 
\def\thp{\vrule height 0.4pt width 0.45em}
\def\cc#1{\hfill#1\hfill}
\setbox3=\vbox{\offinterlineskip
\+ \thp&\thp&\thp&\cr
\+ \tvp\cc{}&\tvp\cc{}&\tvp\cc{}&\tvp\cr
\+ \thp&\thp&\thp&\cr  }
\setbox21=\vbox{\offinterlineskip
\cleartabs
\+ \thp&\thp&\cr
\+ \tvp\cc{}&\tvp\cc{}
                 &\tvp\cr
\+ \thp&\thp&\cr
\+ \tvp\cc{}&\tvp\cr
\+ \thp&\cr  }
\setbox111=\vbox{\offinterlineskip
\cleartabs
\+ \thp&\cr
\+ \tvp\cc{}&\tvp\cr 
\+ \thp&\cr
\+ \tvp\cc{}&\tvp\cr 
\+ \thp&\cr
\+ \tvp\cc{}&\tvp\cr 
\+ \thp&\cr  }
\setbox77=\vbox{\offinterlineskip
\cleartabs
\+ \thp&\cr
\+ \tvp\cc{}&\tvp\cr
\+ \thp&\cr}
\setbox2=\vbox{\offinterlineskip
\cleartabs
\+ \thp&\thp&\cr
\+ \tvp\cc{}&\tvp\cc{}&\tvp\cr
\+ \thp&\thp&\cr}
\setbox11=\vbox{\offinterlineskip
\cleartabs
\+ \thp&\cr
\+ \tvp\cc{}&\tvp\cr
\+ \thp&\cr
\+ \tvp\cc{}&\tvp\cr
\+ \thp&\cr}
\setbox222=\vbox{\offinterlineskip
\cleartabs
\+ \thp&\thp&\thp&\cr
\+ \tvp\cc{}&\tvp\cc{}&\tvp\cc{}&\tvp\cr
\+ \thp&\thp&\thp&\cr
\+ \tvp\cc{}&\tvp\cc{}&\tvp\cc{}&\tvp\cr
\+ \thp&\thp&\thp&\cr}
\setbox221=\vbox{\offinterlineskip
\cleartabs
\+ \thp&\thp&\thp&\cr
\+ \tvp\cc{}&\tvp\cc{}&\tvp\cc{}&\tvp\cr
\+ \thp&\thp&\thp&\cr
\+ \tvp\cc{}&\tvp\cc{}&\tvp\cr
\+ \thp&\thp&\cr}
\setbox22=\vbox{\offinterlineskip
\cleartabs
\+ \thp&\thp&\cr
\+ \tvp\cc{}&\tvp\cc{}&\tvp\cr
\+ \thp&\thp&\cr
\+ \tvp\cc{}&\tvp\cc{}&\tvp\cr
\+ \thp&\thp&\cr}
\setbox211=\vbox{\offinterlineskip
\cleartabs
\+ \thp&\thp&\thp&\cr
\+ \tvp\cc{}&\tvp\cc{}&\tvp\cc{}&\tvp\cr
\+ \thp&\thp&\thp&\cr
\+ \tvp\cc{}&\tvp\cr
\+ \thp&\cr}
\setbox31=\vbox{\offinterlineskip
\cleartabs
\+ \thp&\thp&\cr
\+ \tvp\cc{}&\tvp\cc{}&\tvp\cr
\+ \thp&\thp&\cr
\+ \tvp\cc{}&\tvp\cr
\+ \thp&\cr
\+ \tvp\cc{}&\tvp\cr
\+ \thp&\cr}
\setbox32=\vbox{\offinterlineskip
\cleartabs
\+ \thp&\thp&\cr
\+ \tvp\cc{}&\tvp\cc{}&\tvp\cr
\+ \thp&\thp&\cr
\+ \tvp\cc{}&\tvp\cc{}&\tvp\cr
\+ \thp&\thp&\cr
\+ \tvp\cc{}&\tvp\cr
\+ \thp&\cr}
\setbox33=\vbox{\offinterlineskip
\cleartabs
\+ \thp&\thp&\cr
\+ \tvp\cc{}&\tvp\cc{}&\tvp\cr
\+ \thp&\thp&\cr
\+ \tvp\cc{}&\tvp\cc{}&\tvp\cr
\+ \thp&\thp&\cr
\+ \tvp\cc{}&\tvp\cc{}&\tvp\cr
\+ \thp&\thp&\cr
}
\setbox44=\vbox{\offinterlineskip
\cleartabs
\+ \thp&\thp&\thp&\thp&\cr
\+ \tvp\cc{}&\tvp\cc{}&\tvp\cc{}&\tvp\cc{}&\tvp\cr
\+ \thp&\thp&\thp&\thp&\cr
\+ \tvp\cc{}&\tvp\cc{}&\tvp\cr
\+ \thp&\thp&\cr}
\setbox55=\vbox{\offinterlineskip
\cleartabs
\+ \thp&\thp&\thp&\cr
\+ \tvp\cc{}&\tvp\cc{}&\tvp\cc{}&\tvp\cr
\+ \thp&\thp&\thp&\cr
\+ \tvp\cc{}&\tvp\cc{}&\tvp\cr
\+ \thp&\thp&\cr
\+ \tvp\cc{}&\tvp\cr
\+ \thp&\cr}
\setbox66=\vbox{\offinterlineskip
\cleartabs
\+ \thp&\thp&\thp&\thp&\cr
\+ \tvp\cc{}&\tvp\cc{}&\tvp\cc{}&\tvp\cc{}&\tvp\cr
\+ \thp&\thp&\thp&\thp&\cr
\+ \tvp\cc{}&\tvp\cr
\+ \thp&\cr
\+ \tvp\cc{}&\tvp\cr
\+ \thp&\cr}

$$\Delta^2(x_0,x_1)=-2 ch_{\copy2}+3 ch_{\copy77} \
ch_{\copy77} -6 ch_{\copy11}
$$

$$=-2 ch_{\copy2} + 3(ch_{\copy2} + ch_{\copy11} ) -6 ch_{\copy11}
$$

$$=ch_{\copy2} - 3 ch_{\copy11}
= <0,3>-3 <1,2>$$

and the slightly less obvious one,

$$ \Delta^2(x_0,x_1,x_2)=-6 ch_{\copy222}+8 ch_{\copy77} \
ch_{\box221}
-10 ch_{\copy2} \ ch_{\copy22}$$

$$ -12 ch_{\copy11} \
ch_{\box211}+15 ch_{\copy21} \  ch_{\box21}-20
ch_{\box22} \  ch_{\box11} +24 ch_{\box111}
\  ch_{\box3}$$

$$-30 ch_{\box31} \  ch_{\box2} +40 ch_{\box32}
\  ch_{\box77}
-60 ch_{\copy33}  $$

$$=-6 ch_{\copy222} + 8(ch_{\copy44} +ch_{\copy222} + ch_{\copy55})
-10 (ch_{\copy44}+ch_{\copy55}+ch_{\copy33})$$

$$-12 (ch_{\copy44}+ch_{\copy66} + ch_{\copy55})+15
(ch_{\box44}+ch_{\copy66}+ch_{\copy222}+2 ch_{\copy55}+ch_{\copy33})
$$

$$-20( ch_{\box222}+ch_{\copy55})+24 ch_{\copy66}-30
(ch_{\box66}+ch_{\copy55}) + 40 (ch_{\box55}+ch_{\copy33})
$$

$$-60 ch_{\box33}$$

$$=<0,3,6>-3 <1,2,6>-3 <0,4,5>+6 <1,3,5>-15 <2,3,4>$$

Since for given $N$ the Clebsch Gordan rule yields,
\eqn\cg{ch_{Y_1} ch_{Y_2}=\sum_{Y_3} N_{Y_1,Y_2}^{Y_3} ch_{Y_3}}
with $N_{Y_1,Y_2}^{Y_3}$ non negative multiplicities, and
$\vert Y_3 \vert=\vert Y_1 \vert + \vert Y_2 \vert$, we rewrite
equation \final\ as
\eqn\equv{\eqalign{
&\Delta^2(x_0,\cdots ,x_{N-1})= \cr
&\sum_{Y} \left( \sum_{Y_1\in N\times (N-1)} N_{Y_1,Y_1^c}^{Y}
(-1)^{{N(N-1)\over 2}+\Sigma_0^{N-1}f_k(Y_1)}
(f_0(Y_1)+1) \cdots (f_{N-1}(Y_1)+N)\right) \cr
&\times ch_Y(x_0,\cdots , x_{N-1}) }}
Equivalently,
\eqn\equv{g_Y=\sum_{Y_1\in N\times (N-1)} N_{Y_1,Y_1^c}^Y
(-1)^{{N(N-1)\over 2}+\sum_0^{N-1} f_k(Y_1)} (f_0(Y_1)+1) \cdots
(f_{N-1}(Y_1)+N) }
The computation of $g_Y$ is here reduced to the application of the
Hall-Richardson rule to yield the Clebsch Gordan series over the
${\cal N}=\sum_f \Pi_{N,N-1} (f)$ products of characters pertaining
respectively to a Young tableau inscribed in an $N\times (N-1)$
rectangle and its complementary.

\newsec{Sum rules}

A famous identity due to Dixon (1891) states that \cinq\
\eqn\cI{\sum_{0\leq s\leq 2p} (-1)^{s+p}
{2p \choose s}^3={(3p)!\over (p!)^3}}
A considerably generalized form first conjectured by Dyson
was proved by Gunson and Wilson independently.  Let
as before,
\eqn\cII{P(x)=\prod_{0\leq i \leq N-1} (x-x_i)}
Set
\eqn\cIII{\CP (x_0 ... x_{N-1} ; a_0 ... a_{N-1})
= \prod_{0\leq i \leq N-1} \left( {P'(x_i)\over x_i^{N-1}}
\right) ^{a_i}
=\prod_{0\leq i \neq j\leq N-1} (1-{x_j\over x_i})^{a_i}}
Consider the constant term in $\CP$ understood  as a Laurent series
in the variables $x_i$
\eqn\cIV{G_N(a_0,\cdots , a_{N-1})=\oint \prod_{0 \leq i \leq N-1}
{d\theta_i\over 2 \pi} \ \CP (e^{i \theta_0}, ... ,e^{i\theta_{N-1}}
; a_0, ... , a_{N-1} )}
then

\noindent{\bf Theorem}(Dyson, Gunson, Wilson \refsix\ ):
\eqn\cV{G_N(a_0, \cdots , a_{N-1})={(\sum a_i)!\over \prod a_i!}}
Dixon's identity is nothing else than the particular case where
$N=3$ and $a_0=a_1=a_2=p$.  The remarkable proof due to Good \refsept\
goes
as follows. Remarking from Cauchy's formula that
\eqn\cauchy{\sum_i {x_i^{N-1}\over P'(x_i)} =
\oint {dz\over 2 \pi i} {z^{N-1}\over P(z)} =1}
we multiply $\CP (x_0, ...  x_{N-1} ; a_0, ... a_{N-1})$
by 1 written as above to get for the constant term,
\eqn\const{G_N(a_0,\cdots , a_{N-1})=\sum_{0\leq i\leq N-1}
G_N(a_0, \cdots , a_i-1, \cdots , a_{N-1})}
Since when a particular $a_i$ vanishes the variable
$x_i$ only appears with non negative power in
$\CP (x_0, ... x_{N-1}; a_0,..., a_i=0, ... a_{N-1})$
we can let $x_i=0$ in the computation of the constant term, i.e.
\eqn\constII{G_N(a_0, \cdots , a_i=0, \cdots , a_{N-1})
= G_{N-1}(a_0, \cdots , \hat{a_i}, \cdots , a_{N-1})}
Furthermore
\eqn\further{G_N(0,\cdots , 0)=1}
Finally $G_N$ is obviously symmetric in the $a_i$'s.  This
property together with \const\ \constII\ and
\further\ uniquely determines
$G_N$ recursively and it is clear that the r.h.s. in \cV\
satisfies the same conditions.  This suffices to prove the required
equality.

Let us apply Dyson's theorem in the following
form. The integral
\eqn\dys{\oint {d\theta_0 \over 2 \pi} \cdots {d \theta_{N-1}
\over 2 \pi} \vert \Delta (e^{i\theta_0}, \cdots , e^{i\theta_{N-1}})
\vert^{2p}}
is equal for $p=2s+1$ odd to ,
\eqn\dysI{N! \sum_{0\leq l_0 < \cdots < l_{N-1}\leq p(N-1)}
\vert g^{(s)}_{l_0,\cdots , l_{N-1}}\vert^2}
according to its expansion \decslatbis .  On the other hand it
is also equal to
\eqn\ino{\eqalign{&\oint {d\theta_0\over 2 \pi} \cdots {d \theta_{N-1}
\over 2 \pi} {\Delta(e^{i\theta_0} \cdots e^{i\theta_{N-1}})^{2p}
\over e^{i(\theta_0+\cdots +\theta_{N-1})(N-1)p} }
(-1)^{N(N-1)p \over 2} \cr
&=\oint {d \theta_0\over 2 \pi}\cdots {d \theta_{N-1}\over 2 \pi}
\prod_{0\leq k \leq N-1} \left[ {P'(e^{i\theta_k})\over
e^{i(N-1)\theta_k}} \right]^p \cr
&={(Np)!\over (p!)^N}
}}
where the last equality is obtained from \cV\ by setting all $a_i=p$.
Therefore for $p$ odd we conclude that we have the sum rule,

{\bf \noindent Property 6}
\eqn\srule{\sum_{0\leq l_0 < \cdots < l_{N-1} \leq p(N-1)}
\vert g^{(s)}_{l_0,\cdots , l_{N-1}}\vert^2={(Np)!\over N! (p!)^N}}
where both sides are integers (i.e. $N!$ divides $(Np)!\over (p!)^N$).
For example when s=1 we have
\eqn\ex{\sum_{0\leq l_0 < \cdots < l_{N-1}\leq 3(N-1)}
g^2_{l_0,\cdots , l_{N-1}}={(3N)!\over N! 3!^N}}
In a similar manner we can compute the coefficient
$g^{(s)}_{s(N-1),s(N-1)+1,\cdots,(s+1)(N-1)}$ of the most
compact term noting that the corresponding character is
$\sigma_N^{s(N-1)}\equiv (\prod_{0\leq k \leq N-1} x_k^{N-1})
^s$, hence
\eqn\coef{\eqalign{N! \ &g^{(s)}_{s(N-1),\cdots , (s+1)(N-1)}=\cr
&\oint
{d\theta_0\over 2 \pi}\cdots {d \theta_{N-1}\over 2\pi}
{\Delta(e^{i \theta_0},\cdots , e^{i\theta_{N-1}})^{2s+1}
\Delta(e^{-i\theta_0}, \cdots , e^{-i\theta_{N-1}})
\over \left(
\prod_{0\leq k \leq N-1} e^{i\theta_k(N-1)}\right)^s} \cr
&=(-1)^{N(N-1)s\over 2} \oint {d\theta_0\over 2\pi} \cdots
{d\theta_{N-1}\over 2\pi} \prod_{k=0}^{N-1} \left[
{P'(e^{i\theta_k})\over e^{i(N-1)\theta_k}}\right]^{s+1}
}}
This gives
\eqn\cg{g^{(s)}_{s(N-1),\cdots , (s+1)(N-1)}=(-1)^{N(N-1)s\over 2}
{[(s+1)N]!\over N! (s+1)!^N}}
In particular when $s=1$,
\eqn\cgI{g_{N-1,N,\cdots , 2N-2}=(-1)^{N(N-1) \over 2} (2N-1)!!}
{}From equation \partfunc\ and \srule\ it follows that,
\eqn\suup{{(Np)!\over (p!)^N} \inf_{Y adm} \ l_0! \cdots l_{N-1}!
\leq Z_N(p) \leq {(Np)!\over (p!)^N} \sup_{Y adm} \ l_0! \cdots
l_{N-1}!}
The sup and inf are taken over admissible tableaux.  We suspect that
these correspond respectively to the most extended and most compact
terms, in other words that
\eqn\bounds{{(Np)!\over (p!)^N} \prod_{k=s(N-1)}^{(s+1)(N-1)} k!
\leq Z_N(p) \leq {\prod_{k=1}^N (pk)!\over p!^N}}
Referring to equation (A.18) and our expectation \extexc\ we
see that these bounds are unfortunately quite miserable.
We add the following remark. Consider for simplicity in
$Z_N(3)$ the ratio $u_N$ of the contribution of the
most extended to the most compact term,
\eqn\ratio{u_N={\prod_{k=0}^N (3k)!\over \prod_{k=N-1}^{2(N-1)} k!
((2N-1)!!)^2}}
Then,
\eqn\ratioI{{u_{N+1}\over u_N}=2{  N! (3N)!\over (2N+1)!^2 }
\sim_{N\rightarrow \infty} {\sqrt{3} \over (2N+1)^2}
({27\over 16})^N}
Therefore for $N\rightarrow \infty$, $u_N$ goes to infinity
which suggests that in the large $N$ limit, $Z_N(3)$ is dominated
by the contributions from {\it extended} states in agreement
with "physical intuition".

\newsec{Linear system for the coefficients}

We now come to linear equations satisfied by the
expansion coefficients $g$, where for simplicity we will
set $s=1$ although all that will be said admits a generalization
to arbitrary $s$.  We start with the following obvious observation.
Let $Q(x_0,\cdots , x_{N-1})$ be an arbitrary symmetric polynomial
in $x_0,\cdots , x_{N-1}$, homogeneous in these variables of
degree $N(N-1)$ and vanishing when $x_0=x_1$.  Then
$Q(x_0,\cdots , x_{N-1})$ is proportional to $\Delta(x_0,\cdots
,x_{N-1})^2$.  Indeed it admits the factor
$x_0-x_1$, hence from symmetry is divisible by $\Delta$.  But
$Q\over \Delta$ is antisymmetric, thus is divisible by $\Delta$.
{}From the comparison of degrees, $Q\over \Delta^2$ is a constant.
Similarly an antisymmetric polynomial in $x_0, \cdots , x_{N-1}$,
of degree ${3\over 2} N(N-1)$ vanishes obviously at $x_0=x_1$.
If its derivatives also vanish at $x_0=x_1$ then it is proportional
to $\Delta (x_0,\cdots , x_{N-1})^3$.  Let us apply this to the
expansion \decslat\ - \decslatbis .  By requiring that the sum
\eqn\sixI{\sum_{0\leq l_0 < \cdots < l_{N-1} \leq
3(N-1) \atop \Sigma l_i=3N(N-1)/2}
g_{l_0,\cdots , l_{N-1}} \vert x^{l_0} \cdots x^{l_{N-1}}\vert
}
obviously antisymmetric in $x_0,\cdots ,x_{N-1}$ has vanishing
derivative at $x_0=x_1$, we will obtain a quantity proportional
to $\Delta^3$.  The condition reads (by multiplying it by $x_0$
assumed different from zero),
\eqn\cond{\sum_{0\leq l_0 < \cdots < l_{N-1} \leq
3(N-1) \atop \Sigma l_i=3N(N-1)/2} g_{l_0,\cdots , l_{N-1}}
\left\vert \matrix{l_0 x_0^{l_0} & l_1 x_0^{l_1} & \cdots & \cdots &
l_{N-1}
x_0^{l_{N-1}} \cr
x_0^{l_0} & x_0^{l_1} &\cdots & \cdots & x_0^{l_{N-1}} \cr
x_2^{l_0} & x_2^{l_1} & \cdots & \cdots & x_2^{l_{N-1}} \cr
\vdots & \ddots & \ddots & \vdots & \vdots \cr
x_{N-1}^{l_0} & x_{N-1}^{l_1} & \cdots & \cdots & x_{N-1}^{l_{N-1}}}
\right\vert =0	}
Let us expand this in powers of $x_0$ with coefficients which are
determinants in $x_2, \cdots , x_{N-1}$.  We set equal to zero
the coefficient of
$$x_0^{l_a+l_b} \vert x^{l_0} \cdots \hat{x^{l_a}}
\cdots \hat{x^{l_b}} \cdots x \vert  (x_2, \cdots , x_{N-1})$$
extending $g_{l_0,\cdots , l_{N-1}}$ as an antisymmetric tensor
in its indices to get

\noindent{\bf Property 7}

{\it
Given non negative integers $0\leq l_0 < \cdots < l_{N-3} \leq
3(N-1)$ with sum equal to ${3N(N-1)\over 2}-L$, $L>0$, the
coefficients g. extended as antisymmetric tensors are uniquely
defined by the conditions
\eqna\co{\eqalignno{\sum_{3(N-1) \geq l > l' \geq 0 \atop l+l'=L}
(l-l') &g_{l_0,\cdots ,l_{N-3},l',l} =0 &\co a \cr
&g_{0,3,\cdots ,3(N-1)}=1. &\co b }}
}
\vskip 2iN
\noindent {\bf Examples:}

$N=2$

\eqn\egII{\sum_{l > l'\geq 0 \atop l+l'=3} (l-l') g_{l',l}=
3 g_{0,3}+g_{1,2} \equiv 0}
i.e. $g_{0,3}=1$ and $g_{1,2}=-3$.
\vskip .2in

$N=3$

\eqn\exemp{\eqalign{
&(a) \ 3 g_{0,3,6}+g_{0,4,5}=0 \cr
&(b) \ 4 g_{1,2,6}+2 g_{1,3,5}=0 \cr
&(c) \ 5 g_{2,1,6}+g_{2,3,4}  \cr & \ \ \ \
\equiv -5 g_{1,2,6}+g_{2,3,4}=0 \cr
&(d) \ 6 g_{3,0,6}+4 g_{3,1,5}+2 g_{3,2,4}  \cr & \ \ \ \
\equiv -6 g_{0,3,6}-4 g_{1,3,5}-2 g_{2,3,4}=0 \cr
&(e) \ 5 g_{4,0,5}+g_{4,2,3} \cr & \ \ \ \
\equiv -5 g_{0,4,5}+g_{2,3,4}=0 \cr
&(f) \ 4 g_{5,0,4}+2 g_{5,1,3} \cr & \ \ \ \ \equiv
4 g_{0,4,5}+2 g_{1,3,5}=0 \cr
&(g) \ 3 g_{6,0,3}+g_{6,1,2} \cr & \ \ \ \ \equiv
3 g_{0,3,6}+g_{1,2,6}=0 \cr }}

Conditions (a) and (g) yield from the normalization
$$
g_{0,3,6}=1,\ g_{0,4,5}=g_{1,2,6}=-3$$
Conditions (b) and (c) give the remaining
coefficients
$$g_{1,3,5}=6\ ,  \ g_{2,3,4}=-15$$
One checks that (d), (e) and (f) are identically fulfilled.
This example illustrates the fact that we have an overdetermined
set of equations.  To make the algorithm really useful would
require to extract a subset of equations of co-rank equal to 1.
This means that if $A_N\equiv A_N^{(1)}$ is the number of admissible Young
tableaux to find a subset of $A_N$ equations of rank $A_N-1$.
Should this be possible we could solve for the $g$'s in terms
of minors of the corresponding matrix up to normalization.

A weak consequence follows from the fact
that $\Delta(x_0,\cdots ,x_{N-1})^2$ vanishes when all $x_i=1$
hence, since $ch_Y(1,\cdots ,1)\equiv dim_Y$.
\eqn\dim{\sum_{Y \ adm} g_Y \ dim_Y = \sum_{0\leq l_0 <
\cdots < l_{N-1} \leq 3(N-1)} g_{l_0,l_1,\cdots , l_{N-1}}
\prod_{i>j} \left( {l_i-l_j\over i-j} \right) =0}
where $dim_Y$ is the dimension of the corresponding representation
of the linear group.  When $N=2$ this coincides with our previous
condition, while for $N=3$ one finds
\def\tvp{\vrule height 3.2pt depth 1pt} 
\def\thp{\vrule height 0.4pt width 0.45em}
\def\cc#1{\hfill#1\hfill}

\setbox222=\vbox{\offinterlineskip
\cleartabs
\+ \thp&\thp&\thp&\cr
\+ \tvp\cc{}&\tvp\cc{}&\tvp\cc{}&\tvp\cr
\+ \thp&\thp&\thp&\cr
\+ \tvp\cc{}&\tvp\cc{}&\tvp\cc{}&\tvp\cr
\+ \thp&\thp&\thp&\cr}
\setbox33=\vbox{\offinterlineskip
\cleartabs
\+ \thp&\thp&\cr
\+ \tvp\cc{}&\tvp\cc{}&\tvp\cr
\+ \thp&\thp&\cr
\+ \tvp\cc{}&\tvp\cc{}&\tvp\cr
\+ \thp&\thp&\cr
\+ \tvp\cc{}&\tvp\cc{}&\tvp\cr
\+ \thp&\thp&\cr
}
\setbox44=\vbox{\offinterlineskip
\cleartabs
\+ \thp&\thp&\thp&\thp&\cr
\+ \tvp\cc{}&\tvp\cc{}&\tvp\cc{}&\tvp\cc{}&\tvp\cr
\+ \thp&\thp&\thp&\thp&\cr
\+ \tvp\cc{}&\tvp\cc{}&\tvp\cr
\+ \thp&\thp&\cr}
\setbox55=\vbox{\offinterlineskip
\cleartabs
\+ \thp&\thp&\thp&\cr
\+ \tvp\cc{}&\tvp\cc{}&\tvp\cc{}&\tvp\cr
\+ \thp&\thp&\thp&\cr
\+ \tvp\cc{}&\tvp\cc{}&\tvp\cr
\+ \thp&\thp&\cr
\+ \tvp\cc{}&\tvp\cr
\+ \thp&\cr}
\setbox66=\vbox{\offinterlineskip
\cleartabs
\+ \thp&\thp&\thp&\thp&\cr
\+ \tvp\cc{}&\tvp\cc{}&\tvp\cc{}&\tvp\cc{}&\tvp\cr
\+ \thp&\thp&\thp&\thp&\cr
\+ \tvp\cc{}&\tvp\cr
\+ \thp&\cr
\+ \tvp\cc{}&\tvp\cr
\+ \thp&\cr}

\eqn\zero{\eqalign{27 g_{\box44}+10 g_{\box66}+
10 g_{\box222} +8 g_{\box55} + g_{\box33} \cr
\equiv 27 - 10 \times 3 - 10 \times 3 + 8 \times 6 - 15=0}}
This may be interpreted as saying that $\Delta(x_0,...,x_{N-1})^2$
is the difference of two characters of $GL_N$ operating in two
vector spaces of equal dimension
$$\sum_{Y \ adm, g_Y>0} g_Y \ dim_Y$$

\newsec{Specializations}

In this section we study specializations of the
discriminant leading to sum rules on the coefficients $g_Y^{(s)}$.  For
simplicity we stick again to the case s=1.

As we saw when $x_k=e^{2 i \pi k\over N}$,
\eqn\ddel{\Delta^2(1,e^{2 i \pi\over N},\cdots , e^{2 i \pi (N-1)
\over N})=(-1)^{(N-1)(N-2)\over 2} N^N}
For any integer $l$ let $<l>$ denote its representative
mod N in the range $0,1,\cdots ,N-1$. If the sequence $0\leq l_0
< \cdots < l_{N-1} \leq 3(N-1)$ is such that in the sequence
$<l_0>,\cdots ,<l_{N-1}>$ two numbers coincide, the
corresponding character $ch_{l_0,\cdots ,l_{N-1}} (1,e^{2 i\pi\over N}
,\cdots )$ vanishes.  If all $<l_0>,\cdots ,<l_{N-1}>$ are distinct
let $\epsilon (l_0,\cdots ,l_{N-1})$ denote the sign of
the permutation
\eqn\perm{\pmatrix{<l_0> & \cdots & <l_{N-1}> \cr
		   0     & \cdots & N-1 }}
in which case we have
\eqn\chp{ch_{l_0,\cdots ,l_{N-1}} (1,e^{2 i\pi\over N},\cdots
,e^{2 i \pi (N-1)\over N})=\epsilon (l_0,\cdots ,l_{N-1}).}
Thus,
\eqn\chpI{N^N (-1)^{(N-1)(N-2)\over 2}=\sum_{
Yadm  \atop \{ <l_o> \cdots <l_{N-1}> \} perm. \ of \
\{ 0 \cdots N-1 \} } g_{l_0,\cdots , l_{N-1}}
\ \epsilon (l_0,\cdots , l_{N-1})}
The reader can check this sum rule using the listings in
appendix D.

More
generally we can specialize the $x_i$'s to be the successive
powers of a single variable $q$, the so called principal
specialization,
\eqn\qvar{x_i=q^i \ \ \ 0\leq i \leq N-1}
Thus
\eqn\qva{\eqalign{\Delta(1,q,\cdots , q^{N-1})&=(-1)^{N(N-1)\over 2}
\prod_{0\leq i < j \leq N-1} (q^i-q^j) \cr &=
(-1)^{N(N-1)\over 2} q^{\sum_0^{N-1} n(N-1-n)} \times
[1]_q [2]_q \cdots [N-1]_q }}
where the $q$-factorial is defined as
\eqn\qfact{[n]_q=(1-q)(1-q^2)\cdots (1-q^n); \ \ \lim_{q
\rightarrow 1} {[n]_q\over (1-q)^n} = n!}
and
\eqn\qfactI{\sum_{n=0}^{N-1} n(N-1-n)={N(N-1)(N-2)\over 6}}
Note that the symbol $[\infty ]_q$ makes sense for $\vert q \vert
< 1$ being equal to $q^{-1/24} \eta (q)$ where $\eta$ is Dedekind's
function, and that Jacobi's identity reads
\eqn\jaco{\prod_{n>0} (1-q^n)^3
=\sum_{j>0} (-1)^j (2j+1) q^{j(j+1)\over 2}}
Under the same specialization,
\eqn\spe{\eqalign{&\vert x^{l_0} \cdots x^{l_{N-1}} \vert =
det \ q^{a l_b} \bigg{\vert}_{0\leq a , b \leq N-1} \cr
&=\Delta(q^{l_0}, \cdots ,q^{l_{N-1}}) = (-1)^{N(N-1)\over 2}
\prod_{0\leq i < j \leq N-1} (q^{l_i}-q^{l_j}) \cr
&=(-1)^{N(N-1)\over 2} q^{\sum_{k=0}^{N-1} l_k (N-1-k)}
\prod_{0\leq i<j \leq N-1} (1-q^{l_j-l_i})}}
It follows from \qva\ and \spe\ that
\eqn\septI{\eqalign{(&[1]_q [2]_q \cdots  [N-1]_q)^{2s+1}
=\cr & \sum_{0\leq l_0 < \cdots < l_{N-1} \leq (2s+1)(N-1)
\atop \Sigma l_i=(2s+1){N(N-1)\over 2}}
g^{(s)}_{l_0,\cdots ,l_{N-1}}
\ q^{\Sigma_{k=0}^{N-1} [l_k-(2s+1)k] (N-1-k)}
\prod_{0\leq i<j\leq N-1} (1-q^{l_j-l_i}) }}
Set
\eqn\septII{\mu_k=l_{N-1-k}-(2s+1)(N-1-k) \ \ \ 0\leq k\leq N-1}
which represents, counted from the "top" the departures of
the $l$'s from their values in the most extended case.  The
exponent in the explicit $q$-factor on the r.h.s. of \spe\ reads
\eqn\expo{E(l_0,\cdots , l_{N-1}) =\sum_{0}^{N-1}
[l_k-(2s+1)k](N-1-k)= \sum_{k=1}^{N-1} k \mu_k
=\sum_{k=1}^{N-1} n_k \geq 0}
in terms of the notation introduced in section 4.  Its extreme
values are,
\eqn\extreme{\eqalign{&E(0,2s+1,2 (2s+1),\cdots , (N-1)(2s+1))=0 \cr
&E(s(N-1),\cdots , (s+1)(N-1))=\sum_{k=1}^{N-1} s k (N-k)
= {s(N-1)N(N+1)\over 6}}}
Since the power of $(1-q)$ occurring on the l.h.s. is higher than the
one in any term
on the r.h.s. we recover \dim\ and its generalization to
all $s$ by letting $q\rightarrow 1$ on both sides of \septI .
Of course one can equate further terms to zero as $q\rightarrow 1$
(in fact as many as $sN(N-1)$ ) to require a similar leading behavior
on both sides, but this would be less effective than the equations
discussed in section 6.

Finally we can rewrite \septI\ as
\eqn\rewr{([1]_q \cdots [N-1]_q)^{2s}=\sum_{Y \ admiss} g_Y^{(s)}
q^{\sum_{k=1}^{N-1} n_k(Y)} \prod_{b\in Y} \left[
{1-q^{c(b)}\over 1-q^{h(b)}} \right] }
where following Macdonald  \quatre\
$b$ runs over the boxes of a tableau,
$h(b)$ is the hook length at $b$, i.e. one plus the number of
boxes at the right or below $b$, and $c(b)$ is its
"label" or "content", i.e. the sum $i+j$ where $i$ is the line index
starting with zero from the $N$-th bottom line and $j$ its
column index starting at 1 at the left as shown in the
illustration for $N=4$ where the shaded box has $h(b)=7$ and
$c(b)=6$.
Equivalently the label of the left uppermost box
is $N$ and labels increase (decrease) by a unit when moving
one step to the right (down).

\fig{Hook length and label of
a box}{tableau2.eps}{5cm}
\figlabel\tabb
The
expressions occurring on the right hand side of equation \rewr\
generalize the $q$-binomial coefficients.  For a tableau with a
unique column $Y_k$ of height $k\leq N$ we have,
\eqn\qbin{\prod_{b\in Y_k} \left(
{1-q^{c(b)}\over 1-q^{h(b)}}\right) =
{(1-q^N)(1-q^{N-1})\cdots (1-q^{N-k+1})\over (1-q) (1-q^2)
\cdots (1-q^k)}={[N]_q\over [k]_q [N-k]_q}}
while for a single row of length $k$, call it $\widetilde{Y_k}$,
\eqn\sow{\prod_{b\in \tilde{Y_k}} \left(
{1-q^{c(b)}\over 1-q^{h(b)}} \right)
={(1-q^N)(1-q^{N+1})\cdots
(1-q^{N+k-1})\over (1-q) (1-q^2)\cdots (1-q^k)}=
{[N+k-1]_q \over [k]_q [N-1]_q}}
and as $q\rightarrow 1$, they reduce to
\eqn\redu{\prod_{b\in Y} {c(b)\over h(b)}=dim Y(GL_N)}
which gives one of the fastest means to compute these dimensions.  Even
though equations \septI\ or \rewr\ depend on a single variable $q$
it is still a tremendous task to extract from these
equalities the coefficients $g$  for instance by expanding
at small $q$.

\newsec{\bf The number of admissible tableaux}

The number of admissible tableaux $A_N^{(s)}$ is according to
property 3 of section 4. the number of
$N+1$--uples of non--negative integers
$(n_0=0,n_1 \geq 0, ..., n_{N-1} \geq 0, n_N=0)$ such that
\eqn\polito{
2s-2 n_k +n_{k-1} + n_{k+1} \geq 0 \qquad k=1,2,..,N-1.}
\fig{The polytope $\Pi_{4}^{(1)}$ and its $16$ integer points. The
black dots belong to vertices or edges, the half--filled to
faces, and the empty one is strictly inside the polytope.}{politope.eps}{8cm}
\figlabel\polit
This counts the number of points with integral coordinates in the
closed polytope
$\pi_N^{(s)}$ defined by the equations \polito.
For illustration, we represent the $3$--dimensional
polytope $\pi_{4}^{(1)}$ and its $16$
points with integer coordinates on Fig.\polit.
We first concentrate on the number $A_N \equiv A_N^{(1)}$, for $s=1$.

Let $f_N(x_0,...,x_N)$ be the Laurent series
$$ f_N(x_0,...,x_N)= {x_1^2 x_2^2 ... x_{N-1}^2 \over
\prod_{j=1}^{N-1} (1 - {x_{j-1} x_{j+1} \over x_j^2}) },$$
and expand its denominators as infinite power series to find
$$\eqalign{
f_N &=\sum_{n_1,...,n_{N-1} \geq 0}
\prod_{j=1}^{N-1} x_j^2 \left( {x_{j-1} x_{j+1} \over x_j^2} \right)^{n_j}\cr
&= \sum_{n_1,...,n_{N-1} \geq 0 \atop n_0=n_N=0}
x_0^{n_1} x_N^{n_{N-1}}
\prod_{k=1}^{N-1} x_k^{2-2 n_k +n_{k-1} + n_{k+1}}, \cr}$$
therefore the desired number $A_N$ is equal to the number of
monomials appearing in $[f_N]_+$, the
polynomial part of $f_N$.

\vskip 2in
{\bf Examples}

1) N=2:

We have
$$ f_2(x_0,x_1,x_2)= {x_1^2 \over (1- {x_0 x_2 \over x_1^2})}$$
and
$$ [f_2(x_0,x_1,x_2)]_+= x_1^2 +x_0 x_2 ,$$
thus $A_2=2$.

2) N=3:

We have
$$ f_3(x_0,x_1,x_2,x_3)= {x_1^2 x_2^2 \over
(1-{x_0 x_2 \over x_1^2})(1-{x_1 x_3 \over x_2^2})} $$
and
$$[f_3(x_0,x_1,x_2,x_3)]_+=x_1^2 x_2^2+ x_0 x_2^3 +x_3 x_1^3
+x_0 x_1 x_2 x_3 + x_0^2 x_3^2 ,$$
which implies $A_3=5$.

3) N=4:

$$\eqalign{
[f_4]_+&=x_1^2 x_2^2 x_3^2+ x_0 x_2^3 x_3^2+x_1^3 x_3^3+x_1^2 x_2^3 x_4
+x_0 x_1 x_2 x_3^3+ x_0^2 x_3^4 \cr
&+ x_0 x_2^4 x_4+ x_1^3 x_2 x_3 x_4+x_0 x_1 x_2^2 x_3 x_4
+x_0 x_1^2 x_3^2 x_4+x_0^2 x_2 x_3^2 x_4 \cr
&+x_1^4 x_4^2
+x_0 x_1^2 x_2 x_4^2 +x_0^2 x_2^2 x_4^2+x_0^2 x_1 x_3 x_4^2+
x_0^3 x_4^3, \cr}$$
hence $A_4=16$.

We now derive an integral formula for $A_N$.
Let $f$ be a Laurent series of the form
$$f=\sum_{k=- \infty }^m a_k x^k,$$
and $f_+=\sum_{k=0}^m a_k x^k$.
By Cauchy's theorem, we can write
$$f_+(1)= \oint_{|x| >1} {dx \over 2i \pi} {f(x) \over x-1}=\sum_{k=0}^m a_k.$$
This gives an integral formula for $A_2$, by replacing $f$ by $f_2$.,
and taking $x_0=x_2=1$.
More generally, we have a similar formula for
$$A_N =[f_N]_+(1,1,...,1), $$
which reads
$$A_N= \oint_{
|x_i|=\rho^{ i(N-i) }}
{dx_1 \over 2i \pi} ..{dx_{N-1} \over 2i \pi} {f_N(1,x_1,..,x_{N-1},1) \over
\prod_{j=1}^{N-1} (x_j-1 ) },$$
for any $\rho >1$.
The choice of contours ensures that
$$ \big\vert {x_{k-1} x_{k+1} \over x_k^2} \big\vert <1 \ \ {\rm and} \ \
|x_k|>1 \ \ \ k=1,...,N-1.$$

The number of states $A_N^{(s)}$ corresponding to the expansion of
$\Delta^{2s}$ is obtained by using as generating function
$$f_N^{(s)}(x_0,x_1,..,x_N)= {x_1^{2s} x_2^{2s} ... x_{N-1}^{2s} \over
\prod_{k=1}^{N-1} (1-{x_{k-1} x_{k+1} \over x_k^2}) },$$
and
$$\eqalign{A_N^{(s)}&=[f_N^{(s)}]_+(1,1,...,1)\cr
&= \oint_{
|x_i|=\rho^{ i(N-i)/2s }}
{dx_1 \over 2i \pi} ..{dx_{N-1} \over 2i \pi}
{f_N^{(s)}(1,x_1,..,x_{N-1},1) \over
\prod_{j=1}^{N-1} (x_j-1 ) },}$$
for arbitrary $\rho >1$. The computation of $A_N^{(s)}$ is consequently
reduced to a straightforward but tedious extraction of residues.
The result for $A_N^{(s)}$ is a polynomial of degree $(N-1)$ in $s$
with rational coefficients.  This is a general property as discussed
below.

We list below the first few polynomials $A_N^{(s)}$.
\eqn\lispol{\eqalign{
A_1^{(s)}&=1 \cr
A_2^{(s)}&=s+1 \cr
A_3^{(s)}&=2s^2+2s+1\cr
A_4^{(s)}&={16 \over 3} s^3 +{13 \over 2} s^2 +{19 \over 6} s + 1 \cr
A_5^{(s)}&={50 \over 3} s^4 + 24 s^3 +{40 \over 3} s^2+4s  +1\cr
A_6^{(s)}&={288 \over 5} s^5 + 96 s^4 +{385 \over 6} s^3
+23 s^2+{157 \over 30}s  +1\cr
A_7^{(s)}&={9604 \over 45} s^6 +{6076 \over 15}s^5
+{2858 \over 9}s^4 + 134 s^3+{1531 \over 45} s^2+{267 \over 45}s+1. \cr
A_8^{(s)}&={262144 \over 315} s^7 +{26624 \over 15} s^6 +{71992 \over 45}
s^5+{6371 \over 8} s^4 +  {43657 \over 180}s^3+{5783 \over 120} s^2+
{971 \over 140} s +1 \cr
A_9^{(s)}&={118098 \over 35}s^8+{279936 \over 35}s^7+{367144 \over 45}s^6+
4696 s^5+{75724 \over 45}s^4+{5911 \over 15}s^3 \cr
&+{19927 \over 315}s^2+{163 \over 21}s+1 \cr
A_{10}^{(s)}&={8000000 \over 567}s^9+{2320000 \over 63}s^8+
{39666608 \over 945}s^7+{1236328 \over 45}s^6+
{12340889 \over 1080}s^5  \cr
&+ {228025 \over 72}s^4+{13641133 \over 22680}s^3+
{204457 \over 2520}s^2+{10957 \over 1260}s+1 \cr
A_{11}^{(s)}&={857435524 \over 14175}s^{10}+{162983612 \over 945}s^{9}+
{205427098 \over 945}s^8+{50279276 \over 315}s^7 \cr
&+{51136132 \over 675}s^6+{365944 \over 15}s^5+{61904509 \over 11340}s^4+
{1628849 \over 1890}s^3  \cr
&+{622849 \over 6300}s^2+{ 323 \over 35}s+1 \cr}}

These quantities take a neat form when expressed
in terms of the polynomials
$$ \Gs_k(s) ={s(s-1)(s-2)...(s-k+1) \over k!}. $$
instead of powers $s^k$. We have
\eqn\listsig{\eqalign{
A_1^{(s)}&= 1 \cr
A_2^{(s)}&=\Gs_1+ 1 \cr
A_3^{(s)}&= 4 \Gs_2 + 4 \Gs_1+ 1 \cr
A_4^{(s)}&= 32 \Gs_3 + 45 \Gs_2 + 15 \Gs_1+ 1 \cr
A_5^{(s)}&=400 \Gs_4 + 744 \Gs_3 + 404 \Gs_2 + 58 \Gs_1+ 1 \cr
A_6^{(s)}&= 6912 \Gs_5 + 16128 \Gs_4 + 12481 \Gs_3 + 3503 \Gs_2
+ 246 \Gs_1+ 1 \cr
A_7^{(s)}&=153664 \Gs_6 + 432768 \Gs_5 + 437776 \Gs_4
+ 188244 \Gs_3 + 30702 \Gs_2 + 1110 \Gs_1+ 1 \cr
A_8^{(s)}&=4194304 \Gs_7 + 13860864 \Gs_6 + 17367872 \Gs_5
+ 10162473 \Gs_4 + 2731521 \Gs_3 \cr
&+ 275599 \Gs_2 + 5301 \Gs_1+ 1 \cr
A_9^{(s)}&=136048896 \Gs_8 + 516481920 \Gs_7 + 773038912 \Gs_6
+ 574771360 \Gs_5 + 218829232 \Gs_4 \cr
&+ 39174798 \Gs_3 + 2537596 \Gs_2 + 26375 \Gs_1+ 1 \cr
A_{10}^{(s)}&=5120000000 \Gs_9+21964800000 \Gs_8+38261688576 \Gs_7+
34587246976 \Gs_6 \cr
&+17164897361 \Gs_5+4532740981 \Gs_4+561999521 \Gs_3
+23932596 \Gs_2+135669 \Gs_1+1\cr
A_{11}^{(s)}&=219503494144 \Gs_{10}+1050351430656 \Gs_9+
2088303502080 \Gs_8+2225685070528 \Gs_7 \cr
&+1366844046336 \Gs_6+482748121856 \Gs_5+92013182474 \Gs_4 \cr
&+8109500738 \Gs_3+230617119 \Gs_2+716541 \Gs_1 + 1 .\cr}}

These triangular expressions for $A_N^{(s)}$
in terms of the $\sigma_k(s)$ are easily inverted as follows.
If
$$A_N^{(s)}=\sum_{k=0}^{N-1} a_{N,k} \Gs_k(s),$$
then one has
$$a_{N,k}= \sum_{s=0}^k A_N^{(s)} {k \choose s} (-1)^{k-s}$$
due to the orthogonality relation:
$$
\sum_{k=m}^s {s \choose k} {k \choose m} (-1)^{k-m} =\delta_{m,s}
$$

This shows why the coefficients in (8.3) are integers.  We have
no clue as to why they turn out to be positive, implying inequalities
among $A_N^{(s)}$ for fixed N.

The problem of counting the number of integral points in an integral
polytope (i.e. a polytope with all its vertices at integer points)
is a classical (and difficult) one.  We base the following on
the work of Ehrhart
\ref\EHRH{E. Ehrhart, Sur un probl\`eme de g\'eom\'etrie diophantienne
lin\'eaire I et II, J. reine angew. Math. t.226 (1967), p.1-29,
t.227 (1967), p.25-49 and C.R.A.S. t.265 (1967), A5, A91 and A160. },
 who, starting from an integral
$d$--dimensional
polytope $\Pi$ with vertices in $\IZ^d$ considered the number of
lattice (integral) points in the dilated polytope $s \Pi$, $s$ integer,
$$ E_{\Pi}(s) = {\rm card}\big( \IZ^d \cap s \Pi \big).$$
He showed that this number is a polynomial of degree $d$ in $s$,
hence deserves to be called the Ehrhart polynomial of $\Pi$.
Our number $A_N^{(s)}$ of admissible tableaux is precisely the
Ehrhart polynomial of the $(N-1)$--dimensional polytope $\Pi_N^{(1)}$,
defined by the equations \polito, with $n_i \in \IR_+$.
In the following, we derive a simple expression for the Ehrhart
polynomial of a specific class of polytopes, which we conjecture
is valid for $\Pi_N^{(1)}$.
We will make use of the Ehrhart reciprocity theorem, which reads as follows.
Let ${\bar E}_{\Pi}(s)$ denote the number of lattice points which lie strictly
inside the dilated polytope $s \Pi$, i.e. the number $E_{\Pi}(s)$ minus
the number of lattice points on the boundary $\partial (s \Pi)$.
Then one has the reciprocity relation

{\noindent {\bf Theorem (Ehrhart):}}
\eqn\ehrecip{ {\bar E}_{\Pi}(s) =(-1)^d E_{\Pi}(-s).}

Given an integral
polytope $\Pi$ in $d$ dimensions,
we wish to perform a decomposition
into ``elementary cells", namely non--flat polyhedral integral
simplices with $d+1$ vertices,
which do not contain any other lattice point (the intersection
of an elementary cell with $\IZ^d$ is reduced to its $d+1$ vertices).
We have the following

{\bf Proposition 1.} Such a decomposition exists but is in general not unique.

\noindent{} To prove the existence, we rely on a classical theorem which
guarantees
the existence of a decomposition in any integral polytope.
Let us consider a simplex in this decomposition. If it is not an elementary
cell, it contains at least one lattice point $P$. If we draw all the edges
linking this point $P$ to the $(N+1)$ vertices of the simplex, this defines
a decomposition of the simplex into smaller simplices.

\fig{The three possible new decompositions of a tetrahedron, according
to the position of the point $P$ on an edge (2 tetrahedra), on a face
(3 tetrahedra) or strictly inside (4 tetrahedra).}
{tetra.eps}{9cm}
\figlabel\tetra

Retaining only the
non--flat ones, we end up with $(d+1)$ simplices if the point $P$ is strictly
inside the initial simplex, $d$ simplices if it lies strictly inside a $(d-1)$
dimensional face,..., $2$ simplices if it lies strictly inside a $1$
dimensional edge of the initial simplex (we illustrate the case $d=3$
on Fig.\tetra ).
We can iterate this procedure with the new decomposition, until the
decomposition is into elementary cells.

For any such elementary cell decomposition of $\Pi$,
the total numbers of vertices ($0$--faces),
edges ($1$--faces),..., $k$--faces,..., cells ($d$--faces) are not
in general invariant. However, the situation gets much better in the
following particular case

{\bf Proposition 2.}
If the polytope $\Pi$ is such that one
of its elementary cell decompositions is only made of ``basic cells",
i.e. elementary cells with volume equal to the minimal value $1 \over d!$,
then the numbers of $j$--faces of the decomposition, which we denote by
$F_0$, $F_1$,..., $F_d$, are arithmetical invariants
of the polytope $\Pi$.
Such a polytope will be called basic in the following.

To prove the above, we will derive a formula for the Ehrhart
polynomial of $\Pi$, involving only the $F_j$, $j=0,...,d$, which
will consequently appear as invariants of $\Pi$.
The issue of whether an elementary cell is basic or not becomes relevant
in $3$ or more dimensions. For $d=2$, an elementary cell is a triangle
with vertices in $\IZ^2$, which contains no other lattice point.
Any two of its edges define two basis vectors of $\IZ^2$, giving
rise to a parallelogram of area $1$, therefore the triangle, half of the
parallelogram, has area $1 \over 2$.
So in $2$ dimensions, all elementary cells are basic.
This is no longer the case in $3$ dimensions.
Take for instance the tetrahedron with vertices $(0,0,0)$, $(1,0,0)$,
$(0,1,0)$, $(1,p,q)$, $p$ and $q$ co-prime, it defines an elementary cell,
with volume $q \over 6$, in general not basic. Nevertheless one
checks that the closed tetrahedron intersects the lattice $\IZ^3$ only
at its vertices!
However, in the particular case of $\Pi_4^{(1)}$ ( of Fig.\polit\ ),
it is possible to show that it decomposes into $32$ elementary cells
(tetrahedra) which turn out to be all basic (volume $1 \over 6$).
We are thus led to the conjecture that the polytopes
$\Pi_N^{(1)}$ are basic, for all $N$.

Before treating the general case of a basic polytope, let us first concentrate
on the simplest example of the basic simplex $\Sigma_d$, with
vertices at $(n_1,...,n_d)$, where either all $n_i=0$, or all $n_i=0$
except one, which is $1$. The simplex itself is its own basic decomposition,
and we compute very easily the numbers $F_j^{(d)}$ of $j$--faces
$$ F_j^{(d)} = {d+1 \choose j+1} \qquad j=0,1,...,d.$$
On the other hand, the computation of the Ehrhart polynomial
$E_{\Sigma_d}(s)={\rm card}(s\Sigma_d \cap \IZ^d)$ is very simple,
we find
$$ E_{\Sigma_d}(s)= \sum_{k=0}^s \sum_{n_1,n_2,...,n_d \geq 0 \atop
n_1+n_2+...+n_d=k} 1.$$
The number of partitions of the integer $k$ into
$d$ arbitrary integers is just
${d+k-1 \choose k}$, and
$$\eqalign{ E_{\Sigma_d}(s) &= \sum_{k=0}^s {d+k-1 \choose k} \cr
&={(s+1)(s+2)(s+3)...(s+d) \over d!} , \cr}$$
which can be recast into
$$ E_{\Sigma_d}(s) = \sum_{j=0}^d {d+1 \choose j+1} {s-1 \choose j}.$$
In this particular case, we get a simple expression for the Ehrhart polynomial
in terms of the $F_j^{(d)}$
\eqn\sigeh{ E_{\Sigma_d}(s)= \sum_{j=0}^d F_j^{(d)} {s-1 \choose j}.}

We are now ready to generalize this to any basic polytope $\Pi$. The
result reads

{\bf Proposition 3.}  Let $\Pi$ be a basic $d$--dimensional
integral polytope, and $F_j$,
$j=0,1,...,d$ denote the numbers of $j$--faces in a
given basic decomposition. Then the Ehrhart polynomial of $\Pi$
reads
\eqn\univers{\encadremath{E_{\Pi}(s)= \sum_{j=0}^d F_j  {s-1 \choose j}.} }

Proposition 2 above appears as a simple corollary: due to
the invariant definition of $E_{\Pi}(s)$, the numbers $F_j$ are
also invariant, i.e. do not depend on the particular basic decomposition
performed. Note also that the intuitive fact that $E_{\Pi}(s=0)=1$
for any polytope $\Pi$,
because the zero--dilated polytope is reduced to a point,
translates into the non--trivial Euler's relation linking the numbers $F_j$
to the zero genus of the $d$--dimensional decomposed space
$$ \sum_{j=0}^d (-1)^j F_j = E_{\Pi}(0)=1.$$
Let us proceed to the proof of proposition 3.
Equation \univers\ was just derived in the case of the particular
basic simplex $\Sigma_d$, and
extends trivially to any elementary basic cell.
For a given polytope $\Pi$, we first perform an elementary cell decomposition
as in sect.1.  We want to count the number $E_{\Pi}(s)$ of lattice points
in the dilated polytope $s \Pi$.
This number is equal to a sum of $(d+1)$ terms.

($0$) the number of dilated vertices, $F_0$.

($1$) the number of dilated ($1$ dimensional) edges of the initial
elementary basic cell decomposition, $F_1$,
multiplied by the number of lattice points which lie strictly inside each
dilated edge. This last number is $(s-1)$, the same for all the dilated edges.

$\vdots$

($j$) the number of dilated $j$--dimensional faces of the initial elementary
basic cell decomposition, $F_j$, multiplied by the number of lattice points
which lie strictly inside each dilated $j$--face.
By Ehrhart reciprocity \ehrecip, the number of lattice points inside
any dilated $j$--face is
$$\eqalign{(-1)^j E_{\Sigma_j}(-s)&=
(-1)^j {(-s+1)(-s+2)...(-s+j) \over j!} \cr
&={(s-1)(s-2)...(s-j) \over j!} \cr
&={s-1 \choose j} . \cr }$$

$\vdots$

($d$) the number of dilated $d$--dimensional basic cells of the elementary
basic decomposition, $F_d$, multiplied by the number of lattice points strictly
inside each cell, namely ${(s-1)(s-2)...(s-N) \over N!}$.

Summing all these contributions, we get the desired result \univers\
$$E_{\Pi}(s)= \sum_{j=0}^d  { s-1 \choose j} F_j.$$

When applied to our case of (conjecturally) basic polytopes
$\Pi_N^{(1)}$, the above formula \univers\ suggests to rewrite
the Ehrhart polynomials $A_N^{(s)}$ in the basis
$$\nu_k(s) = {(s-1)(s-2)...(s-k) \over k!} ={s-1 \choose k}, \ \
\nu_0(s)=1$$
instead of $\sigma_k(s)$ as in (8.3).
We get again {\bf positive} integers, which reinforce our hope
that our conjecture might be true.  We obtain the following table

\eqn\lisigf{\eqalign{
A_1^{(s)}&= 1 \cr
A_2^{(s)}&=\nu_1+ 2 \cr
A_3^{(s)}&= 4 \nu_2 + 8 \nu_1+ 5 \cr
A_4^{(s)}&= 32 \nu_3 + 77 \nu_2 + 60 \nu_1+ 16 \cr
A_5^{(s)}&=400 \nu_4 + 1144 \nu_3 + 1148 \nu_2 + 462 \nu_1+ 59 \cr
A_6^{(s)}&= 6912 \nu_5 + 23040 \nu_4 + 28609 \nu_3 + 15984 \nu_2
+ 3749 \nu_1+ 247 \cr
A_7^{(s)}&=153664 \nu_6 + 586432 \nu_5 + 870544 \nu_4
+ 626020 \nu_3 + 218946 \nu_2 + 31812 \nu_1+ 1111 \cr
A_8^{(s)}&=4194304 \nu_7 + 18055168 \nu_6 + 31228736 \nu_5
+ 27530345 \nu_4 + 12893994 \nu_3 \cr
&+ 3007120 \nu_2 + 280900 \nu_1+ 5302 \cr
A_9^{(s)}&=136048896 \nu_8 + 652530816 \nu_7 + 1289520832 \nu_6
+ 1347810272 \nu_5 + 793600592 \nu_4 \cr
&+ 258004030 \nu_3 + 41712394 \nu_2 + 2563971 \nu_1+ 26376 \cr
A_{10}^{(s)}&=5120000000 \nu_9+27084800000 \nu_8+ 60226488576 \nu_7+
72848935552 \nu_6 \cr
&+51752144337 \nu_5+21697638342 \nu_4+5094740502 \nu_3
+585932117 \nu_2+24068265 \nu_1+135670 \cr
A_{11}^{(s)}&=219503494144 \nu_{10}+1269854924800 \nu_9+
3138654932736\nu_8+4313988572608\nu_7 \cr
&+ 3592529116864\nu_6+1849592168192 \nu_5+574761304330 \nu_4 \cr
&+100122683212\nu_3+8340117857 \nu_2+231333660\nu_1 + 716542 .\cr}}

For $N=3$, we read that the $2$--dimensional polytope $\Pi_3^{(1)}$
has a basic decomposition into $4$ basic triangles, with
$8$ edges and $5$ vertices. For $N=4$, we have $32$ basic tetrahedra,
with $77$ triangular faces, $60$ edges, and $16$ vertices, etc...

In particular, if our conjecture is true, the volume of the polytope
$\Pi_N^{(1)}$ should read
$$V_N = {F_{N-1} \over (N-1)!},$$
where $F_{N-1}$ is the number of basic cells of any basic decomposition of
$\Pi_N^{(1)}$.
In the remainder of this section, we will derive explicitly the first few
leading coefficients of $A_N^{(s)}$ when $s$ is large.
The leading coefficient is intuitively the volume $V_N={2^{N-1} N^{N-3}
\over (N-1)!}$ (see appendix E for the calculation) of the
polytope $\Pi_N^{(1)}$.  A more surprising feature is that more can
be said on the subleading coefficients.

We apply a particular case
of the combinatorial Riemann--Roch theorem
of Kantor and Khovanski
\ref\KAKO{J.M. Kantor and A. Khovanskii,
``Integral points in convex polyhedra,
combinatorial Riemann-Roch theorem and generalized Euler-MacLaurin
Formula", IHES-Preprint 1992, and ``Une application du th\'eor\^eme
de Riemann-Roch combinatoire au polyn\^ome d'Ehrhart des polytopes
entiers de $\IR^d$", Preprint 1992. }.
The latter relates the number of points with integer
coordinates in a given integral polytope to
the volume of the polytope deformed by slightly shifting its faces,
parallelly to themselves. To simplify notations
it proves more convenient to work in $N$
dimensions, i.e. to consider the polytope $\Pi_{N+1}^{(1)}$ and its
Ehrhart polynomial $A_{N+1}^{(s)}$.

Let $V_{N+1}^{(s)}(h_1,...,h_N;\Ge_1,...,\Ge_N)$ denote
the volume of the deformed polytope defined by the inequalities
$$  -2 \Ge_k \leq 2 n_k \leq n_{k-1}+n_{k+1}+2s +h_k \qquad k=1,...,N,$$
with as usual
$n_0=n_{N+1}=0$\foot{This definition is only valid for $N>1$.
When $N=1$, one has to take the deformation $-\Ge_1 \leq n_1 \leq s+h_1$,
in which $h_1$ is multiplied by $2$ in the definition.},
and ${\cal T}(x)$ the generating function of Bernoulli numbers
$$ {\cal T}(x) = {x \over 1 - e^{-x} }=1+{1 \over 2} x +{1 \over 12} x^2
-{1 \over 720 }x^4 +{1 \over 30240} x^6 + ... $$
Consider the polynomial $B_N(s)$ of degree $N$ in $s$
\eqn\bNdef{
B_N(s)\equiv \bigg[ \prod_{k=1}^N {\cal T}({\partial \over \partial h_k})
{\cal T}({\partial \over \partial \Ge_k}) \bigg]
V_{N+1}^{(s)}(h_i;\Ge_i) \bigg\vert_{h_i=\Ge_i=0}
=\sum_{k=0}^{N} b^{(N)}_k s^k,}
On the other hand, the desired number of points, $A_{N+1}^{(s)}$, is also
a polynomial in $s$
\eqn\www{A_{N+1}^{(s)}=\sum_{k=0}^N a_k^{(N)} s^k.}

The theorem relates the coefficients of both polynomials, we have

\eqn\wwww{\eqalign{
a_N^{(N)}&=b_N^{(N)}=V_{N+1}\cr
a_{N-1}^{(N)}&=b_{N-1}^{(N)}=\mu_{N-1} \cr
a_{N-2}^{(N)}&=b_{N-2}^{(N)} + \sum_{F_{N-2}}
\mu_{N-2}(F_{N-2}) \tau_k({\hat F}_{2}), \cr}}
where $V_{N+1}$ is the volume of the polytope
$\Pi_{N+1}^{(1)}$,
the last sum extends over the $(N-2)$--dimensional faces $F_{N-2}$
of the polytope $\Pi_{N+1}^{(1)}$, $\mu_{N-1}$ is the so--called
$(N-1)$--dimensional
relative measure of $\Pi_{N+1}^{(1)}$ (the sum of the $(N-1)$--dimensional
measures of the $(N-1)$--dimensional faces of $\Pi_{N+1}^{(1)}$, such that
the unit cell of the restriction of $\IZ^N$
to the given face be taken as unity),
$\mu_{N-2}(F_{N-2})$ the relative
$(N-2)$--dimensional measure of the face $F_{N-2}$ defined analogously, and
$\tau_2$ a certain invariant of the two--dimensional polar cone
${\hat F}_2$ to $F_{N-2}$ (the cone generated by the normals to the two
$(N-1)$--dimensional faces of which $F_{N-2}$ is the intersection).

The crucial hypothesis which enables to apply the above mentioned theorem
is that the polytopes we are considering have so--called
$2$--primitive fans. The fan of a polytope is the set of cones generated by
subsets of the (integer valued) normal vectors to its $(N-1)$--dimensional
faces: it is a kind of dual envelope of the polytope.
The $k$--primitivity means that for all
$m$--dimensional cones of the fan, $m<k$, the generators of the cone, i.e.
a set of $m$ (integer valued) normals to the
$(N-1)$--dimensional faces of the
polytope, can be completed into a $\IZ$--basis of the lattice $\IZ^N$.
In our particular case, the normals to the $2N$ $(N-1)$--dimensional faces
read
$$\eqalign{ (0,..,0,-1,0,...,0)&  \ \ -1 \ {\rm in} \ i^{th} \
{\rm position} \ i=1,..,N \cr
(0,..,-1,2,-1,0,..,0)& \ \ \ \ 2 \ {\rm in} \ i^{th} \
{\rm position} \ i=1,..,N \cr }$$
It is clear that for instance the two normals
$$(2,-1,0,...,0) \ \ \ {\rm and } \ \ \ (0,-1,0,..,0)$$
cannot be completed into a $\IZ$--basis of $\IZ^N$. Otherwise
one could find $(N-2)$ vectors of $\IZ^N$ $v_1,v_2,..,v_{N-2}$,
with integer valued coordinates $v_i^{(j)}\in \IZ$,
such that the determinant (volume of the unit cell associated to
the basis)
$$\left\vert\matrix{2&0&v_1^{(1)}&\cdots&v_{N-2}^{(1)} \cr
-1&-1&v_1^{(2)}&\cdots&v_{N-2}^{(2)} \cr
0&0&v_1^{(3)}&\cdots&v_{N-2}^{(3)} \cr
\vdots&\vdots&\vdots& &\vdots \cr
0&0&v_1^{(N)}&\cdots&v_{N-2}^{(N)} \cr}\right\vert, $$
would be equal to one.
But after subtracting the second column from the first one,
we can factor out $2$, hence the determinant
cannot be equal to $1$. Therefore our polytopes are
$2$--primitive, as any single normal can be completed into a
$\IZ$--basis of $\IZ^N$.

The above proves rigorously the intuitively obvious fact that the volume
$V_{N+1}$ of the polytope $\Pi_{N+1}^{(1)}$
appears as leading $s^N$ term in $A_{N+1}^{(s)}$, and gives us
compact formulae for the next two subleading coefficients.

We summarize the conjectures in the following proposition, and leave the
detailed arguments to appendix E.

{\bf Proposition 4.} For large s the polynomial $A_{N+1}^{(s)}$
behaves like
\eqn\ttt{A_{N+1}(s)
= a_N^{(N)} s^N + a_{N-1}^{(N)} s^{N-1} + a_{N-2}^{(N)} s^{N-2}
+ \ O(s^{N-3}),}
where
\eqn\tnt{\eqalign{a_N^{(N)} &= V_{N+1} =  {2^N (N+1)^{N-2} \over N!} \cr
a_{N-1}^{(N)} &= {N \over 4} V_{N+1}+{1 \over 2} \sum_{k=0}^N
V_k V_{N-k} \cr
&= {2^{N-1} (N+1)^{N-4} \over (N-1)! 2! } (N^2+3 N+8) \cr
a_{N-2}^{(N)} &= b_{N-2}^{(N)}+{1 \over 4} V_{N-1} \cr
&= {2^{N-2} \over (N-2)! 4!}
\big[ (N+1)^{N-6} (3N^4+17 N^3+72 N^2+144N +266)+6 (N-1)^{N-4}
\big].}}

Our conjecture that $\Pi_{N+1}^{(1)}$ is basic
would imply that
the various numbers $F_N$, $F_{N-1}$ and $F_{N-2}$ of
respectively basic cells, $(N-1)$--faces and $(N-2)$--faces in the
basic cell decomposition of $\Pi_{N+1}^{(1)}$, are given by
\eqn\tttt{\eqalign{ F_N &= 2^N (N+1)^{N-2} \cr
F_{N-1} &= 2^{N-2} (N+1)^{N-4} (2 N^3 +7 N^2 + 9 N + 10) \cr
F_{N-2} &= {2^{N-4} \over 6} \big[ (N+1)^{N-6}(12 N^6+64 N^5+143 N^4 \cr
&+237 N^3+ 264 N^2 +212 N+258) + 6(N-1)^{N-4} \big] . \cr }}

\noindent{ \bf Acknowledgments}

It is a pleasure to thank D. Bessis who raised up our interest for this
problem,
B. Derrida who contributed table II,
T. Shiota who found the general expression of appendix C,
the participants of a seminar at Luminy who made us aware of
Dyson's conjecture (now a theorem), B. Jancovici who shared
with us his insight in Coulomb gases, and J.M. Kantor who
introduced us to lattice point counting.  F. Lesage is supported
by a Canadian NSERC 67 scholarship.


\appendix{A}{Asymptotic evaluation of $ln Z_N(1)$ using Euler-
Mac Laurin's formula}

As indicated in eqn.\partden\ we have
\eqn\againone{ Z_N(1)=\prod_{j=1}^N j!}
This section is devoted to a proof that asymptotically as $N \to \infty$
\eqn\expanone{\eqalign{
ln Z_N(1)&={N^2 \over 2} ln N -{3 \over 4}N^2
+N ln N+N({1 \over 2}ln 2 \pi -1)+{5 \over 12}ln N \cr
&+{1-\gamma+5 ln 2  \pi \over 12}+
{\zeta'(2) \over 2 \pi^2}+{1 \over 12 N}
-{1 \over 720 N^2}-{1 \over 360 N^3} +O({1 \over N^4}).\cr}}
Here $\gamma$ is Euler's constant, $\zeta(t)=\sum_{n \geq 1} n^{-t}$
is Riemann's zeta--function, and the numerical constant is
\eqn\numcons{{1-\gamma+5 ln 2  \pi \over 12}+
{\zeta'(2) \over 2 \pi^2}=.75351738...}
Such an  expression vindicates the estimate \extexc\ with
\eqn\firstf{ -2 f(1)={1 \over 2} ln 2 \pi -1.}
Note that beyond the term of order $N$ we find corrections of order
$ln N$ (and not $\sqrt{N}$ as  crudely expected from the thermodynamic
reasoning).
We guess that this is not peculiar to the case $p=1$.

We rederive Euler--Mac Laurin's formula for our purpose \EVA . Let
$f(z)$ be an analytic function for $Re(z)>0$ such
that in this half plane $|f(x+iy)|={\it o}(e^{2 \pi y})$.
We have in mind a function such as $f(z)=z^{\alpha}$,
$\alpha$ real positive, and we take the principal determination.

\fig{The contour of integration for Euler--Mac Laurin's formula.}
{contour.eps}{7cm}
\figlabel\contour

Remarking that in the vicinity of any integer $k$, $\cotan \pi z$
behaves as
$$1/\pi (z-k)$$
we have for a sum $I_N$,
\eqn\eulmac{ I_N=f(1)+f(2)+...+f(N)={1 \over 2i} \oint_{{\cal C}}
dz f(z) \cotan \pi z,}
where the contour $\cal C$ is indicated on figure \contour,
$\theta$ being an arbitrary real parameter between $0$ and $1$
and $\Gamma$ a half circle of arbitrary small radius $\rho$.
Call ${\cal C}_+$ and ${\cal C}_-$ the upper and lower parts of the
contour $\cal C$, excluding the arc $\Gamma$ (see  fig.\contour\ ).
Noting that for any fixed $x$
$$\lim_{|y| \to \infty} \cotan \pi (x+i y)={1 \over i}\  sgn(y),$$
where $sgn(y)$ is the sign of $y$, we add and subtract the integrals
${1 \over 2} \int_{{\cal C}_{\pm}} f(z) dz$.
Thus
\eqn\intermI{\eqalign{
I_N&={1 \over 2i} \int_{{\cal C}_+}dz f(z)(\cotan \pi z +i)+
{1 \over 2i} \int_{{\cal C}_-}dz f(z)(\cotan \pi z -i)+\cr
&+{1 \over 2i} \int_{\Gamma}dz f(z)\cotan \pi z
-{1 \over 2} \int_{{\cal C}_+} dz f(z) +{1 \over 2}
\int_{{\cal C}_-} dz f(z) .\cr}}
As $\rho$ shrinks to zero, the third integral over $\Gamma$ tends to
${1 \over 2} f(N)$, while the fourth and fifth combine to yield
the integral of $f$ along the real axis from $\theta$ to $N$.
Letting $M \to \infty$, the two horizontal branches in the first and second
integral vanish and we are left with
\eqn\leftwith{\eqalign{ I_N&=\int_{\theta}^N dx f(x) +
{1 \over 2} f(N)+\int_{\theta}^{\theta+i \infty} dz
{f(z) \over e^{-2i\pi z} -1} \cr
&+\int_{\theta}^{\theta-i \infty} dz
{f(z) \over e^{2i\pi z} -1}+\int_0^{\infty}{dy \over i}
{f(N+iy)-f(N-iy) \over e^{2 \pi y}-1} \cr}}
In the sense of asymptotic series we replace $[f(N+iy)-f(N-iy)]/i$
in the last integral by its Taylor series at $y=0$ to the
effect that
\eqn\expfinI{ \eqalign{ I_N \simeq \int_{\theta}^N f(x) dx
+C(\theta)&+{1 \over 2}f(N)+\sum_{k \geq 0} {2 (-1)^k \over (2k+1)!}
f^{(2k+1)}(N) \int_0^{\infty}dy {y^{2k+1} \over e^{2 \pi y}-1},\cr
C(\theta) &= \int_{\theta}^{\theta+i \infty} dz {f(z) \over e^{-2i\pi z} -1}
+\int_{\theta}^{\theta -i \infty} dz {f(z) \over e^{2i \pi z} -1}.\cr}}
This is indeed an asymptotic expansion provided
$$\eqalign{\lim_{N \to \infty}
f^{(k+1)}(N)/f^{(k)}(N)&=0 \cr
\vert f^{(k)}(N+iy) \vert &\leq A_{\epsilon} f^{(k)}(N)
e^{(2 \pi - \epsilon)|y|},\cr}$$
which will be satisfied in our applications.
By using a specific function $f(z)=e^{tz}$ and comparing $I_{N+1}-I_N=f(N)$
to its asymptotic expansion, we find for the coefficient of $f^{(2k+1)}(N)$
\eqn\coefff{ {2(-1)^k \over (2k+1)!} \int_{0}^{\infty} dy
{y^{2k+1} \over e^{2 \pi y} -1} = {B_{2k+2} \over (2k+2)!} \ ,}
where the Bernoulli numbers $B_k$ are defined through
\eqn\bernou{ {t \over e^t -1} = \sum_{k=o}^{\infty} t^k {B_k \over k!}}
and vanish for $k$ odd larger than $1$.
If $f(0)$ exists, one can let $\theta \to 0$ in which case
\eqn\valuec{ C(0)= -{1 \over 2} f(0) -\int_0^{\infty} {dy \over i}
{f(iy) -f(-iy) \over e^{2 \pi y} -1} \ .}
On applying eqn.\expfinI\ to $f(z)=z^{\alpha}$, $\alpha >0$, one
obtains
\eqn\appli{ \sum_{j=1}^N j^{\alpha} \simeq {N^{\alpha+1} \over \alpha +1}
+{N^{\alpha} \over 2} +C_{\alpha}(\theta) -{\theta^{\alpha+1} \over
\alpha +1} + \sum_{k \geq 0} {B_{2k+2} \over (2k+2)!}
\alpha(\alpha-1)...(\alpha-2k) N^{\alpha-2k-1},}
where
\eqn\defcalpha{\eqalign{ C_{\alpha}(\theta)-{\theta^{\alpha+1} \over
\alpha +1} &=C_{\alpha}(0)=-2 \sin {\pi \alpha \over 2}
\int_0^{\infty} dy { y^{\alpha} \over e^{2 \pi y} -1} \cr
&=-2 \sin {\pi \alpha \over 2} \sum_{m=1}^{\infty} \int_0^{\infty}
dy y^{\alpha} e^{-2m\pi y}\cr
&=-2 \sin {\pi \alpha \over 2} \sum_{m=1}^{\infty} {\Gamma(\alpha+1) \over
(2 \pi m)^{\alpha+1}} \cr
&=-2 \sin {\pi \alpha \over 2}  {\Gamma(\alpha+1) \over (2 \pi)^{\alpha+1}}
\zeta(\alpha +1) \cr
&= \zeta(- \alpha),\cr}}
since the zeta function satisfies the reciprocity relation
\eqn\recizet{ \zeta(s) = 2 \Gamma(1-s) (2 \pi)^{s-1}
\sin {\pi s \over 2} \zeta(1-s).}
The case $f(z)=ln z$ obtained by taking a derivative at $\alpha=0$
of the previous estimate yields Stirling's formula as well as
$\partial_{\alpha} C_{\alpha}(0)\vert_{\alpha=0}
=-\zeta'(0)={1 \over 2} ln 2 \pi$. Thus
\eqn\applilog{
ln N! \simeq (N+{1 \over 2})ln N - N +{1 \over 2} ln 2 \pi +
\sum_{k \geq 0} { B_{2k+2} N^{-2k-1} \over (2k+1)(2k+2)},}
where of course the asymptotic series diverges !
Similarly when $f(z)= z ln z$ we take a derivative at $\alpha=1$,
use $\Gamma(2)=1$, $\Gamma'(2)=1-\gamma$, $\zeta(2)={\pi^2 \over 6}$ and find
\eqn\applizlog{\eqalign{
\sum_{j=1}^N j ln j &\simeq {N^2 \over 2} ln N
-{N^2 \over 4} +{N \over 2} ln N +{1 \over 12} ln N+
{\gamma + ln 2 \pi \over 12} -{\zeta'(2) \over 2 \pi^2}\cr
&-\sum_{k \geq 1} {B_{2k+2} N^{-2k} \over 2k(2k+1)(2k+2)} \ .\cr}}
Finally from equations \applilog\ and \applizlog\
\eqn\finappli{\eqalign{ ln Z_N(1)&= ln \prod_{j=1}^N j! \cr
&=\sum_{j=1}^N (N+1-j) ln j \cr
&=(N+1)ln N! -\sum_{j=1}^N j ln j \cr
&={N^2 \over 2} ln N -{3 \over 4}N^2+N ln N + N({ln 2 \pi \over 2}-1)
+{5 \over 12} ln N \cr
&+{1-\gamma+5 ln 2 \pi \over 12}+{\zeta'(2) \over 2 \pi^2}+{1 \over 12 N}
-{1 \over 720 N^2}-{1 \over 360 N^3} +O({1 \over N^4}), \cr}}
as stated in equation \expanone.

In view of the above remarks,
we add here the following estimate
\eqn\estlog{ ln \prod_{j=1}^N (jp)!={1 \over p} ln \prod_{j=1}^{Np} j!
+{p+1 \over 2} ln (p^N N!) -{1\over p} ln ((Np)!) +O(ln N).}
The first term on the r.h.s. is ${1 \over p} ln Z_{Np}(1)$.
Indeed
\eqn\indeed{ \eqalign{ ln \prod_{j=1}^N (jp)!&=
\sum_{r=0}^{N-1} \sum_{k=1}^p (N-r) ln (k+rp) \cr
&={1 \over p} [ Np\sum_{q=1}^{Np} ln q -\sum_{r=0}^{N-1}
\sum_{k=1}^p (rp+k-k) ln (k+rp) ]\cr
&={1 \over p} [\sum_{q=1}^{Np} (Np-q) ln q +\sum_{k=1}^p k
\sum_{r=0}^{N-1} ln (k+rp) ]\cr
&= {1 \over p} ln \prod_{j=1}^{Np} j! + R_N(p), }}
where
\eqn\reste{ R_N(p)={ 1 \over p} \sum_{k=1}^p k \sum_{r=0}^{N-1} ln (k+rp) .}
With $1 \leq k \leq p$ we have
\eqn\square{ \sum_{r=1}^{N-1} ln (rp) \leq \sum_{r=0}^{N-1}
ln (k + rp) \leq \sum_{r=1}^N ln (rp) }
thus
\eqn\cadre{ {p+1 \over 2} ln (p^{N-1} (N-1)!) \leq R_N(p)\leq
{p+1 \over 2} ln (p^N N!) }
which yields the required result \estlog.

\appendix{B}{Instability of the cuboctahedron configuration}

In the stereographic representation of the sphere on the
plane $\IC$, the equilibrium configurations, stable or unstable,
$\{ z_j \}$ are solutions of the equations ${\partial V\over
\partial z_i}={\partial V\over \partial \bar{z}_i}=0$,
$\forall i$, that is
\eqn\diai{
\sum_{j=1\atop j\neq i}^N {1\over z_i-z_j}=(N-1) {\bar{z}_i\over
1+z_i \bar{z}_i} , \ \ \ i=1,2,...,N
}
expressing that the potential
\eqn\diaii{
V=-\sum_{i<j} ln z_{ij} \bar{z}_{ij}+(N-1) \sum_j ln
(1+\bar{z}_j z_j)
}
is stationary.  Since $V$ is invariant under an overall
rotation of the configuration, that is by a unitary homographic
substitution
\eqn\diaiii{\eqalign{&
z_j\rightarrow \xi_j={\alpha z_j+\gamma \over -\bar{\gamma} z_j +\bar{
\alpha}}\cr
&\alpha \bar{\alpha}+\gamma \bar{\gamma}=1}
}
the system \diai\ is formally invariant in this substitution.
We can easily then show the existence of equilibrium configurations
on the sphere defined not only by the regular
polyhedron but by
their composites in diverse polyhedra, so that we have
by symmetry
\eqn\diaiv{
\sum_{j=1}^{N-1} {1\over \xi_j}=0
}
by putting the vertex $\xi_N$ at the pole ($\xi_N=0$).

Around a solution $\{ z_j \}$ of \diai\ the potential is
approximated, up to third order in the variations $\{ \delta
z_j \}$ by the real quadratic form $Q(\delta z, \delta \bar{z})$
\eqn\diav{
2Q\equiv \sum_{i,j} A_{ij} \delta z_i \delta z_j +
\overline{A}_{ij} \delta \bar{z}_i \delta \bar{z}_j
+2 D_i \delta z_i \delta \bar{z}_i
}
with
\eqn\diavi{
A_{ij}={\partial^2 V\over \partial z_i \partial z_j}=
-{1\over (z_i-z_j)^2}+\delta_{ij} \left(
\sum_{l\atop l\neq i} {1\over (z_i-z_l)^2}-(N-1)
{\bar{z}_i^2\over (1+z_i \bar{z}_i)^2} \right)
}
\eqn\diavii{
D_i={\partial^2 V\over \partial z_i \partial \bar{z}_i}=
{N-1\over (1+z_i \bar{z}_i)^2}
}
This is a real form in the $2N$ variables $\delta x_j, \delta y_j$
($\delta z=\delta x+i \delta y$) with rank not greater
than $2N-3$ since the variations $\{ \delta z_i \}$
associated with an overall
rotation  of the configuration $\{ z_i \}$ annihilate $Q$ and
depends on 3 real parameters as shown by \diaiii .
The configuration is stable if $Q$ is semi-positive.

To prove that the configuration $12_{(24)}$ of the
cuboctahedron is unstable we must show the existence
of a negative eigenvalue of the real $2N\times 2N$ symmetric matrix
\eqn\diaviii{
\left\Vert \matrix{D+Re A & -Im A \cr -Im A & D-Re A} \right\Vert
}
but the spectrum of this matrix is not invariant under
the rotations. However the signature is invariant
and this is all we need.
To obtain the intrinsic modes of vibration we consider
the form $Q(\delta u, \delta \bar{u})$
\eqn\diaix{
2Q\equiv \sum B_{ij} \delta u_i \delta u_j + c.c.+2 \delta u_i
\delta \bar{u}_i
}
with
\eqn\diax{
B_{ij}={1\over \sqrt{D_i D_j}} A_{ij}=B_{ji}
}
and
\eqn\diaxi{
\delta u_i = \sqrt{D_i} \delta z_i = \sqrt{N-1} {\delta z_i
\over 1+ z_i \bar{z}_i}
}
where
\eqn\diaxii{
{4\over N-1} \delta u \delta \bar{u} = {1\over 4}
{\delta z \delta \bar{z} \over (1+z \bar{z})^2}= \sin^2 \theta
d \phi^2+d \theta^2
}
is the invariant metric on the sphere.  The secular equation
associated with \diax\ reads
\eqn\diaxiii{
B_{ij} \delta u_j + (1-\lambda ) \delta \bar{u}_i =0
}
and its conjugate.  Denoting by $\rho^2$ the spectrum of
the non-negative hermitian matrix $B \bar{B}$ of order $N$,
we have the $2N$ eigenvalues
\eqn\diaxiv{
\lambda_n=1-\rho_n \ \ {\rm and} \ \ 2-\lambda_n=1+\rho_n
}
and the diagonal form of $Q$
\eqn\diaxv{
Q\simeq \sum_{n=1}^N (1+\rho_n) \delta \xi_n^2+ (1-\rho_n) \delta
\eta_n^2
}
With the exception of the three zero modes ($\rho_{1,2,3}=1$), the
configuration is stable if $\rho_n^2 <1$, $\forall n>3$ and it
is unstable if $\max_n \rho_n^2 >1$.

The spectrum of $B \bar{B}$ is invariant under \diaiii\ since
the hessian, at the stationary point, is transforming as
\eqn\diaxvi{
{\partial^2 V\over \partial z_i \partial z_j} =
{\partial^2 V\over \partial \xi_i \partial \xi_j}
{d \xi_i\over dz_i} {d \xi_j\over d z_j}
}
with
\eqn\diaxvii{
\left\vert {d\xi \over dz} \right\vert = {1\over
\vert \bar{\alpha} - \bar{\gamma} z\vert^2}=
{1+\xi \bar{\xi}\over 1+z \bar{z}}
}
Consequently if we call $\chi_j$ the argument of
${d \xi_j\over d z_j}$, the similarity property
\eqn\diaxviii{
B\bar{B}(x)=e^{i \chi} B \bar{B}(z) e^{-i \chi}
}
implies the invariance of the spectrum.  The eigenvalue
equation can also be written
\eqn\diaxix{
A \delta z+(1-\lambda) D \delta \bar{z}=0, \ \ \ {\rm and \ \ c.c.}
}
and differs from the naive form related to \diaviii\ which was
written $A \delta z+(D-\lambda ) \delta \bar{z}=0$.
Explicitly, equation \diaxix\ reads
\eqn\diaxx{
\sum_{j=1}^N {\delta z_i-\delta z_j\over z_{ij}^2}+ {N-1\over
(1+z_i \bar{z}_i)^2} ((1-\lambda) \delta \bar{z}_i -
\bar{z}_i^2 \delta z_i)=0
}
With the help of \diai ,
it can be reexpressed using $\delta z_j = z_j \xi_j$ as
\eqn\diaxxi{
\sum_{j \atop j\neq i} {z_i z_j\over z_{ij}^2} (\xi_i-\xi_j)
+ {N-1 z_i \bar{z}_i \over (1+ z_i \bar{z}_i)^2} (\xi_i-\rho
\bar{\xi}_i) =0
}
The 3 zero modes $(\rho=1)$ are associated with the 3
eigenvectors
\eqn\diaxxii{\eqalign{
& \delta z_j=i z_j \delta (Arg \alpha) \ \ {\rm or} \ \
\xi_j\propto i  \cr
& \delta z_j = (z_j^2+1) \delta (Re \gamma ) \ \ {\rm or} \ \
\xi_j\propto (z_j+z_j^{-1})  \cr
& \delta z_j=i (z_j^2-1 ) \delta (Im \gamma ) \ \ {\rm or}
\ \ \xi_j\propto i (z_j-z_j^{-1})
}}
Let us exhibit one negative eigenvalue in the case of the
cuboctahedron.
The invariance of
the system \diaxxi\ by the octaedral group reduces the
degree of the equations to 3 or 4.  In the cyclic representation
of order $m=4$, that is taking as a polar axis $m$ the axis
having a quaternary symmetry, the $z_j$'s separate in
4 cosets: $\sqrt{i} i^\nu , a i^\nu, a^{-1} i^\nu$ with
$\nu=1,2,3,4$(mod 4) belonging to 3 modules
$$
1, \  a, \  a^{-1}, \  \  a=\sqrt{2}-1=tg {\pi \over 8}$$
so that
\eqn\diaxxiii{\eqalign{
f(z)=\prod_1^{12} (z-z_j )&\equiv (z^4+1)(z^4-a^4)(z^4-a^{-4})
\cr &= z^{12}+1-33 (z^8+z^4)
}}
We can also choose the ternary representation in 4 cosets
\eqn\diaxxiv{\eqalign{
&i j^\nu , \ -i j^\nu , \ -b j^\nu , \ b^{-1} j^\nu , \ \ \
\nu=0,1,2 \  {\rm mod \ 3} \cr
&b=\sqrt{3}-\sqrt{2} =tg {\chi \over 2}}
}
with $\chi$ the angle between ternary and binary axis. This
yields the polynomial
\eqn\diaxxv{\eqalign{
g(\xi )&=(\xi^6+1)(\xi^3+b^3) (\xi^3-b^{-3})\cr
&\equiv \xi^{12}-1-22 \sqrt{3} (\xi^9+\xi^3)}
}
We switch from $z$ to $\xi$ by the rotation
\eqn\diaxxvi{
\xi={z \sqrt{i} + \rho\over 1- \rho z \sqrt{i}}, \ \ \
\rho=tg{\psi\over 2}={\sqrt{3}-1\over \sqrt{2}}=2 \sin {\pi
\over 12}
}
where $\psi$ is the angle between ternary and quaternary axis.
Although we did not want to do the complete reduction
according to the octahedral group, let us note that this is the
group of permutations of the 4 diagonals of
the cube named 1,2,3,4.  The elements of the
proper group fall in various classes according
to the table
$$\vbox{\offinterlineskip
\halign{\tv\quad # & \quad\tv \quad
# & \quad \tv \qquad  # & \quad \tv \qquad  #
& \quad \tv \qquad  # & \quad \tv \qquad  # &  \quad \tv #\cr
\noalign{\hrule}
\tvi ${\rm classes}$ & $(1)$ & $(13)$ & $(134)$ & $(13)(24)$ & $(1234)$& \cr
\tvi ${\rm order}$ & $1$ & $2$ & $3$ & $2$ & $4$ &  \cr
\tvi ${\rm number \ of \ elements}$ & $1$ & $6$ & $8$ & $3$ & $6$& \cr
\tvi ${\rm name}$ & $E$ & & $J^{-1}$& $I^2$ & $I$& \cr
\tvi ${\rm realization}$ &  & $z\rightarrow {i\over z}$ &
$z\rightarrow {z-i\over z+i}$ & $z\rightarrow -z$ &
$z\rightarrow iz$&\cr
\noalign{\hrule} }} $$
And the reflection with respect to the center
is $z\rightarrow -{1\over \bar{z}}$.

We are now going to seek the proper modes \diaxxi\ of the
system corresponding to invariant vectors by $I (J)$. The
form \diaxxi\ is adapted to Fourier transformation.  We limit
ourselves to the zero axial moment $\xi_j\equiv \xi_c$
where $c$ is the index of the coset of $j$
$$\eqalign{
&c=\sqrt{i} , \ a , \ a^{-1} \ \ {\rm for} \ I \ m=4 \cr
&c= i, \ i^{-1}, \ -b , \ b^{-1} \ \ {\rm for} \ J \ m=3}$$
We only need to compute
\eqn\diaxxvii{
\sum_{j \in b} {z_i z_j\over (z_i-z_j)^2} = m^2 {a^m b^m \over
(a^m-b^m)^2} , \ \ i\in a
}
where $a,b$ distinguishes here 2 indices of arbitrary cosets.
The system \diaxxi\ is then reduced to
\eqn\diaxxviii{
\sum_b (\xi_a - \xi_b) {m^2 a^m b^m \over (a^m-b^m)^2}+
{N-1 \over (\vert a \vert + \vert a^{-1} \vert )^2}
(\xi_a - \rho \bar{\xi}_a)=0
}
In the case $m=4$ the coefficients are all real because of
\diaxxiii\ -\diaxxv\ and the system decouples in
$Re \xi_a=0$ or $Im\xi_a=0$ according to the sign of
$\rho$.  We are left with one equation of second
degree in $\lambda$, after elimination of the root
$\lambda=0$ in the space of cosets
\eqn\diaxxix{\eqalign{
&A=16 \left\Vert \matrix{{1\over 18} & {-1\over 36} &
{-1\over 36} \cr {-1\over 36} & {31\over 36 \times 32} &
{1\over 2 \times 24^2} \cr {-1\over 36} & {1\over 2 \times
24^2} & {31\over 36 \times 32} } \right\Vert \cr & \ \cr
&D={11\over 8} \left\vert
\matrix{2 & . & . \cr . & 1 & . \cr . & .& 1 } \right\vert
}}
If we write $\lambda '={9 \times 11\over 2^5} \lambda$, the
secular equation
\eqn\diaxxx{
\left\vert \matrix{2-2\lambda ' & -1 & -1 \cr
-1 & 1-\lambda '-2^{-5} & 2^{-5} \cr
-1 & 2^{-5} & 1-\lambda '-2^{-5} } \right\vert =0
}
has the three roots $\lambda '=0,2,1-2^{-4} $ which
give the five stable modes
$$\vbox{\offinterlineskip
\halign{\tv\quad # & \quad\tv \quad
# & \quad \tv \qquad  # &  \quad \tv #\cr 
\noalign{\hrule}
\tvi $\lambda$ &$ 2-\lambda $& $ \rho $& \cr
\noalign{\hrule}
\tvi $*$  &$2$ & $1$ &\cr
\tvi ${10\over 33}$ & ${56\over 33} $ & ${23\over 33}$ &\cr
\tvi ${64\over 99}$ & ${134\over 99}$ & ${35\over 99}$ &\cr
\noalign{\hrule} }} $$
We would obtain the other modes relative to the moments
$\vert k \vert = 2,1 $ with the ansatz
\eqn\diaxxxii{
\xi_j=\xi_c i^{k \nu} , \ \ j\equiv (c,\nu) , \ \ \nu=0,1,2,3
({\rm mod 4})
}
but we will now turn to the ternary representation $m=3$
limiting ourselves again to the total zero moment
(states invariant under $J$).
Taking into account \diaxxvi ,
naming $\xi_{1,2,3,4}$ the components of the vector
$\xi$ relative to the 4 cosets (axial triangles)
$i, -i, -b, b^{-1}$, the equations \diaxxviii\ reads
\eqn\diaxxxiii{\eqalign{
-{1\over 4} (\xi_1-\xi_2) - {1\over 2 \ 3^5} (1+11 i \sqrt{2})
(2 \xi_1-\xi_3-\xi_4 )+{11\over 36} (\xi_1-\rho \bar{\xi}_1)&=0\cr
 {1\over 4} (\xi_1-\xi_2)-{1\over 2 \ 3^5} (1-11 i \sqrt{2})
(2\xi_2-\xi_3-\xi_4)+{11\over 36} (\xi_2-\rho \bar{\xi}_2)&=0\cr
-{1\over 4 \ 3^5} (\xi_3-\xi_4)-{1\over 2 \ 3^5} ((2 \xi_3-\xi_1
-\xi_2)-11 i \sqrt{2}(\xi_1-\xi_2))+{11\over 3 \ 36}(\xi_3
-\rho \bar{\xi}_3)&=0
\cr
{1\over 4 \ 3^5} (\xi_3-\xi_4)-{1\over 2 \ 3^5} ((2 \xi_4-\xi_1
-\xi_2)-11 i\sqrt{2} (\xi_1-\xi_2))+{11\over 3 \ 36} (\xi_4
-\rho \bar{\xi}_4)&=0
}}
Setting
\eqn\diaxxxiv{\eqalign{
&\xi_1+\xi_2=\xi , \ \ \ \xi_3+\xi_4=\eta \cr
&i(\xi_1-\xi_2)=\xi ',  \ \ \ \xi_3-\xi_4=\eta '
}}
The ternary system \diaxxxiii\ decouples, as predicted,
according to whether the variables \diaxxxiv\ are
all real or all imaginary corresponding to 2 values of
opposite sign of $\rho$ and can be written
\eqn\diaxxxv{\eqalign{
-{1\over 3^5} (\eta-\xi-11 \sqrt{2} \xi')&= {11\over 4\times 3^2}
(\xi-\rho \bar{\xi}) \cr
{1\over 3^5} (\eta-\xi-11 \sqrt{2} \xi ')&={11\over 4 \times 3^3}
(\eta-\rho \bar{\eta}) \cr
({1\over 2}+{1\over 3^5})\xi '-{11 \sqrt{2}\over 3^5} (\xi-\eta )
&={11\over 4\times 3^2}(\xi '+\rho \bar{\xi}')\cr
{1\over 2 \ 3^4} \eta '&={11\over 4\times 3^3} (\eta '-\rho
\bar{\eta}')
}}
which gives assuming that $Re (\xi , ... , \eta ')=0$, the zero
root $\lambda=0 \ (\rho=-1)$, the obvious root $\lambda={11\over 3 \times
11}$ and two other solutions of an equation of second degree
in $\mu={11\times 3^3\over 4} \lambda$
\eqn\diaxxxvi{
(\mu-4)(\mu-26)-8\times 11^2=0
}
So $\mu=15 \pm 33$.  We therefore obtain 7 eigenvalues of which
one is negative
$$\vbox{\offinterlineskip
\halign{\tv\quad # & \quad\tv \quad
# &\quad\tv#\cr
\noalign{\hrule}
\tvi $\lambda$ &$ 2-\lambda $& \cr
\noalign{\hrule}
\tvi $*$  &$2$ & \cr
\tvi ${2\over 33}=0.0606...$ & $1.9393... $ &\cr
\tvi ${2^6\over 11\times 3^2}=0.6464...$ & $1.3535...$ &\cr
\tvi $-{2^3\over 11\times 3}=-0.2424...$ & $2.2424...$ &\cr
\noalign{\hrule} }} $$
We note that the 15 eigenvalues computed (from the 21 non
zero ones) are rational numbers.  Would this be a general
property of hessian matrices of equilibrium configuration
on the sphere? We should first check this on the
icosahedron.

It is interesting to know the direction of the unstable
ternary mode $\lambda={-8 \over 33}=-0.2424...$ and also that of
the stable mode $\lambda={2\over 33}=0.0606...$.  With the help
of \diaxxxv\ we obtain for the unstable ternary mode
the proportion
\eqn\diaxxxvii{
Im (\xi: \xi ' : \eta : \eta ')=1 : -\sqrt{2}: -3 : 0
}
and for the less stable only $Im \eta '$ is non zero.
This gives according to \diaxxxiv\ for the unstable
mode
\eqn\diaf{\eqalign{
&\xi_3=\xi_4=-1.5 i \cr
&\xi_1=0.5 i-0.707... \ \ \ \xi_2=0.5 i+0.707...}}
and for the stable mode
\eqn\diafi{
\xi_1=\xi_2=0 , \ \ \xi_3=-\xi_4=i}
The elementary displacements $\delta z$ in $C$ are such that
$\delta z=\xi z$ according to \diaxxi . To interpret
the result \diaxxxvi\ we see that it is possible, by a
global rotation around the ternary axis $(\delta z_j /z_j=
1.5 i)$, to keep fixed the two cosets $-b$ and $b^{-1}$
made of the two opposite triangles of the cuboctahedron centered
on the ternary axis.  Finally this gives us the displacements
\eqn\diafii{\eqalign{
&\delta z_3=\delta z_4=0 \cr
& \delta z_1 / z_1=2 i-{1\over \sqrt{2}} \cr
&{\delta z_2\over z_2}=2 i+{1\over \sqrt{2}}
}}
which describe a deformation of the plane hexagon of the
cuboctahedron made of i) a rotation of the hexagon
(${dz\over z}=2i$) ii) a deformation of the latter
by dedoubling of the plane of the two cosets $i$ and $-i$.
The value of the angle ($Arctg(
1/(2\sqrt{2})$) and the direction indicated on the figure
suggests that this unstable ternary mode describes well
the beginning of the displacement of the cuboctahedron towards
the stable icosahedron, the two opposite triangles
staying the same to first order.  The less stable
mode \diafi\ is related to the vibration of torsion (opposite
rotations) of only these two triangles, the plane hexagon
remaining fixed.
\fig{Going from the cuboctahedron to the icosahedron}{geom.eps}{12cm}
\figlabel\tabb
\vfill

\appendix{C}{Shiota's Formula}

It is possible to write a formula for $\Delta^{2s}$
in terms
of the traces $t_l=\sum_{1\leq i \leq N} z_i^l$.
Noting that
\eqn\shi{
\Delta^{2s}=\prod_{1\leq i <j\leq N} (z_i-z_j)^{2s}
}
is $(-1)^{sN(N-1)/2}$ times the coefficient of $\epsilon^{N(N-1)}$
in
\eqn\shii{
\Pi:=
\prod_{1\leq i \leq N \atop 1 \leq j \leq N} (1-\epsilon (z_i
- z_j)^s )
}
where $\epsilon$ is a formal scalar parameter close to zero.
Let us compute this quantity
\eqn\shiii{\eqalign{
\Pi&=\exp \left( \sum_{1\leq i \leq N\atop 1\leq j \leq N} \log
(1-\epsilon(z_i-z_j)^s ) \right) \cr
&=\exp \left( -\sum_{n=1}^\infty {\epsilon^n \over n} \sum_{
1\leq i \leq N\atop 1\leq j \leq N} (z_i-z_j)^{ns} \right) \cr
&=\exp \left( - \sum_{n=1}^\infty {\epsilon^n\over n}
\sum_{l=0}^{ns} {ns \choose l} \sum_{1\leq i \leq N\atop 1\leq j \leq N}
z_i^{ns-l} (-z_j)^l \right) \cr
&=\exp \left( -\sum_{n=1}^\infty {\epsilon^n \over n} \sum_{l=0}^{ns}
{ns \choose l} (-1)^l t_{ns-l} \ t_l \right)
}}
Taylor
expanding the exponential we have
\eqn\shiv{
\Pi=\sum_{k=0}^\infty {1\over k!} \left( - \sum_{n=1}^\infty
{\epsilon^{n}\over n} \sum_{l=0}^{ns} {ns \choose l}
(-1)^l t_{ns-l} \ t_l \right)^k
.}
Taking $(-1)^{sN(N-1)/2}$ times the coefficient of $\epsilon^{N(N-1)}$
we obtain
\eqn\shivi{
\Delta^{2s}=(-1)^{sN(N-1)/2} \sum_{k\geq 1}
{(-1)^k\over k!} \sum_{n_1\geq 1,..., n_k\geq 1 \atop
n_1+...+n_k=N(N-1)} \prod_{i=1}^k {1\over 2 n_i}
\sum_{l=0}^{n_i s} {n_i s \choose l} (-1)^l t_{n_i s-l} \ t_l
}
(summing over $k\leq N(N-1)$ suffices since the $n_i \geq 1$ must
add up to $N(N-1)$ ), or
\eqn\shivii{\encadremath{\eqalign{ \Delta^{2s}=
&(-1)^{sN(N-1)/2} \times \cr \times
\sum_{k_1,k_2,...\geq 0 \atop k_1+2 k_2+3 k_3
+...=N(N-1)}& {(-1)^{k_1+k_2+...}\over k_1! k_2 ! \cdots}
\prod_{n\geq 1} \left( {1\over n} \sum_{l=0}^{ns}
{ns \choose l} (-1)^l t_{ns-l} \ t_l \right)^{k_n}
\cr}}}

\appendix{D}{Decomposition in terms of characters for small $N$}

The decomposition of the square of the Vandermonde
determinant in terms of
characters can be written in the following form,
\eqn\decomp{
\Delta^2= \prod
(1-\tau_i)^3 (1-\tau_i \tau_{i+1})^3 \cdots (1-\tau_i ...
\tau_{N-1})^3 \bigg\vert_{adm} \ ch_{0,3,...,3(N-2),3(N-1)}
}
where the operators $\tau_i$ are
acting on the labels of the characters at position i shifting
$l_{i-1}$ upward by one and $l_i$ downward by one.
Doing the decomposition explicitly we find

\noindent N=2 (2 terms)
\eqn\ndeux{
\Delta^2=-3  ch_{1,2}  +  ch_{0,3} .
}

\noindent N=3 (5 terms)
\eqn\ntrois{\eqalign{
\Delta^2=-15  ch_{2,3,4}  + 6 & ch_{1,3,5} -
  3 ch_{0,4,5}  \cr &- 3 ch_{1,2,6}  + ch_{0,3,6} .
}}

\noindent N=4 (16 terms)
\eqn\nquatre{\eqalign{
&\Delta^2=105  ch_{3,4,5,6} - 45 ch_{2,4,5,7}  -
  6 ch_{2,3,6,7}  + 27 ch_{1,4,6,7}  \cr & -
  15 ch_{0,5,6,7}
  + 27 ch_{2,3,5,8}  -
  9 ch_{1,4,5,8}   - 12 ch_{1,3,6,8} \cr & +
  6 ch_{0,4,6,8}  + 9 ch_{1,2,7,8}    -
  3 ch_{0,3,7,8}  - 15 ch_{2,3,4,9}  \cr & +
  6 ch_{1,3,5,9}
  - 3 ch_{0,4,5,9}  -
  3 ch_{1,2,6,9}  + ch_{0,3,6,9} .
}}

\noindent N=5 (59 terms)
\eqn\ncinq{\eqalign{
&\Delta^2=945 ch_{4,5,6,7,8}  - 420 ch_{3,5,6,7,9}  -
  75 ch_{3,4,6,8,9}  \cr & + 270 ch_{2,5,6,8,9}  +
  45 ch_{2,4,7,8,9}  - 180 ch_{1,5,7,8,9}  \cr &+
  105 ch_{0,6,7,8,9}  + 270 ch_{3,4,6,7,10}   -
  90 ch_{2,5,6,7,10}  \cr & + 45 ch_{3,4,5,8,10}  -
  144 ch_{2,4,6,8,10}  + 72 ch_{1,5,6,8,10}  \cr &-
  18 ch_{2,3,7,8,10}  + 81 ch_{1,4,7,8,10}  -
  45 ch_{0,5,7,8,10}  \cr & - 18 ch_{2,4,5,9,10}  +
  111 ch_{2,3,6,9,10}  - 27 ch_{1,4,6,9,10}  \cr &-
  6 ch_{0,5,6,9,10}  - 54 ch_{1,3,7,9,10}   +
  27 ch_{0,4,7,9,10}  \cr &+ 45 ch_{1,2,8,9,10}  -
  15 ch_{0,3,8,9,10}  - 180 ch_{3,4,5,7,11}  \cr & +
  72 ch_{2,4,6,7,11}  - 36 ch_{1,5,6,7,11}  +
  81 ch_{2,4,5,8,11}  \cr &- 27 ch_{2,3,6,8,11}   -
  36 ch_{1,4,6,8,11}  + 27 ch_{0,5,6,8,11}  \cr &+
  18 ch_{1,3,7,8,11}  - 9 ch_{0,4,7,8,11}    -
  54 ch_{2,3,5,9,11}  \cr &+ 18 ch_{1,4,5,9,11}  +
  24 ch_{1,3,6,9,11}  - 12 ch_{0,4,6,9,11} \cr &  -
  18 ch_{1,2,7,9,11}  + 6 ch_{0,3,7,9,11}   +
  45 ch_{2,3,4,10,11}  \cr & - 18 ch_{1,3,5,10,11}   +
  9 ch_{0,4,5,10,11}  + 9 ch_{1,2,6,10,11} \cr & -
  3 ch_{0,3,6,10,11}  + 105 ch_{3,4,5,6,12}   -
  45 ch_{2,4,5,7,12}  \cr &- 6 ch_{2,3,6,7,12}  +
  27 ch_{1,4,6,7,12}  - 15 ch_{0,5,6,7,12} \cr &   +
  27 ch_{2,3,5,8,12}  - 9 ch_{1,4,5,8,12}  -
  12 ch_{1,3,6,8,12}  \cr &+ 6 ch_{0,4,6,8,12}  +
  9 ch_{1,2,7,8,12}  - 3 ch_{0,3,7,8,12} \cr & -
  15 ch_{2,3,4,9,12}  + 6 ch_{1,3,5,9,12}   -
  3 ch_{0,4,5,9,12}  \cr &- 3 ch_{1,2,6,9,12}  +
   ch_{0,3,6,9,12}.
}}

If we label the coefficients
by $C(\tau_i ... \tau_j)$ the following simple rules
can be found directly by looking at \decomp ,
\eqn\ruleI{
C(\tau_i \tau_{i+1}...\tau_{i+n})=C(\tau_j \tau_{j+1}...\tau_{j+n})
}
\eqn\ruleIII{
C(\tau_i...\tau_{i+p})=-2 C(\tau_{i+1}...\tau_{i+p})
}
We also read from these examples the following list of
coefficients,
\eqn\coefI{
C(\tau_i)=-3
}
\eqn\coefII{
C(\tau_i^2)=3
}
\eqn\coefIII{
C(\tau_i...\tau_{i+n})=-3 (-2)^{n}
}
\eqn\coefIV{
C(\tau_i^2...\tau_{i+n}^2)=3 (4^{n}-9 n)
}
\eqn\coefV{
C(\tau_i \tau_{i+1}^2...\tau_{n+i-1}^2)=
 -3 (2^{2n-3}-{27\over 2} n+{99\over 4}+{1\over 4} (-5)^{n-2})
}
\eqn\coefVI{
C(\tau_i \tau_{i+1}^2...\tau_{i+n-2}^2 \tau_{i+n-1})=
162+(24-{27\over 4} n) (-5)^{n-4}+3 \ \ 2^{2n-4}-{243\over 4} n
}

For higher powers of the Vandermonde determinant
we found the following decompositions in terms of characters.

\vskip .2in
\noindent {\bf Fourth power}

\noindent N=2  (3 terms)
\eqn\ei{\Delta^4=ch_{0,5}-5 ch_{1,4}+10 ch_{2,3}}

\noindent N=3 (13 terms)
\eqn\eii{\eqalign{&\Delta^4=ch_{0,5,10}-5 ch_{1,4,10}-5 ch_{0,6,9}
+10 ch_{2,3,10}+10 ch_{0,7,8}-15 ch_{2,5,8}\cr &+20 ch_{1,5,9}
-25 ch_{1,6,8}-25 ch_{2,4,9}+100 ch_{3,4,8}+100 ch_{2,6,7} \cr &
-160 ch_{3,5,7}+280 ch_{4,5,6}}}

\noindent N=4 (76 terms)
\eqn\eiii{\eqalign{&\Delta^4=ch_{0,5,10,15}-5 ch_{0,5,11,14}
+10 ch_{0,5,12,13} \cr & -5 ch_{0,6,9,15}+20 ch_{0,6,10,14}
-25 ch_{0,6,11,13}+10 ch_{0,7,8,15} \cr &
-25 ch_{0,7,9,14}-15 ch_{0,7,10,13}+100 ch_{0,7,11,12}
+100 ch_{0,8,9,13} \cr & +280 ch_{0,9,10,11}
-160 ch_{0,8,10,12}-5 ch_{1,4,10,15}+25 ch_{1,4,11,14} \cr
& -50 ch_{1,4,12,13}+20 ch_{1,5,9,15}-80 ch_{1,5,10,14}
+100 ch_{1,5,11,13} \cr &
-25 ch_{1,6,8,15}+50 ch_{1,6,9,14}+100 ch_{1,6,10,13}
-375 ch_{1,6,11,12} \cr &
+75 ch_{1,7,8,14}-300 ch_{1,7,9,13}+375 ch_{1,7,10,12}
+300 ch_{1,8,9,12} \cr & -600 ch_{1,8,10,11} +10 ch_{2,3,10,15}
-50 ch_{2,3,11,14}+100 ch_{2,3,12,13} \cr &
-25 ch_{2,4,9,15}+100 ch_{2,4,10,14}
-125 ch_{2,4,11,13}-15 ch_{2,5,8,15} \cr &
+100 ch_{2,5,9,14}
-290 ch_{2,5,10,13} +475 ch_{2,5,11,12}+100 ch_{2,6,7,15} \cr &
-300 ch_{2,6,8,14}+200 ch_{2,6,9,13}
+100 ch_{2,6,10,12}+825 ch_{2,7,8,13}\cr & -1125 ch_{2,7,9,12}
+150 ch_{2,7,10,11}
+1200 ch_{2,8,9,11}+100 ch_{3,4,8,15}\cr & -375 ch_{3,4,9,14}
+475 ch_{3,4,10,13} -250 ch_{3,4,11,12}-160 ch_{3,5,7,15} \cr &
+375 ch_{3,5,8,14}+100 ch_{3,5,9,13}
-685 ch_{3,5,10,12}+300 ch_{3,6,7,14}\cr &-1125 ch_{3,6,8,13}
+1425 ch_{3,6,9,12} -750 ch_{3,6,10,11}-200 ch_{3,7,8,12} \cr &
+800 ch_{3,7,9,11}-4400 ch_{3,8,9,10}
+280 ch_{4,5,6,15}-600 ch_{4,5,7,14}\cr &+150 ch_{4,5,8,13}
-750 ch_{4,5,9,12} +3180 ch_{4,5,10,11}
+1200 ch_{4,6,7,13}\cr &+800 ch_{4,6,8,12}-3800 ch_{4,6,9,11}
-1200 ch_{4,7,8,11}+6600 ch_{4,7,9,10}\cr &-4400 ch_{5,6,7,12}
+6600 ch_{5,6,8,11}-880 ch_{5,6,9,10}
-9240 ch_{5,7,8,10}\cr &+15400 ch_{6,7,8,9}
}}

\vskip .2in
\noindent {\bf Sixth power}

\noindent N=2  (4 terms)
\eqn\eiv{\Delta^6=ch_{0,7}-7 ch_{1,6}+21 ch_{2,5}-35 ch_{3,4}}

\noindent N=3 (25 terms)
\eqn\ev{\eqalign{&\Delta^6=ch_{0,7,14}-7 ch_{0,8,13}+21 ch_{0,9,12}
\cr & -35 ch_{0,10,11}-7 ch_{1,6,14}+42 ch_{1,7,13}-98 ch_{1,8,12}
\cr & +98 ch_{1,9,11}+21 ch_{2,5,14}-98 ch_{2,6,13}+119 ch_{2,7,12}
\cr & +147 ch_{2,8,11}-539 ch_{2,9,10}-35 ch_{3,4,14}
+98 ch_{3,5,13} \cr & +147 ch_{3,6,12}-868 ch_{3,7,11}+1078 ch_{3,8,10}
-539 ch_{4,5,12}\cr & +1078 ch_{4,6,11}+231 ch_{4,7,10}
-2695 ch_{4,8,9}-2695 ch_{5,6,10} \cr & +3850 ch_{5,7,9}
-5775 ch_{6,7,8}
}}

\vskip .2in
\noindent {\bf Eighth power}

\noindent N=2 (5 terms)
\eqn\evi{\eqalign{\Delta^8=ch_{0,9}&-9 ch_{1,8}+36 ch_{2,7} \cr &
-84 ch_{3,6}+126 ch_{4,5}}}

\noindent N=3 (41 terms)
\eqn\evii{\eqalign{&\Delta^8=ch_{0,9,18}-9 ch_{0,10,17}
+36 ch_{0,11,16} \cr & -84 ch_{0,12,15}+126 ch_{0,13,14}
-9 ch_{1,8,18}+72 ch_{1,9,17} \cr & -243 ch_{1,10,16}
+432 ch_{1,11,15}-378 ch_{1,12,14}+36 ch_{2,7,18} \cr &
-243 ch_{2,8,17}+603 ch_{2,9,16}-432 ch_{2,10,15}
-972 ch_{2,11,14} \cr & +2646 ch_{2,12,13}-84 ch_{3,6,18}
+432 ch_{3,7,17}-432 ch_{3,8,16} \cr & -1776 ch_{3,9,15}
+5724 ch_{3,10,14}-6048 ch_{3,11,13}+126 ch_{4,5,18} \cr &
-378 ch_{4,6,17}-972 ch_{4,7,16}+5724 ch_{4,8,15}
-7650 ch_{4,9,14} \cr & -3402 ch_{4,10,13}+19656 ch_{4,11,12}
+2646 ch_{5,6,16}-6048 ch_{5,7,15}\cr & -3402 ch_{5,8,14}
+28224 ch_{5,9,13}-34398 ch_{5,10,12}+19656 ch_{6,7,14} \cr &
-34398 ch_{6,8,13}-3822 ch_{6,9,12}+68796 ch_{6,10,11}
+68796 ch_{7,8,12}\cr & -91728 ch_{7,9,11}
+126126 ch_{8,9,10}
}}

{\noindent \bf Coefficients $C_{\{ n\} }^{(s)}$ for the vertices
of the admissible polytope.}

The admissible monomials in the expansion (D.1) are of the
form $\tau_1^{n_1} ... \tau_{N-1}^{N-1}$, where the
$(n_0=0, n_1, ..., n_{N-1}, n_N=0)$ are subject to
\eqn\polii{
2s+n_{i+1}+n_{i-1}-2n_i \geq 0; \ \ n_i>0; \ \ i=1,...,N-1.}
The equations \polii\ define a polytope or $\IR^{N-1}$, which
is a certain deformation of an hypercube.  It can be shown
\ref\gelfand{I.M. Gelfand, M.M. Kapranov and A.V. Zelevinsky,
Adv. in Math. {\bf 84} (1990) 237.}\
that its $2^{N-1}$ vertices are in one to one
correspondence with strictly increasing sequences of integers:
$$
i_0=0 < i_1 <i_2 ... <i_k <i_{k+1} =N$$
denoted $[0,i_1,...,i_k,N]$ here. The corresponding points
read
\eqn\pointss{
n_i=s (i-i_k) (i_{k+1}-i), \ \ \forall i \in [i_k,i_{k+1}],
\ \ l=1,...,k-1.
}
These points are easily identified as the only ones which
vanish at $i=i_1,...,i_k$ (saturate the second inequality
of \polii\ at these points), and have $n_{i+1}-n_{i-1}
-2 n_i+2s=0$ (saturate the first inequality of \polii\ )
between those.

The farthest vertex from the origin corresponds to
$[0,N]$, and $n_i=s i(N-i)$ for $i=1,...,N-1$. It
yields the most compact character $ch_{s(N-1),s(N-1)+1,
...,(s+1)(N-1)}$ with coefficient already given
in (5.16)
\eqn\coecinq{
C^{(s)}_{\{ si(N-i)\}} = {[(s+1)N]!\over [(s+1)!]^N N!}
(-1)^{sN(N-1)/2} }
Thanks to the factorization property 5 (eqn \facto\ )
the coefficient pertaining to the vertex
$[0,i_1,...,i_k,N]$ reads
\eqn\ccco{
C^{(s)}_{[0,i_1,...,i_k,N]}=\prod_{l=0}^k (-1)^{s(i_{l+1}
-i_l)(i_{l+1}-i_{l-1})/2} \times {[(s+1)(i_{l+1}-i_l)]!
\over [(s+1)!]^{i_{l+1}-i_l} (i_{l+1}-i_l)!}}

\appendix{E}{More on the number of terms in the expansion}

{\bf Computing the volume $b_N^{(N)}=a_N^{(N)}=V_{N+1}$ }

If we reorganize the inequalities \polito\ for $s=1$, we find that
$$\eqalign{
sup(0,2n_{N-2}-n_{N-3}-2) \leq n_{N-1} &\leq 1+{n_{N-2} \over 2} \cr
sup(0,2n_{N-3}-n_{N-4}-2) \leq n_{N-2} &\leq 2+{2 n_{N-3} \over 3} \cr
\cdots & \cdots \cr
sup(0,2n_2 -n_1 -2) \leq n_3 &\leq N-3+{(N-3)n_2 \over N-2} \cr
sup(0,2n_1 -2) \leq n_2 &\leq N-2+{(N-2)n_1 \over N-1} \cr
0 \leq n_1 &\leq N-1 \cr}$$
Let us define $\phi_{N-1-k}$ as $\phi_{N-1}=1$, and
$$\phi_{N-1-k}=\theta(k+1+{(k+1)n_{N-k-2} \over k+2}-n_{N-1-k})
\int_{sup(0,2n_{N-1-k}-n_{N-k-2}-2)}^{k+{k n_{N-1-k}
\over k+1}} dn_{N-k} \phi_{N-k},$$
where
$$\eqalign{ \theta(x) &=1 \quad {\rm if} \ \ x \geq 0 \cr
&=0 \quad { \rm otherwise}. \cr}$$
For instance,
$$ \phi_{N-2}=\theta(2+{2 n_{N-3} \over 3}-n_{N-2})
inf(1+{n_{N-2} \over 2},{3 \over 2}(2+{2 n_{N-3} \over 3}-n_{N-2})). $$
The volume of the polytope $\pi_N^{(1)}$ is
$$V_N= \phi_0 .$$

Let us rewrite
$$\phi_{N-2}={3 \over 2}(2+{2 n_{N-3} \over 3}-n_{N-2})
\theta(2+{2 n_{N-3} \over 3}-n_{N-2})   -
2( 1+{n_{N-3} \over 2} -n_{N-2} ) \theta( 1+{n_{N-3} \over 2} -n_{N-2} ),$$
and introduce the notation
$$x_k^{(j)}=j+{j n_{k-1} \over j+1} -n_{k}$$
($x_1^{(j)}=j-n_1$ and $x_0^{(j)}=j$)
then
$$\phi_{N-2}=
{3 \over 2} x_{N-2}^{(2)}\theta(x_{N-2}^{(2)})
-2 x_{N-2}^{(1)} \theta(x_{N-2}^{(1)})$$

Suppose we know that $\phi_{N-k}$ has the form
\eqn\rechyp{\phi_{N-k}=
\sum_{j=1}^k p_{j}^{(k-1)} (x_{N-k}^{(j)}) \theta(x_{N-k}^{(j)})}
where the $p^{(k-1)}$'s are some polynomials of degree $k-1$.
Then by definition,
$$\eqalign{\phi_{N-k-1}&=\theta(x_{N-k-1}^{(k+1)})
\int_{sup(0,2n_{N-1-k}-n_{N-k-2}-2)}^{k+{k n_{N-1-k}
\over k+1}} dn_{N-k} \ \phi_{N-k}\cr
&=\theta(x_{N-k-1}^{(k+1)})
\sum_{j=1}^k\int_{sup(0,2n_{N-1-k}-n_{N-k-2}-2)}^{k+{k n_{N-1-k}
\over k+1}} dn_{N-k} \ p_j^{(k-1)}(x_{N-k}^{(j)})
\theta(x_{N-k}^{(j)}) \cr}$$
which becomes after the changes of variables
$n_{N-k} \to x_{N-k}^{(j)}$
$$\eqalign{\phi_{N-k-1}=\theta(x_{N-k-1}^{(k+1)})\sum_{j=1}^k
\theta(inf(j+{jn_{N-1-k} \over j+1}, &
{j+2 \over j+1} x_{N-k-1}^{(j+1)})) \times \cr & \times
\int_0^{inf(j+{jn_{N-1-k} \over j+1},
{j+2 \over j+1} x_{N-k-1}^{(j+1)})} du \ p_j^{(k-1)}(u)}$$

Note that
$${j+2 \over j+1}x_{N-k-1}^{(j+1)}-(j+{jn_{N-1-k} \over j+1})=
2 x_{N-k-1}^{(1)},$$
so that
$$\eqalign{
\phi_{N-k-1}&=\sum_{j=1}^k [1-\theta(x_{N-k-1}^{(1)})]
\theta(x_{N-k-1}^{(j+1)})
\int_0^{{j+2 \over j+1}x_{N-k-1}^{(j+1)}} du \ p_j^{(k-1)}(u) \cr
&+\sum_{j=1}^k \theta(x_{N-k-1}^{(1)})\int_0^{j(1+{n_{N-1-k} \over j+1})}
du \ p_j^{(k-1)}(u) \cr
&=\theta(x_{N-k-1}^{(1)}) \sum_{j=1}^k \int_{j(1+{n_{N-1-k} \over j+1})+
2x_{N-k-1}^{(1)}}^{j(1+{n_{N-1-k} \over j+1})} du \ p_j^{(k-1)}(u) \cr
&-\sum_{j=1}^k \theta(x_{N-k-1}^{(j+1)})
\int_0^{{j+2 \over j+1}x_{N-k-1}^{(j+1)}} du \ p_j^{(k-1)}(u).\cr}$$
where we used the inequality $x_{N-k-1}^{(1)} \leq x_{N-k-1}^{(j+1)}$
for $j=1,...,k$ to rewrite
$$\theta(x_{N-k-1}^{(1)})\theta(x_{N-k-1}^{(j+1)}) =
\theta(x_{N-k-1}^{(1)}).$$
The form of $\phi_{N-k-1}$ will be that of the recursion hypothesis \rechyp\
iff
\eqn\trial{p_1^{(k)}(x)=\sum_{j=1}^k
\int_{j(1+{y \over j+1})+2x}^{j(1+{y \over j+1})}
du \ p_j^{(k-1)}(u) }
is independent on $y$, and a polynomial of $x$ only.
Then we have, for $j=2,...,k+1$
$$p_j^{(k)}(x) = \int_0^{{j+1 \over j}x} du \ p_{j-1}^{(k-1)}(u).$$
Thanks to these relations, we can write $p_1^{(k)}$ as
$$p_1^{(k)}(x)=\sum_{j=2}^{k+1}
\bigg[ p_j^{(k)}({j(j-1) \over j+1}(1+{y \over j})) -
p_j^{(k)}({j(j-1) \over j+1}(1+{y \over j})+{2xj \over j+1}) \bigg].$$

Let us prove by recursion that this expression is independent on $y$.
Suppose the property true for $k-1$,
then in the above we can use it to rewrite \trial\
$$
p_1^{(k)}(x)=\sum_{j=1}^k
\int_{j(1+{y \over j+1})+2x}^{j(1+{y \over j+1})} du \ p_j^{(k-1)}(u)$$
by replacing
$$ p_1^{(k-1)}(u)= \sum_{j=2}^{k}
\bigg[ p_j^{(k-1)}({j(j-1) \over j+1}(1+{z \over j})) -
p_j^{(k-1)}({j(j-1) \over j+1}(1+{z \over j})+{2uj \over j+1}) \bigg]$$
valid for any $z$.
We get
$$\eqalign{
p_1^{(k)}(x)&=\sum_{j=2}^k \bigg[
\int_{j(1+{y \over j+1})+2x}^{j(1+{y \over j+1})} du \ p_j^{(k-1)}(u) \cr
&+\int_{1+{y \over 2}+2x}^{1+{y \over 2}} \big[
 p_j^{(k-1)}({j(j-1) \over j+1}(1+{z \over j})) -
p_j^{(k-1)}({j(j-1) \over j+1}(1+{z \over j})+{2uj \over j+1})\big]
\bigg] \cr}$$
We take the range of the integrations to be $[2x,0]$ by suitable
changes (shifts) of variables, to get
$$\eqalign{
p_1^{(k)}(x)&=\int_{2x}^0 du \sum_{j=2}^k
\bigg[
p_j^{(k-1)}(u+j(1+{y \over j+1})+  \cr
&+p_j^{(k-1)}({j(j-1) \over j+1}(1 +{z \over j}))
-p_j^{(k-1)}({j(j-1) \over j+1}(1 +{z \over j})
+{2j \over j+1}(u+1+{y \over 2})) \bigg].\cr}$$
But this is valid for any $z$, let us take $z=-u$, then the
first and last term in the sum cancel out exactly, leaving us with
$$p_1^{(k)}(x)=\sum_{j=2}^k \int_{2x}^0 du \
p_j^{(k-1)}({j(j-1) \over j+1}(1 -{u \over j})).$$
clearly independent of $y$. This completes the proof of
the general formula
$$ \phi_{N-k-1}=\sum_{j=1}^{k+1} \theta(x_{N-k-1}^{(j)})
p_j^{(k)}(x_{N-k-1}^{(j)}), $$
with the recursions
$$\eqalign{
p_1^{(0)}(x)&=1 \cr
p_j^{(k)}(x) &= \int_0^{{j+1 \over j} x}
du \ p_{j-1}^{(k-1)}(u) \quad j=2,3,...,k+1 \cr
p_1^{(k)}(x) &= \sum_{j=2}^{k+1}
\bigg[ p_j^{(k)}({j(j-1) \over j+1}(1+{y \over j})) -
p_j^{(k)}({j(j-1) \over j+1}(1+{y \over j})+{2xj \over j+1})
\bigg],\cr}$$
where the last expression is independent of $y$.

The volume reads
$$V_N=\phi_0=\sum_{j=1}^{N} p_j^{(N-1)}(j).$$
The expression for $p_1^{(k)}(x)$ gives for $y=0$
$$p_1^{(k)}(x)=\sum_{j=2}^{k+1}
\bigg[ p_j^{(k)}({j(j-1) \over j+1}) -
p_j^{(k)}({j(j-1) \over j+1}+{2xj \over j+1}) \bigg],$$
hence for $x=1$,
$$\sum_{j=1}^{k+1} p_j^{(k)}(j)= \sum_{j=2}^{k+1}
p_j^{(k)}({j(j-1) \over j+1}).$$

The proof of independence of $y$ gives us another expression for
$p_1^{(k)}$
$$p_1^{(k)}(x)=\sum_{j=3}^{k+1} {j \over j-2} \big[
p_j^{(k)}({(j-1)(j-2) \over j+1}(1-{2x \over j-1}))-
p_j^{(k)}({(j-1)(j-2) \over j+1}) \big]$$

The first few $p$'s are listed below.

$$\eqalign{
k=0 \ \ \ p_1^{(0)}&= 1 \cr
k=1 \ \ \ p_2^{(1)}&={3 \over 2} x \cr
          p_1^{(1)}&= -2 x \cr
k=2 \ \ \ p_3^{(2)}&={4 \over 3}x^2 \cr
          p_2^{(2)}&=-{9 \over 4}x^2 \cr
          p_1^{(2)}&=x^2 -2x \cr
k=3 \ \ \ p_4^{(3)}&={125 \over 144}x^3 \cr
          p_3^{(3)}&=-{16 \over 9}x^3 \cr
          p_2^{(3)}&={9 \over 8}x^3-{9 \over 4}x^2\cr
          p_1^{(3)}&=-{2 \over 9}x^3+2x^2 -4x \cr
k=4 \ \ \ p_5^{(4)}&={9 \over 20}x^4 \cr
          p_4^{(4)}&=-{625 \over 576}x^4\cr
          p_3^{(4)}&={8 \over 9}x^4 - {16 \over 9}x^3 \cr
          p_2^{(4)}&=-{9 \over 32}x^4
+{9 \over 4}x^3-{9 \over 2}x^2 \cr
          p_1^{(4)}&={1 \over 36}x^4-{2 \over 3}x^3
+5 x^2-{32 \over 3}x \cr
k=5 \ \ \ p_6^{(5)}&={16807 \over 86400}x^5 \cr
          p_5^{(5)}&=-{27 \over 50}x^5 \cr
          p_4^{(5)}&={625 \over 1152}x^5-{625 \over 576}x^4 \cr
          p_3^{(5)}&=-{32 \over 135}x^5+{16 \over 9}x^4
-{32 \over 9}x^3 \cr
          p_2^{(5)}&={27 \over 640}x^5-{27 \over 32}x^4
+{45 \over 8}x^3-12 x^2 \cr
          p_1^{(5)}&=-{1 \over 450}x^5+{1 \over 9}x^4
-2 x^3+{44 \over 3}x^2 -{100 \over 3}x \cr
}$$

which lead to the volumes
$$\eqalign{
N \qquad &{\rm Volume}\ V_N \cr
2 \qquad &1 \cr
3 \qquad &2 \cr
4 \qquad &{16 \over 3} \cr
5 \qquad &{50 \over 3} \cr
6 \qquad &{288 \over 5} \cr
7 \qquad &{9604 \over 45} \cr
8 \qquad &{262144 \over 315}.\cr}$$


It is easy to read the actual exact result
\eqn\qqqqq{a_N^{(N)}=b_N^{(N)}=V_{N+1}=2^N {{(N+1)}^{N-2} \over N!} }
hence the dilated polytope $\Pi_{N+1}^{(s)}$ has
volume
\eqn\volqqq{V_{N+1}^{(s)}=a_N^{(N)} s^N =(2s)^N {(N+1)^{N-2}\over N!}.}
We guess formula (E.3) on the basis of the previous
calculations.

{\bf Computing $b_{N-1}^{(N)}=a_{N-1}^{(N)}$}

Let us now concentrate on the subleading coefficient
$a_{N-1}^{(N)}=b_{N-1}^{(N)}$.
By definition, we have
\eqn\subcoef{
b_{N-1}^{(N)}s^{N-1}=
\sum_{k=1}^N {1 \over 2}({\partial \over \partial h_k}+
{\partial \over \partial \Ge_k})
V_{N+1}^{(s)}(h_i;\Ge_i) \bigg\vert_{h_i=\Ge_i=0}.}
Let us write the deformed volume
\eqn\voldefor{ V_{N+1}^{(s)}(h_i;\Ge_i) =\int_{u \in \IR^N} d^N u
\prod_{k=1}^N \theta(u_k+\Ge_k) \theta(-2 u_k+u_{k+1}+u_{k-1}+2s+h_k),
}
${d \over dx} \theta(x)= \delta(x)$.
Therefore we have two contributions to \subcoef, the one coming from
$\Ge$ derivatives
$$\eqalign{
\partial_{\Ge_k} V_{N+1}^{(s)}(h_i;\Ge_i) \bigg\vert_{h_i=\Ge_i=0}&=
\int_{u \in \IR^N} d^N u \delta(u_k)
\prod_{j \neq k} \theta(u_j) \prod_{l=1}^N
\theta(-2 u_l+u_{l+1}+u_{l-1}+2s) \cr
&= V_{k}^{(s)} V_{N-k}^{(s)} \cr }$$
and the one coming from $h$ derivatives
$$\eqalign{
\sum_{k=1}^N \partial_{h_k} V_{N+1}^{(s)}(h_i;\Ge_i)
\bigg\vert_{h_i=\Ge_i=0}&= {d \over d(2s)} V_{N+1}^{(s)}\cr
&={N \over 2s} V_{N+1}^{(s)}.\cr}$$
Putting all contributions together, we find
$$ b_{N-1}s^{N-1}=a_{N-1}s^{N-1}=
{N \over 4s} V_{N+1}^{(s)} +{1 \over 2}
\sum_{k=0}^N V_k^{(s)} V_{N-k}^{(s)}.$$
The latter sum turns out to have a simple form
$$ \sum_{k=0}^N V_k^{(s)} V_{N-k}^{(s)} =
{(2s)^{N-1} \over (N-1)!} (N+1)^{N-4} (N+7),  $$
which leads to
\eqn\qqq{a_{N-1}^{(N)}=b_{N-1}^{(N)}= {2^{N-1} \over (N-1)! 2!}
(N+1)^{N-4} (6+(N+1)(N+2)).}
Applying this to $N=2,3,4,5,6,7,8,9$ yields respectively
$2$, ${13 \over 2}$, $24$, $96$, ${6076 \over 15}$,
${26624 \over 15}$, ${279936 \over 35}$,
${2320000 \over 63}$,
so we get all the subleading coefficients in the list \lispol.
Note that the general expression fails to reproduce the corresponding $N=1$
term, due to the different definition of the deformation for $N=1$.
We also get the subleading coefficients in the expression of the
$A$'s in terms of the $\Gs$'s \listsig, namely the
coefficient of
$\Gs_{N-1}$ in $A_{N+1}^{(s)}$ reads
\eqn\aqq{ 2^{N-2} (N+1)^{N-4} [6+N(N+1)(2N+1)],   }
which yields for $N=2,3,4,5,6,7,8,9,10$ respectively
$4$, $45$, $744$, $16128$, $432768$,
$13860864$, $516481920$, $21964800000$, $1050351430656$.

{\bf Computing $b_{N-2}^{(N)}$ and $a_{N-2}^{(N)}$}

The next to leading coefficient $a_{N-2}^{(N)}$ is trickier to obtain,
because it differs from $b_{N-2}$ by a highly non--trivial
quantity. The latter involves the two--dimensional cone invariant
$\tau_2$ computed in [10] and function of the Dedekind sum of
the two integers defining the (rational) cone.
However in our very particular case, we are able to guess a simple
answer.
The first task is to compute $b_{N-2}$ then to proceed and compare
its value to that of $a_{N-2}$ which we read from \lispol.
Thanks to the definition \bNdef, we find a closed formula for $b_{N-2}$
$$\eqalign{
b_{N-2}s^{N-2}&=\bigg[ {1 \over 12} \sum_i (\partial_{h_i}^2+
\partial_{\Ge_i}^2)
+{1 \over 4} \sum_{i<j} (\partial_{h_i} \partial_{h_j} +
\partial_{\Ge_i} \partial_{\Ge_j}) \cr
&+{1 \over 4} \sum_{i,j} \partial_{h_i} \partial_{\Ge_j} \bigg]
V_{N+1}^{(s)}(h_i,\Ge_i)\bigg\vert_{h_i=\Ge_i=0}.}$$
We use the fact that
$$\eqalign{
\sum_i \partial_{h_i} V &={d \over d(2s)} V \cr
\sum_{i<j} \partial_{h_i} \partial_{h_j} V &=
{1 \over 2}(\left({d \over d(2s)}\right)^2 -\sum_i \partial_{h_i}^2)
V, \cr}$$
to rewrite
$$\eqalign{
b_{N-2}s^{N-2}&=\bigg[ {1 \over 12} \sum_i ( \partial_{\Ge_i}^2
-{1 \over 2} \partial_{h_i}^2)
+{1 \over 4} \sum_{i<j}
\partial_{\Ge_i} \partial_{\Ge_j}+{1 \over 32}{d^2 \over ds^2} \cr
&+{1 \over 8} {d \over ds}\sum_{j} \partial_{\Ge_j} \bigg]
V_{N+1}^{(s)}(h_i,\Ge_i)\bigg\vert_{h_i=\Ge_i=0}.\cr}$$
Each of the above terms is easily derived by using the integral
expression for the deformed volume \voldefor. We find respectively
$$\eqalign{
\sum_{i<j} \partial_{\Ge_i} \partial_{\Ge_j} V_{N+1}&= \sum_{p,q \geq 1
\atop p+q \leq N-1} V_p V_q V_{N-p-q} \cr
{d^2 \over ds^2} V_{N+1} &= N(N-1) V_{N+1} /s^2 \cr
{d \over ds}\sum_{j} \partial_{\Ge_j} V_{N+1} &=
{N-1 \over s} \sum_{p=1}^{N-1} V_p V_{N-p}. \cr}$$
The only difficult part involves the double derivatives.
For $\Ge$ derivatives, we have
$$\eqalign{
\partial_{\Ge_i}^2 V_{N+1} &=\int_{u \in \IR^N} d^N u
\delta'(u_i) \prod_{j \neq i}
\theta(u_j) \prod_j \theta(-2u_j+u_{j+1}+u_{j-1}+2s) \cr
&=-\int_{R^N} du \delta(u_i) \prod_{j \neq i}\theta(u_j) \varphi_i(u_j),
\cr}$$
where
$$\eqalign{ \varphi_{i}(u_j) &= \delta(u_i+u_{i+2}+2s-2u_{i+1})
\prod_{j \neq i+1} \theta(-2u_j+u_{j+1}+u_{j-1}+2s) \cr
&+ \delta(u_i+u_{i-2}+2s-2u_{i-1})
\prod_{j \neq i-1} \theta(-2u_j+u_{j+1}+u_{j-1}+2s) \cr
&-2 \delta(u_{i+1}+u_{i-1}+2s-2u_i)
\prod_{j \neq i} \theta(-2u_j+u_{j+1}+u_{j-1}+2s).\cr}$$
The last term does not contribute because it multiplies
$\delta(u_i) \theta(u_{i-1})\theta(u_{i+1})$
and for $s>0$ the various constraints are incompatible.
So we are finally left with
$$
\partial_{\Ge_i}^2 V_{N+1} = -\sum_{p=1}^{N-1}(
 V_{N-i} \partial_{h_{i-1}} V_i+V_i \partial_{h_{i+1}} V_{N-i}).$$
Using again the integral definition \voldefor, we find
$$\partial_{h_{i-1}} V_i = {1 \over 2} \sum_{p=1}^{i-1}
V_p V_{i-p},$$
so that
$$\sum_i \partial_{\Ge_i}^2 V_{N+1}=-\sum_{p,q \geq 1
\atop p+q \leq N-1} V_p V_q V_{N-p-q}.$$
For $h$ derivatives, we find\foot{ See next section for a detailed
proof}
$$\sum_i  \partial_{h_i}^2 V_{N+1} =
{(2s)^{N-2} (N+1)^{N-3} \over (N-2)!}.$$
Finally, we use
$$ \sum_{p,q \geq 1 \atop p+q \leq N-1} V_p^{(s)} V_q^{(s)}
V_{N-p-q}^{(s)}={3 \over 4}{(2s)^{N-2} (N+1)^{N-6} \over (N-2)!}
[N^2+15N+74],$$
so that $b_{N-2}$ reads
\eqn\gggg{b_{N-2} = {2^{N-2} (N+1)^{N-6} \over 4! (N-2)!} [3 N^4+
17 N^3 + 72 N^2 + 144 N + 266].}

Now comparing this to the values of $a_{N-2}$ that we read off from
\lispol, we conjecture the following relation
\eqn\dddd{a_{N-2}s^{N-2} = b_{N-2}s^{N-2} +{1 \over 4} V_{N-1}^{(s)},}
leading to
\eqn\ffff{a_{N-2}= {2^{N-2}\over 4! (N-2)!} \big[ (N+1)^{N-6} (3 N^4+
17 N^3 + 72 N^2 + 144 N + 266) + 6 (N-1)^{N-4} \big].}
We check that for $N=4,5,6,7,8,9,10$, we get
$a_{N-2}={40 \over 3}$, ${385 \over 6}$, ${2858 \over 9 }$,
${71992 \over 45}$, ${367144 \over 45}$,
${39666608 \over 945}$, ${205427098 \over 945}$.
Note that the above formula fails to reproduce the $N=2,3$ cases.
This, we suspect, is related to the difference in the definition
of the deformation for the boundary case $N=1$.

This leads to the coefficient of $\Gs_{N-2}$ in
$A_{N+1}^{(s)}$ of \listsig\
\eqn\rrrr{ {2^{N-2}\over 4!} \big[ (N+1)^{N-6}
[12N^6+16N^5-25N^4+21N^3+16N^2-100N+114]+6(N-1)^{N-4} \big].}
For $N=4,5,6,7,8,9,10$, we get respectively
$404$, $12481$, $437776$, $17367872$, $773038912$,$38261688576$,
$2088303502080$.

{\bf Computing the polynomials $B_N(s)$}

The computation of the polynomials $B_{N-1}(s)$
resembles very much that of the
volume $V_{N}(s)$ of previous appendix.
The only difference is that the defining relations for the polytope are
decorated with the parameters $h_i$ and $\Ge_i$, $i=1,2,...,N-1$.
For simplicity, let us first compute the deformed volume
$W_N(h_i,\Ge_i)$ defined by the relations
$$ -2 \Ge_k \leq 2 n_k \leq n_{k-1}+n_{k+1} +h_k.$$
The desired volume is
$$V_N^{(s)}(h_i,\Ge_i)=W_N(h_i+2s,\Ge_i).$$
Let us again introduce $\phi_{N-1}(h_i,\Ge_i)=1$ and
$\phi_{N-k-1}(h_i,\Ge_i)$ by the recursion
$$\eqalign{ \phi_{N-1-k}(h_i,\Ge_i;x=n_{N-1-k})&=
\theta(\sum_{j=1}^{k+1} j h_{N-j} +(k+1)n_{N-k}-(k+2)x) \times \cr
&\int_{\sup(-\Ge_{N-k},
2n_{N-1-k}-n_{N-2-k}-h_{N-1-k})}^{{k \over k+1}x+{1 \over k+1}
\sum_{j=1}^k j h_{N-j}}
\ du \ \phi_{N-k}(h_i,\Ge_i;u). \cr}$$
With these definitions, the volume $W$ reads
$$W_N(h_i,\Ge_i)=\phi_0(h_i,\Ge_i;0).$$
Just as in the non--deformed case, the $\phi$'s turn out to have a
simple polynomial form
\eqn\polformphi{\phi_{N-k}(h_i,\Ge_i;x)=\sum_{j=1}^k
p_j^{(k-1)}(x_{N-k}^{(j)}(h_i,\Ge_i;x))
\theta(x_{N-k}^{(j)}(h_i,\Ge_i;x)),}
where the $p_j^{(l)}(u)$, $j=1,2,..,l+1$ are some polynomials of $u$, with
coefficients themselves polynomial in $h_i$ and $\Ge_i$,
and we used some reduced variables
$$ x_{N-k}^{(j)}(h_i,\Ge_i;x)\equiv {1 \over j+1}( \sum_{r=1}^j
r h_{N-k+j-r} -\Ge_{N-k+j}) +{j \over j+1} n_{N-k-1} - x,$$
with the convention that $\Ge_N=0$.
The proof of \polformphi\ is identical to that of the
undeformed case and leads to the recursion relations between the $p$'s
\eqn\recupone{
p_j^{(k)}(x)= \int_{0}^{{j+1 \over j}x} \ du \ p_{j-1}^{(k-1)}(u) \qquad
j=2,3,...,k+1,}
and
\eqn\recuptwo{
\eqalign{p_1^{(k)}(u)&=
\sum_{j=2}^{k+1} p_j^{(k)}\bigg({\sum_{r=1}^{j-1} r h_{N-r}
+j \Ge_{N-k}-\Ge_{N-k+j-1}+(j-1)y \over j+1}\bigg) \cr
&- p_j^{(k)}\bigg({\sum_{r=1}^{j-1} r h_{N-r}
+j \Ge_{N-k}-\Ge_{N-k+j-1}+(j-1)y +2ju \over j+1}\bigg), \cr}}
independent on $y$.

We list the first few deformed $p$'s below

$$\eqalign{
k=0 \ \ p_1^{(0)}(u)&=1 \cr
k=1 \ \ p_2^{(1)}(u)&={3 \over 2} u \cr
p_1^{(1)}(u)&=-2u \cr
k=2 \ \ p_3^{(2)}(u)&={4 \over 3}  u^2 \cr
p_2^{(2)}(u)&=-{9 \over 4} u^2 \cr
p_1^{(2)}(u)&=u^2- [h_{N-1}+2  \Ge_{N-1}- \Ge_{N-2}]u \cr
k=3 \ \ p_4^{(3)}(u)&={125 \over 144} u^3 \cr
p_3^{(3)}(u)&=-{16 \over 9} u^3 \cr
p_2^{(3)}(u)&={9 \over 8} u^3
-{9 \over 8}[h_{N-1}+2  \Ge_{N-1}- \Ge_{N-2}]u^2 \cr
p_1^{(3)}(u)&=-{2 \over 9}u^3
+{1 \over 3} [ h_{N-2}+2  h_{N-1}+3  \Ge_{N-1}-\Ge_{N-3}]u^2 \cr
&-{1 \over 6} [ h_{N-1}^2+ h_{N-2}^2+4  h_{N-1}  h_{N-2}
+ h_{N-2} (6  \Ge_{N-1}-2  \Ge_{N-3})\cr
&+ h_{N-1} (6  \Ge_{N-2}
-4  \Ge_{N-3})+ \Ge_{N-3}^2-3 ( \Ge_{N-1}^2+ \Ge_{N-2}^2)+
12  \Ge_{N-1}  \Ge_{N-2}-6  \Ge_{N-1}  \Ge_{N-3}]u \cr}$$

We recover the undeformed $p$'s (at $s=1$) by taking $h_i=2$, $\forall \ i$.
A few remarks are in order in view of this list.
\item{(i)}The coefficients of the polynomials $p_j^{(k)}$ have a simple
structure
for $j>1$, due to the recursive definition \recupone.
Let us denote by $\Ga_{m}^{(1,k)}$ the coefficients of $p_1^{(k)}$
$$p_1^{(k)}(x)= \sum_{m=1}^k \Ga_{m}^{(1,k)} \ x^m ,$$
where we used the definition \recuptwo, to include the fact
that $p_1^{(k)}(0)=0$. Then the coefficients $\Ga_{m}^{(j,k)}$
of $p_j^{(k)}$ satisfy the recursion
$$ \Ga_{m+1}^{(j,k)}= {1 \over m+1} ({j+1 \over j})^{m+1}
\Ga_{m}^{(j-1,k-1)} \ \ \ m=1,2,...,k-1,$$
which implies
$$\eqalign{ \Ga_m^{(k+1,k)}&= \delta_{m,k}
{(k+2)^k \over (k+1)! k!}\Ga_0^{(1,0)} \cr
\Ga_m^{(j,k)}&= \theta(m-j) \theta(k-m) {(j+1)^{m}(m-j+1)! \over
j! m! 2^{m-j+1}} \Ga_{m-j+1}^{(1,k-j+1)} \ \ j=2,3,...,k.}$$
So we have only to compute the coefficients $\Ga_{m}^{(1,k)}$ of
$p_1^{(k)}$.
For instance, due to $\Ga_0^{(1,0)}=1$,
$\Ga_1^{(1,1)}=-2$, $\Ga_2^{(1,2)}=1$,
$\Ga_1^{(1,2)}=-(h_{N-1}+2 \Ge_{N-1}-\Ge_{N-2})$, we have
$$\eqalign{
j=k+1:\ \ \Ga_k^{(k+1,k)}&={(k+2)^k \over (k+1)! k!}\cr
j=k:\ \ \ \ \ \Ga_k^{(k,k)}&=-{(k+1)^k \over (k!)^2} \cr
j=k-1:\ \ \Ga_{k}^{(k-1,k)}&={(k)^k \over k!(k-1)!2!} \cr
\Ga_{k-1}^{(k-1,k)}&=-{(k)^{k-1} \over 2!((k-1)!)^2}
(h_{N-1}+2 \Ge_{N-1}-\Ge_{N-2}) .\cr}$$
The coefficients of $p_1^{(k)}$ satisfy the following
recursion relation
inherited from the definition \recuptwo\
\eqn\recoefp{
\eqalign{\alpha_{m}^{(1,k)} &= -{2^m (k+1)^m \over
m! (k-m)! (k+1)! } [ \sum_{r=1}^k r h_{N-r} +(k+1) \Ge_{N-k}]^{k-m} \cr
&-\sum_{p=m}^k \sum_{j=2}^p {j^m (p-j+1)! 2^{m-p+j-1} \over
j! (p-m)! m!}\Ga_{p-j+1}^{(1,k-j+1)}
[\sum_{r=1}^{j-1} r h_{N-r}+j \Ge_{N-j+1} -\Ge_{N-k+j-1}]^{p-m}, \cr}}
for $m=1,2,...,k$.
As a consequence, we derive the leading coefficient
$\Ga_k^{(1,k)}$ of $p_1^{(k)}$
\eqn\coelead{\eqalign{
\Ga_k^{(1,k)} &= -{2^k (k+1)^k \over k! (k+1)!} +
\sum_{j=2}^k {j^k (k-j+1)! 2^{j-1} \over k! j! } \Ga_{k-j+1}^{(1,k-j+1)}\cr
&= {(-2)^k \over (k!)^2}, \cr }}
thanks to the identity
\eqn\identuseful{\sum_{j=0}^{k+1} (x-j)^{k} (-1)^j {k+1 \choose j} =0,}
valid for any $x$, taken at $x=0$
$$ \sum_{j=1}^{k} (-1)^j {j^k} {k+1 \choose j} = (-1)^k (k+1)^k .$$

\item{(ii)}The dependence on $h_i$ and $\Ge_i$ of the coefficients
$\Ga_{m}^{(1,k)}$ is very peculiar too.
Actually, the coefficients
$\Ga_m^{(1,k)}$ are functions of $h_{N-1}$, ..., $h_{N-k+1}$ and
$\Ge_{N-1}$, ..., $\Ge_{N-k}$ only, as a consequence of \recoefp.

\noindent{}Finally the volume reads
$$\eqalign{
W_N(h_i,\Ge_i)&=\sum_{j=1}^N p_j^{(N-1)}(x_0^{(j)}(h_i,\Ge_i;0) \cr
&=\sum_{j=1}^N p_j^{(N-1)}\bigg(
{ \sum_{r=1}^{j-1} r h_{j-r} -\Ge_j \over j+1}\bigg). \cr}$$
This provides us with a powerful recursive scheme for
computing the deformed volumes.

As an example let us derive exactly the coefficient of $h_i^{N-1}$,
$i=1,2,...,N-1$,
in $W_N$.
The above remark (i) enables to write the leading coefficient of the $p$'s
as
$$\eqalign{ \Ga_k^{(j,k)}&={(j+1)^{k}(k-j+1)! \over
j! k! 2^{k-j+1}} \Ga_{k-j+1}^{(1,k-j+1)} \cr
&= (-1)^{k-j+1} { (j+1)^k \over j! k! (k-j+1)!}.\cr}$$
The contribution to $W_N$ of any subleading piece, say of degree
$m<k$, of  $p_j^{(k)}$
involves, thanks to remark (ii) a coefficient which is only a function
of $h_{N-1}$,...,$h_{N-k+j-1}$, and which multiplies a power
$$ [\sum_{r=1}^{j-1} r h_{j-r} -\Ge_j]^m, $$
but $N-k+j-1 > j-1$, therefore such a term does not contribute to
any $h_i^{N-1}$ term of $W_N$.
The only contributions come from the leading pieces of the $p$'s
$$ \sum_{j=1}^{N} {\Ga_{N-1}^{(j,N-1)} \over (j+1)^{N-1}}
[\sum_{r=1}^{j-1} r h_{j-r} -\Ge_j]^{N-1}, $$
and we get the coefficient of $h_i^{N-1}$ in $W_N$
$$\eqalign{ \sum_{j=i+1}^N (j-i)^{N-1} \Ga_{N-1}^{(j,N-1)} &=
\sum_{j=i+1}^N {(j-i)^{N-1} (-1)^{N-j} \over j! (N-1)! (N-j)!} \cr
&={1 \over N! (N-1)!} \sum_{j=0}^{N-i-1} (-1)^j (N-i-j)^{N-1}
{N \choose j} \cr
&={1 \over N! (N-1)!} \sum_{j=0}^{i-1} (-1)^j (i-j)^{N-1}
{N \choose j}. \cr}$$
The last identity expresses the symmetry of $W_N$ under the interchange
$h_i \leftrightarrow h_{N-i}$ and is a consequence of the identity
\identuseful, taken at $x=i$ and $k=N-1$
$$\sum_{j=0}^N (i-j)^{N-1} (-1)^j {N \choose j} =0.$$
Finally we have at leading order in the $h_i$'s
$$W_N \simeq \sum_{i=1}^{N-1} {h_i^{N-1} \over N! (N-1)!}
\sum_{j=0}^{i-1} (-1)^j (i-j)^{N-1}{N \choose j}.$$
This leads to the following contribution of the $(N-1)$th
derivatives of $W_N$ w.r.t. the $h$'s
$$\eqalign{ \sum_{i=1}^{N-1} \partial_{h_i}^{N-1} W_N &=
{1 \over N!} \sum_{i=1}^{N-1} \sum_{j=0}^{i-1}
(-1)^j (i-j)^{N-1} {N \choose j}\cr
&={1 \over N}. \cr}$$

Using this recursive method, we computed the first few $B_N$'s
$$\eqalign{
B_0(s)&=1 \cr
B_1(s)&= s+1 \cr
B_2(s)&=2s^2+2s+{19 \over 36} \cr
B_3(s)&={16 \over 3}s^3+{13 \over 2}s^2+{8 \over 3}s+{107 \over 288} \cr
B_4(s)&={50 \over 3}s^4+24s^3+{77 \over 6}s^2+{109 \over 36}s
+{641 \over 2400} \cr
B_5(s)&={288 \over 5}s^5+96 s^4+{377 \over 6}s^3+{1453 \over 72}s^2
+{509 \over 160}s+{51103 \over 259200} \cr
B_6(s)&= {9604 \over 45} s^6+{6076 \over 15}s^5+{5641 \over 18}s^4
+{6821 \over 54}s^3 +{20111 \over 720}s^2 +{8663 \over 2700}s
+{1897879 \over 12700800}. \cr }$$

\listrefs

\bye